\documentclass[12pt]{article}
\pdfoutput=1

\usepackage{amsfonts}
\usepackage{amsmath}
\usepackage{amssymb}
\usepackage{array}
\usepackage{bigints}
\usepackage{booktabs}
\usepackage[nosort]{cite}
\usepackage{cancel}
\usepackage{color}
\usepackage{dsfont}
\usepackage{float}
\usepackage{framed}
\usepackage{graphicx}
\usepackage{indentfirst}
\usepackage{mathrsfs}
\usepackage{multirow}
\usepackage{setspace}
\usepackage{subdepth}
\usepackage{subfig}
\usepackage{titlesec}
\usepackage[dotinlabels]{titletoc}
\usepackage{wrapfig}
\usepackage[all]{xy}
\usepackage{young}
\usepackage[vcentermath]{youngtab}
\usepackage{slashed}
\usepackage{pdflscape}

\usepackage{hyperref}
\hypersetup{colorlinks=true}
\hypersetup{linkcolor=black}
\hypersetup{citecolor=black}
\hypersetup{urlcolor=black}

\numberwithin{equation}{section}

\usepackage[left=2.5cm,right=2.5cm,top=2.5cm,bottom=3cm]{geometry}
\DeclareFontShape{OT1}{cmr}{mx}{n}%
    {<->cmr10}{}
\newcommand{\mytitlefont}{\fontseries{mx}\selectfont}
\DeclareMathAlphabet{\titlemath}{OT1}{cmr}{mx}{n}

	\usepackage[dvipsnames]{xcolor} 
	\usepackage{tikz,tikz-3dplot, pgfplots,tkz-graph}
	\usetikzlibrary{arrows,shapes}
    \usetikzlibrary{arrows.meta}
	\usetikzlibrary[positioning,patterns,decorations.markings] 
	\usepackage[rightcaption]{sidecap}
	\tikzstyle arrowstyle=[scale=1]
	\tikzstyle directed=[postaction={decorate,decoration={markings,
    mark=at position .65 with {\arrow[arrowstyle]{stealth}}}}]
	\tikzstyle reverse directed=[postaction={decorate,decoration={markings,
    mark=at position .65 with {\arrowreversed[arrowstyle]{stealth};}}}]
    
\newcommand{\ba}[1]{\begin{align} #1 \end{align} }

\newcommand{\bs}[1]{\begin{split} #1 \end{split} }

\newcommand{\x}[1]{\textcolor{cyan}{[\bf #1 ]}}

\def\tr{{\text{Tr}}}
\def\det{{\rm det}}

\def\CI{{\cal I}}

\def\CL{{\cal L}}

\def\CN{{\cal N}}
\def\CO{{\cal O}}

\def\CS{{\cal S}}

\def\a{{\alpha}}
\def\b{{\beta}}

\newcommand{\Tr}{\mathrm{Tr}}

\usepackage{mathtools}

\begin{document}


\begin{titlepage}

\begin{center}

~\\[1cm]

{\fontsize{24pt}{0pt} \mytitlefont Decoupling Across Dimensions}

\vskip20pt

Param Gogia, Caius Gordon-Bruno, Emily Nardoni\footnote{enardoni@vassar.edu} and Peter Weck\footnote{pweck@vassar.edu}

\bigskip

{\it  Department of Physics and Astronomy, Vassar College}\\[-2pt]
       {\it 124 Raymond Avenue, Poughkeepsie, New York 12604, USA}\\[4pt]

\end{center}

\bigskip
We present a prescription for identifying decoupled sectors of 4d SCFTs obtained from compactifications of branes. Central to our analysis is the comparison of anomaly polynomials obtained via field-theoretic reduction versus anomaly inflow. Our flagship examples are the 4d $\mathcal{N}=1$ SCFTs obtained by compactifying the 6d $(1,0)$ theory from M5-branes probing a $\mathbb{C}^2/\mathbb{Z}_2$ orbifold singularity on a torus with flux. The anomaly polynomial of the interacting fixed point is known to decompose into the $T^2$ reduction of the 6d anomaly polynomial, plus a contribution from flip fields introduced to remove operators from the spectrum. Adopting the inflow perspective, the contribution from the flipped sector is obtained as the difference between inflow anomaly polynomials in the singular and resolved geometries, which we efficiently compute using equivariant localization. Helpfully, the 4d SCFTs admit two dual quiver gauge theory descriptions: $Y^{p,q}$ quiver gauge theories, and class $\mathcal{S}_2$ necklace quivers obtained from trinions. We study the interacting and decoupled sectors on both sides of the duality and construct an explicit map between their chiral rings, identifying a range of parameter space for which the quivers flow to apparently different interacting fixed points. Each chiral ring generator is furthermore matched to an M2-brane wrapping a 2-cycle in the resolved M-theory geometry. 

\vfill


\end{titlepage}


\setcounter{tocdepth}{2} 
\tableofcontents

\newpage


\section{Introduction and Main Results}
\label{sec:intro}


Many interesting Quantum Field Theories (QFTs) are obtained by taking a limit of string theory on a singular space, or on a space with boundary. Generally, matter in the QFT of interest originates from degrees of freedom at special loci of the compactification, such as marked points carrying defect data (as in punctures on the internal space), or singularities probed by the branes.  
Free sectors frequently decouple in the flow to the infrared (IR), leaving an interacting  sector describing the QFT of interest. The interplay  between decoupling and compactification is often intricate, and it is useful  to compare various points of view when available---from  geometric brane engineering, known gauge theoretic descriptions of the low-energy field theories, holographic solutions, {\it etc.}---to determine the physics of the interacting fixed point.

't Hooft anomalies are key observables that capture information about the symmetries and degrees of freedom of a QFT.  Anomalies are invariant under renormalization group flow, and their structure reflects the QFT's geometric origin. 
When the brane system flows to a superconformal fixed point, the 't Hooft anomalies determine the central charges of the low-energy superconformal field theory (SCFT) and dimensions of chiral primary operators \cite{Anselmi:1997am,Kuzenko:1999pi}. 
In the following discussion we will focus  on zero-form continuous symmetries, whose perturbative chiral anomalies are captured by an anomaly polynomial $I_{d+2}$ for $d$ the spacetime dimensionality of the QFT \cite{Alvarez-Gaume:1983ihn,Alvarez-Gaume:1984zlq,Bardeen:1984pm}.\footnote{~It is interesting to 
 extend the discussion to the $(d+1)$-dimensional anomaly field theory in order to track discrete symmetries, or more generally to the $(d+1)$-dimensional  symmetry topological field theory (Symmetry TFT) in order to track generalized symmetry structure, as these objects similarly arise from reducing the topological terms in the effective string theory actions.}

 In this paper, we examine how comparing different viewpoints on obtaining the anomaly polynomial $I_{d+2}^{\text{SCFT}}$ of the lower-dimensional QFT can teach us new physics about the strongly-interacting fixed point. 
 These comparisons are especially interesting at finite $N$, and when degrees of freedom decouple on the flow to the IR. As we explain, combining multiple perspectives can allow one to  isolate the decoupled versus interacting degrees of freedom, and  make predictions about the fixed point that were not accessible from one perspective alone.

For definiteness, we restrict our discussion to the arena of 4d superconformal field theories arising from 6d theories compactified on a 2d surface $\Sigma$, {\it i.e.} $W_6\simeq \Sigma \times W_4$, although we expect that the lessons apply more generally to other compactifications. 
When the system has a UV description in terms of M5-branes, the six-dimensional internal space relative to the 4d worldvolume is a fibration,
\ba{
\label{fib1}
M_4\hookrightarrow M_6 \rightarrow \Sigma\,,\qquad M_{10}\simeq M_4\times W_6
}
where in this expression 11d spacetime splits into the radial direction away from the branes times the 10-dimensional spacetime $M_{10}$.
Many interesting examples are of this type, including: the 4d theories of class $\CS$, obtained by wrapping flat M5-branes on a punctured Riemann surface $\Sigma_{g,n}$ \cite{Gaiotto:2009we,Gaiotto:2009hg}; class $\CS_\Gamma$, in which the branes are taken to probe a $\mathbb{C}^2/\Gamma$ singularity so that $M_4\simeq S^4/\Gamma$ \cite{Gaiotto:2015usa,Razamat:2016dpl,Bah:2017gph,DelZotto:2014hpa,Kim:2018lfo}; or M5-branes wrapped on spindles, where $\Sigma$ is taken to be a two-sphere with deficit angles at the poles \cite{Ferrero:2020laf,Ferrero:2021wvk}. 

We will present and compare the following complementary perspectives on the anomalies of the 4d fixed point, captured by an anomaly polynomial $I_6^{\text{SCFT}}$: first, from reducing the anomalies of the 6d SCFT on the compact surface $\Sigma$; and second, from anomaly inflow of the brane system, evaluated in both the singular and resolved backgrounds, when $M_4$ has singularities. As we explain, the contributions of degrees of freedom coming from special loci on $M_6$ ({\it e.g.}~punctures or orbifold points on $\Sigma$, or orbifold points on the fiber $M_4$) can be isolated by comparing these different perspectives with one another. Furthermore, we will explain how, in a large class of examples, modes which decouple from the IR fixed point can be isolated by subtracting the anomalies computed from these different starting points. 

The rest of this introduction proceeds as follows. First in Section~\ref{sec:complementary}, we introduce the different perspectives on the 4d anomalies which naturally lead to different decompositions \eqref{naiveloc} and \eqref{inflow}, and  describe  the mileage that one can gain by comparing these different perspectives in general. In Section~\ref{sec:flagship} we break down how these ideas play out in the 4d $\CN=1$ theories obtained from class $\CS_k \equiv \CS_{\Gamma = \mathbb{Z}_k}$ on the torus $\Sigma = T^2$ \cite{Gaiotto:2015usa,Razamat:2016dpl,Bah:2017gph,Kim:2018lfo}, which for $k=2$ will constitute our primary subject of analysis in the rest of the paper. The benefit of this particular example is that there are two separate quiver gauge theories known to flow to the IR fixed points, which allow a detailed analysis of the origins of different terms in \eqref{naiveloc} and \eqref{inflow}. As we emphasize, it will be  interesting to extend to a detailed analysis of other classes of examples in which such quiver gauge theory descriptions are not at our disposal, in which the lessons obtained from these different perspectives lead to new predictions.


\subsection{Complementary perspectives on anomalies}
\label{sec:complementary}

 \paragraph{Perspective 1: Reduction from 6d on a compact surface}

\begin{figure}[t]
\centering
\begin{tikzpicture}[x=1.1cm, y=1.1cm, thick,
  arr/.style={-{Stealth[length=7pt,width=6pt]}}]

  \definecolor{boxviolet}{RGB}{ 96, 32,120}
  \definecolor{boxred}   {RGB}{196,  0,  0}
  \colorlet{fillviolet}{violet!15}
  \colorlet{fillred}   {red!15}

  \path (2.1,-1.7)  ++(90:1)  coordinate (Ltop);
  \path (2.1,-1.7)  ++(270:1) coordinate (Lbot);
  \path (5.9,-1.75) ++(72:1)  coordinate (Rlens);
  \path (5.9,-1.75) ++(108:1) coordinate (Rleft);
  \path (5.9,-1.75) ++(252:1) coordinate (Rbotl);
  \path (6.1,-3.75) ++(0.35,0.23) coordinate (Rcross);
  \coordinate (Rtopbigon) at (5.9,-0.75);   

  \tikzset{torus/.pic={
    \draw (0,0) ellipse (0.95 and 0.52);
    \draw (-0.6,0.05) .. controls (-0.3,-0.25) and (0.3,-0.25) .. (0.6,0.05);
    \draw (-0.43,-0.06) .. controls (-0.25,0.12) and (0.25,0.12) .. (0.43,-0.06);
    \draw[boxred] (0.35,0.23) ++(-0.07,-0.07) -- ++(0.14,0.14)
                  (0.35,0.23) ++(-0.07,0.07)  -- ++(0.14,-0.14);
    \draw[boxred] (-0.58,-0.2) ++(-0.07,-0.07) -- ++(0.14,0.14)
                  (-0.58,-0.2) ++(-0.07,0.07)  -- ++(0.14,-0.14);
  }}

  \begin{scope}[shift={(2.1,-1.7)}]
    \draw (0,0) circle (1);
    \node at (-0.1,0.05) {$I_6^{\text{bulk}}$};
    \fill[boxviolet] (90:1)  circle (2.6pt);
    \fill[boxviolet] (270:1) circle (2.6pt);
  \end{scope}
  \node at (0.7,-1.0) {$M_4^{\text{sing.}}$};

  \pic at (2.5,-3.75) {torus};
  \node at (1.2,-3.45) {$\Sigma$};
  \draw[dashed] (Lbot) -- (1.87,-3.5);

  \draw[arr] (3.55,-1.9) -- (4.45,-1.9);

  \node at (4.55,-1.0) {$M_4^{\text{res.}}$};

  \begin{scope}[shift={(5.9,-1.75)}]
    \draw (108:1) arc[start angle=108, end angle=252, radius=1];
    \draw (288:1) arc[start angle=288, end angle=432, radius=1];
    \node at (-0.1,0.05) {$I_6^{\text{bulk}}$};

    \draw[boxred] (108:1) to[out=60, in=120]   (72:1);
    \draw[boxred] (108:1) to[out=-60,in=-120]  (72:1);
    \draw[boxred] (252:1) to[out=60, in=120]   (288:1);
    \draw[boxred] (252:1) to[out=-60,in=-120]  (288:1);

    \foreach \a in {72,108,252,288} \fill[boxviolet] (\a:1) circle (1.8pt);
  \end{scope}

  \pic at (6.1,-3.75) {torus};
  \node at (4.8,-3.45) {$\Sigma$};
  \draw[dashed] (Rbotl) -- (5.4,-3.55);

  \node[anchor=west] (rlab) at (7.25,-1.55)
        {\colorbox{fillred}{$\delta I_6^{\text{loc},i}$}};

  \draw[boxred,arr,shorten >=8pt,shorten <=1pt]
        (rlab.north) to[out=100, in=90] (Rtopbigon);

  \draw[boxred,arr,shorten >=4pt,shorten <=1pt]
        (rlab.south) to[out=-100,in=0] (Rcross);

  \node[anchor=west] (vlab) at (2.75,-0.3)
        {\colorbox{fillviolet}{$I_6^{\text{loc-inher,i}}$}};
  \draw[boxviolet,arr,shorten >=5pt,shorten <=1pt]
        (vlab.east) to[out=-20,in=160] (Rleft);

\end{tikzpicture}
\caption{Schematic representation of which features of the geometry contribute to which terms in the splitting of $I_6^\text{SCFT}$ in \eqref{naiveloc}. The left half represents an internal space with potential orbifold singularities, and the right half represents a resolution of this geometry.
\label{fig:I_6 geometry}}\end{figure}
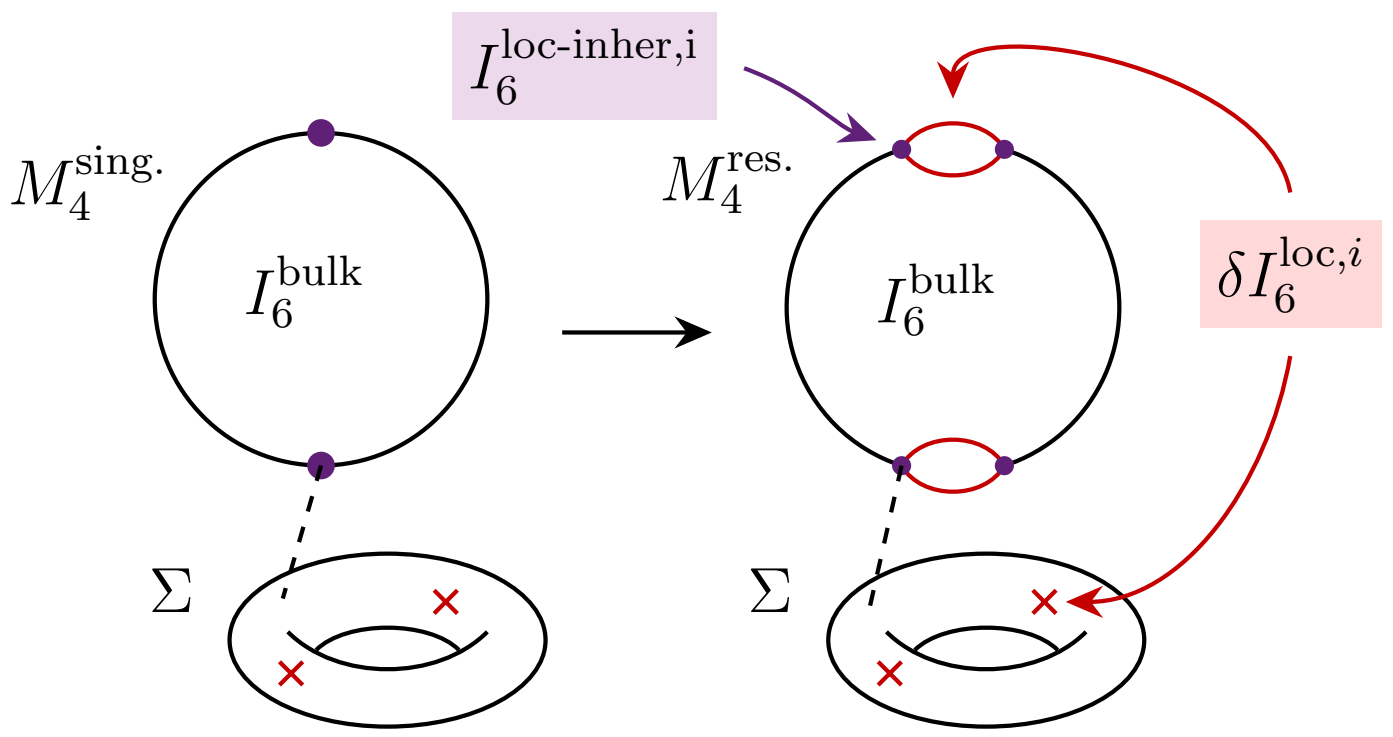

Given a 4d QFT defined as in  \eqref{fib1}, one reduces the anomalies of the interacting 6d SCFT, captured by an anomaly polynomial $I_8$, over the compact surface $\Sigma$, as $\int_\Sigma I_8$ \cite{Benini:2009mz,Alday:2009qq}.  
The contributions
$\int_\Sigma I_8$  furnish a sector of anomalies inherited by the 4d theory, and so we refer to these contributions as the {\it inherited} terms. We can parameterize the different contributions to $\int_\Sigma I_8$, as well as the difference between $I_6^{\text{SCFT}}$ and $\int_\Sigma I_8$, as follows,\footnote{~Throughout, $I_8$ denotes the anomaly polynomial of the interacting 6d SCFT, with the center-of-mass sector already removed.} 
\ba{
\label{naiveloc}
\bs{
I_6^{\text{SCFT}} = & \int_\Sigma I_8 + \sum_i  
\colorbox{red!15}{ 
$\delta I_6^{\,\text{loc},i}$
} \,, \\ \text{with}\quad 
&\int_\Sigma I_8 =  \, \chi(\Sigma)\,  I_6^{\text{bulk}} + \sum_i
\colorbox{violet!15}{ 
$ I_6^{\,\text{loc-inher},i}$
}
}}
Let us discuss each of these contributions in turn. (The colors match those in  Figure~\ref{fig:I_6 geometry}.)  

The first line of \eqref{naiveloc} parameterizes the difference between the inherited terms $\int_\Sigma I_8$ and the rest of the 4d fixed point's anomalies as a sum over contributions $\delta I_6^{\,\text{loc},i}$ (boxed in red), which may be attributed to degrees of freedom supported at special loci of the compactification.  This decomposition is in spirit a \emph{dynamical} split: it separates out the term $ \int_\Sigma I_8 $ inherited from the 6d theory (the 6d UV), from  corrections  which are not counted by the UV-inherited terms, but are nonetheless part of the interacting IR fixed point. In this sense it is the decomposition naturally suited to identifying decoupling, as we will see in examples.

This dynamical split  is distinct from the 
\emph{geometric} split  into a contribution from the smooth bulk of $\Sigma$---the part
proportional to the Euler characteristic
 $\chi(\Sigma)$---and contributions from the special loci. This geometric organization of the inherited terms is made explicit in the second line of \eqref{naiveloc}: the surface integral over $I_8$ picks out one ``bulk'' term whose integral is proportional to $\chi(\Sigma)$, and which survives due to the topological twist with the R-symmetry  (where we've pulled out the overall factor of $\chi(\Sigma)$ from $I_6^{\text{bulk}}$). Then, there are  possibly additional terms linear in fluxes that thread the surface $\Sigma$ that are attached to a special locus, denoted $I_6^{\,\text{loc-inher},i}$ (boxed in purple).\footnote{~To be clear, the label
``loc'' here is a flux/symmetry attribution, not a  statement about the
fixed-point theorem evaluation to be discussed in Section~\ref{sec:inflow}.}

The distinction is best illustrated by examples. For flat M5-branes wrapped on a Riemann surface $\Sigma$ with regular punctures (``standard'' class $\CS$ \cite{Gaiotto:2009we,Gaiotto:2009hg,Gaiotto:2009gz}), the two decompositions coincide: the terms $I_6^{\,\text{loc-inher},i}$ are zero, and the inherited term $\int_\Sigma I_8$ from the parent 6d (2,0) theory {\it is} the bulk piece. 
The corrections $\delta I_6^{\,\text{loc},i}$ are precisely the anomalies due to degrees of freedom localized at the punctures, so that the sum over $i$ is a sum over the punctures. This bulk-plus-punctures structure is understood from a variety of points of view, including from anomaly inflow \cite{Bah:2018gwc,Bah:2018jrv,Bah:2019jts}, and from S-duality and anomaly matching arguments \cite{Gaiotto:2009we,Gaiotto:2009gz,Chacaltana:2010ks,Chacaltana:2012zy,Tachikawa:2015bga}. The latter method  involves identifying  different pair-of-pants decompositions of the surface $\Sigma$ with different S-dual presentations of the 4d SCFT, which arise by gauging a product of basic geometric building blocks (the 3-punctured trinions in class $\CS$).
Given a quiver gauge theory description of these basic building blocks, one can arrive at theories with various numbers of punctures and different genus by gluing (gauging subgroups of flavor symmetries) and closing punctures (Higgsing flavor symmetries), utilizing the quivers for a field-theoretic perspective on the fixed points. This ``tinkertoys'' point of view has been extremely  successful in obtaining properties of the 4d fixed points, and in particular deriving the anomalies $I_6^{\text{SCFT}}$. This perspective also makes it clear that the inherited terms $\int_\Sigma I_8$  satisfy clear geometric consistency conditions; for instance, they must be invariant under different pair-of-pants decompositions that lead to the same $\Sigma$. 

By contrast, one can consider the class $\mathcal{S}_k$ theories corresponding to M5-branes probing a $\mathbb{C}^2/\mathbb{Z}_k$ singularity, leading to parent 6d (1,0) SCFTs,  wrapped on  a Riemann surface $\Sigma$ with fluxes $N_{N_i},N_{S_i}$ threading $\Sigma$ for the $U(1)$ symmetries supported at the north and south poles of the fiber $M_4=S^4/\mathbb{Z}_k$ \cite{Gaiotto:2015usa,Razamat:2016dpl,Bah:2017gph,Kim:2018lfo}. In this case there are  nonzero contributions to both $I_6^{\,\text{loc-inher},i}$ {and} $\delta I_6^{\,\text{loc},i}$ that are proportional to the fluxes $N_{N_i},N_{S_i}$ \cite{Bah:2017gph}, in addition to terms localized at regular punctures on $\Sigma$ which live in  $\delta I_6^{\,\text{loc},i}$ (as in the discussion for flat M5-branes above). In these examples the dynamical and geometric decompositions no longer coincide: at each pole of the fiber, the
inherited flux terms $I_6^{\,\text{loc-inher},i}$  and the corrections  $\delta I_6^{\,\text{loc},i}$ are supported at the \emph{same} locus. The contrast is especially stark on  the torus, where $\chi(T^2)$ vanishes and the geometric split degenerates entirely. This is the situation our flagship example below is designed to dissect.


\paragraph{Perspective 2: Anomaly inflow}

A top-down derivation of the anomalies of the interacting fixed point may be obtained from anomaly inflow. In the full string theory, gauge transformations and diffeomorphisms restricted to the brane worldvolume induce global symmetries for the QFT which might be anomalous. The variation of the effective action of the gravitational theory in the presence of the brane sources must cancel the anomalies of the worldvolume theory, via anomaly inflow \cite{Callan:1984sa}. 

In the context of 4d field theories which arise from M-theory, the anomaly polynomial of the 4d SCFT is extracted by integrating a 12-form $\CI_{12}$, related by descent to the topological terms in the M-theory action, over the  manifold $M_6$ internal to the 4d spacetime \cite{Duff:1995wd,Witten:1996hc,Freed:1998tg,Harvey:1998bx,Bah:2019rgq} (with generalization to other string theories and dimensions): 
\ba{
\label{inflow}
 I_6^{\text{inflow}} = \int_{M_6} \CI_{12} = - I_6^{\text{SCFT}}-I_6^{\text{decouple}}
}
The second equality is the requirement that the resulting $I_6^{\text{inflow}}$ cancels against the interacting SCFT anomalies $I_6^{\text{SCFT}}$ up to decoupling modes---modes retained by the inflow geometry  that decouple from the IR fixed point, including the center of mass mode of the branes, and possibly modes arising from free 6d vector and tensor multiplets.


We will now examine how the inherited versus localized decomposition of  \eqref{naiveloc} is captured in the reduction of the topological terms in the string theory action in the presence of the special loci. A key point to emphasize is that the answer to this question (naturally) depends on the background in which  the inflow computation is done. 
When the internal space  \eqref{fib1} has orbifold loci, there are two natural presentations of the same flux background: the singular $M_6^{\text{sing}}$, where the loci carry non-abelian localized sources; and the resolved $M_6^{\text{res}}$, where the singularity is blown up and  the flux is geometrized on the resolution cycles. In the example of class $\CS_k$ wrapped on $T^2$ with flux, we establish the following
picture. The two evaluations give,
\ba{
\label{t2split}
\text{class }\CS_k \text{ on }T^2:\qquad
I_6^{\text{inf}}\big|_{\text{sing}} = -\int_{T^2} I_8\,, \qquad
I_6^{\text{inf}}\big|_{\text{res}} = -\,I_6^{\text{SCFT}}
}
where we denote by $I_6^{\text{inf}}\big|_{\text{sing}}$ the evaluation of the inflow integral \eqref{inflow} in the singular background, $\int_{M_6^{\text{sing}}}\CI_{12}$, and similarly for the resolved quantities, where the computation in the resolved background first appeared in \cite{Bah:2022a,Bah:2022b}.
Comparing with \eqref{inflow}, we see that the two presentations realize the same
interacting SCFT but with \emph{different} decoupled sectors.\footnote{~Below we will identify the decoupled modes of the singular presentation as  the free flip singlets of the
field-theoretic construction.}
The difference of the two evaluations is identified as
precisely the localized correction of \eqref{naiveloc},\footnote{~To be clear, this difference is attributable entirely to the change of integrand, {\it i.e.}\ to the localized states counted in one presentation and not the other.}
\ba{
\label{deltai}
\Big(-I_6^{\text{inf}}\big|_{\text{res}}\Big) -
\Big(-I_6^{\text{inf}}\big|_{\text{sing}}\Big) = \sum_i \colorbox{red!15}{ 
$\delta I_6^{\,\text{loc},i}$}
}
This equation has a clear physics interpretation: the terms $\delta I_6^{\,\text{loc},i}$ that were missing from the resolved presentation are precisely the contributions of the M2-branes wrapping the new cycles from the blow up, the W-bosons that were massless in the singular presentation and massive in the resolved one. These terms were first reproduced from wrapped M2-branes in \cite{Bah:2022b}, as reviewed in Section \ref{sec:M2}. Furthermore, \eqref{deltai} has an important consequence for decoupling: we see that  we can isolate the modes that decouple from the IR fixed point in the dynamical split~\eqref{naiveloc} by performing the inflow calculation in different backgrounds.  

We expect a similar dichotomy to \eqref{t2split} for general class $\CS_\Gamma$ and general $\Sigma$,
with the following modification: the resolved presentation acquires a nonvanishing decoupled
sector, $I_6^{\text{decouple}}|_{\text{res}} = I_6^{v,t}$, the reduction of the free vector and tensor multiplets of the
resolved phase, with anomalies proportional to $\chi(\Sigma)$ \cite{Bah:2022a,Bah:2022b}.\footnote{~The expectation \eqref{t2split} that inflow in the singular background yields only the inherited terms  $-\int_\Sigma I_8$ rests on flat-brane inflow established for  $\Gamma=$ ADE in \cite{Ohmori:2014kda}, whose extension to the twisted compactification has not been carried out; it would be interesting to clarify this point as a future direction.}


\paragraph{Method of inflow evaluation: equivariant localization}

Next, we turn to the method of evaluation of the inflow integral \eqref{inflow}. In practice, such integrals have often been done by constructing explicit representatives for the various equivariant classes which appear in the integrand $\CI_{12}$ ({\it e.g.} as in our Appendix~\ref{sec:2form}), and then integrating along the internal $M_6$ directions.
Recent results in the context of supergravity emphasized that {\it equivariant localization} can be a powerful tool in performing integrals over equivariantly closed integrands \cite{BenettiGenolini:2023kxp} (also see the review \cite{Sparks:2026taw}). 
In recent years equivariant localization has been used to integrate anomaly polynomials \cite{Hosseini:2020vgl,Martelli:2023oqk, Faedo:2022rqx,BenettiGenolini:2023yfe}, with \cite{Hristov:2026uxs}  considering the generalization to inflow actions in M-theory (see also \cite{BenettiGenolini:2023ndb, Couzens:2025nxw} for more holographic applications, and \cite{Zhang:2026eir} for applications involving orbifolds).
In this work we extend this method to systems involving wrapped M5-branes probing $\mathbb{C}^2/\mathbb{Z}_k$ singularities. 

In our context, equivariant localization amounts to a particularly nice organization of the integral \eqref{inflow}, which computes $I_6^{\text{inflow}}$ as a sum over the fixed loci via a fixed point theorem. 
The idea  is that when the internal space $M_6$ is acted on by a compact torus, the equivariant integration of the anomaly polynomial reduces to a sum over contributions localized at fixed points of the torus action, by the Berline-Vergne-Atiyah-Bott (BVAB) theorem \cite{BerlineVergne:1982,Atiyah:1984px}. 
For the internal manifold $M_6$ given by the fibration $M_4\hookrightarrow M_6 \to \Sigma$ of \eqref{fib1}, schematically,
\ba{
\label{eqsplit}
I_6^{\text{inflow}} =  \int_{M_6} \CI_{12} = \sum_a   \left[\text{locus of }M_4 \right]_a + \sum_p \left[ \text{locus of } \Sigma\right]_p
}
The fixed locus of $M_6$ has two types of terms. The terms counted by the sum over $a$ are  codimension-4 fixed points of $M_4$  fibered over the whole  base $\Sigma$, for instance the two orbifold points at the north and south poles of $M_4=S^4/\mathbb{Z}_k$. The terms counted by the sum over $p$ are codimension-6 isolated fixed points  over  base points in $\Sigma$, such as regular punctures or the spindle poles.

We emphasize that the decomposition \eqref{eqsplit} that arising from the equivariantly-localized inflow integral is distinct from both the geometric and dynamical splits summarized in \eqref{naiveloc}, as well as the desired split into interacting and decoupled modes in \eqref{inflow}. The terms discussed previously fit into \eqref{eqsplit} as follows. 
Each term in $\sum_a$ contributes through two channels: the twisting of the normal bundle over $\Sigma$ supplies terms proportional to $\chi(\Sigma)$, call it the twist channel, which encompasses both the bulk term $\chi(\Sigma)\, I_6^{\rm bulk}$ in \eqref{naiveloc}, and the decoupled modes $I_6^{\rm decouple}$ in \eqref{inflow}; additionally, there can be contributions linear in  fluxes attached to the fixed locus, call these the flux channel.    
For  class $\mathcal{S}_k$, the orbifold poles of the fiber each contributes through {both} channels; its twist channel feeds into
$\chi(\Sigma)\, I_6^{\rm bulk}$ and $I_6^{\text{decouple}}$, while its flux channel
counts both $I_6^{\text{loc-inher},i} + \delta
I_6^{\,\text{loc},i}$ (where whether $\delta
I_6^{\,\text{loc},i}$ is nonzero depends on whether the integral is evaluated in the resolved versus singular geometry, as in \eqref{deltai}). Punctures on $\Sigma$ contribute terms $\delta I_6^{\text{loc},i}$ in \eqref{naiveloc}, counted in the sum over $p$ in \eqref{eqsplit}.  Note that on the smooth torus, the sum $\sum_p$ is empty, and furthermore the
twist channel vanishes due to $\chi(T^2)=0$, so that the entire computation collapses into the flux channel of the $\sum_a$ terms.


What have we gained from comparing the inflow perspective with our previous discussion? The IR fate of the localized degrees of freedom captured by the terms  $\delta I_6^{{\rm loc},i}$ in \eqref{naiveloc} is invisible to the naive reduction of the 6d anomalies on the compact space, $\int_\Sigma I_8$, which as we saw in \eqref{t2split} is obtained from the inflow integral evaluated in the singular presentation. On the other hand, we have seen that the evaluation of  $I_6^{\text{inflow}}$ in the resolved presentation cannot separate $I_6^{\text{loc-inher},i}$ from $\delta I_6^{{\rm loc},i}$---they come packaged together in the flux channel of \eqref{eqsplit}. 
It is only by comparing the two that we can isolate the physics of the localized degrees of freedom captured by $\delta I_6^{{\rm loc},i}$, and learn about the fate of low-energy decoupled sectors:  the difference between $-I_6^{\text{inflow}}|_{\text{res}}$ and $\int_\Sigma I_8$ predicts the charges and multiplicities of the decoupled sector. For compactifications with no known field-theory dual, this is a promising approach towards gaining insight into the physics of the interacting theory.


\subsection{Flagship example: class $\mathcal{S}_{k=2}$ on $T^2$}
\label{sec:flagship}

Let us illustrate these ideas in the flagship example of this paper: the 4d $\CN=1$ SCFTs of class $\CS_{k=2}$ on a torus with flux, with the topology of the internal space 
 encoded by a 6d space $M_6$ as in \eqref{fib1}, comprising the fiber $M_4=S^4/\mathbb{Z}_2$ over the torus.  In a large range of parameter space, these field theories are dual to the $Y^{p,q}$ SCFTs arising from D3-branes on $Y^{p,q}$ spaces \cite{Gauntlett:2004yd,Martelli:2004wu,Benvenuti:2004dy}. While much of the structure we point out below extends to general $k$,  restricting to $k=2$ and $\Sigma=T^2$ allows an in-depth analysis of known quivers on the class $\mathcal{S}_k$ side \cite{Gaiotto:2015usa,Razamat:2016dpl,Bah:2017gph}, and  to utilize and compare with the well-studied dual  $Y^{p,q}$ frames. Below we summarize the main ideas;  details will appear in the main body of the paper.

\paragraph{Class $\CS_2$ anomalies from M5-branes} 
The north and south poles of the $S^4$ are fixed points of the orbifold action, with each pole locally a $\mathbb{C}^2/\mathbb{Z}_2$ singularity, leading to symmetries $SU(2)_N\times SU(2)_S$ of the parent 6d (1,0) theory before compactification on $T^2$. We  consider the resolved geometry obtained by blowing up the fixed points of the orbifold action, with the $SU(2)_{N,S}$ symmetry at each pole broken to its Cartan $U(1)_{N,S}$, supporting fluxes $N_{N,S}$.\footnote{~In Section~\ref{sec:m5intro} we will instead use the flux parameters $N_{\beta,\gamma}$ of the 4d symmetries $U(1)_{\beta,\gamma}$, related as $N_N=2N_\beta$, $N_S=2N_\gamma$.}

A quiver for the interacting SCFT can be obtained from the trinion with two maximal punctures and one minimal puncture as a basic building block (see Figure~\ref{fig:trinion}). In order to obtain the theory with fluxes $N_N$ and $N_S$, one must glue  together some number $2\ell$ ($\ell=1,2,\dots$) of these basic building blocks by gauging subgroups of the maximal puncture flavor symmetries, leaving a geometry with $2\ell$ minimal punctures \cite{Gaiotto:2015usa,Razamat:2016dpl, Bah:2017gph}. Then one closes each  minimal puncture by giving a VEV to a baryon charged under the puncture's $U(1)$ symmetry. Each closure produces one
baryon $B$ which had been charged under the  $U(1)$ puncture symmetry being Higgsed, requiring the introduction of $1$ corresponding flip singlet $\phi_B$ which couples via the superpotential $W=\phi_B B$, and decouples (``flips'') the baryon. 
The choices of which baryons are used to accomplish the closures are mapped to  values of the flux: there are $\beta$-closures and $\gamma$-closures, where the number of closures of different types $n_{\beta,\gamma}$ are related to the flux parameters as,
\ba{
\label{flux1}
n_\beta = N_N\,,\qquad n_\gamma = N_S\,,\qquad n_\beta + n_\gamma = 2\ell 
}
The result is a ``necklace" quiver whose corresponding geometry is a torus with no punctures, and $N_N,N_S$ flux. Aspects of these constructions were studied in \cite{Razamat:2016dpl,Bah:2017gph}.

The anomalies of the interacting fixed point decompose as follows, 
\ba{
\label{flagship1}
I_6^{\text{SCFT}}= \int_{T^2} I_8 \colorbox{red!15}{ $ - I_6^{\text{flip}}$}
}
The inherited term  $\int_{T^2} I_8$ was computed in \cite{Bah:2017gph}, while the terms $I_6^{\text{flip}}$ were identified in \cite{Bah:2022b} as originating from M2-branes wrapping the resolution cycles at the poles. The color coding follows \eqref{naiveloc} (also see Figure~\ref{fig:I_6 geometry}). As we explain below (and as was noted in \cite{Bah:2022b}),
 $I_6^{\text{flip}}$  are the anomalies of the flip singlets $\phi_B$, equal to precisely minus the anomalies of the baryons $B$ flipped in the construction of the necklace quiver.\footnote{\label{fn:flipp}~The anomalies of  flip fields $\phi_\CO$ introduced to remove operators $\CO$ from the chiral ring via $W=\phi_\CO \CO$ satisfy $I_6^{\text{flip}}=-I_6^{\CO}$.} 
 The computation of $I_6^{\text{inflow}}$ in the resolved background was carried out by 
  \cite{Bah:2022a,Bah:2022b}, with the following result, 
\ba{
\label{mains2}
I_6^{\text{SCFT}}= - I_6^{\text{inflow}}\big|_{\text{res}} = \int_{T^2} I_8  - I_6^{\text{flip}}
}
Comparing with \eqref{inflow}, we confirm that $I_6^{\text{decouple}}|_{\text{res}}=0$ on the torus, and the inflow integral in the resolved presentation lands directly on the
interacting anomaly.\footnote{~The fluxes $N_N=2$, $N_S=0$ are a special case, see Footnote~\ref{fn:pq1}.}

Let us now  explain how to understand \eqref{flagship1} from the perspective of the 6d reduction.
As we have stated, $I_6^{\text{flip}}$ are the anomalies of the flip-fields $\phi_B$ introduced to flip the $B$ baryons when closing  punctures to obtain the smooth torus geometries. There are two types of terms depending on the method of closure ($\beta$ vs $\gamma$):
\ba{
\label{split1}
- I_6^{\text{flip}}  = n_\beta \, I_6 (B_\beta) + n_\gamma \, I_6(B_\gamma)
}
It is natural to write the full object that the naive field theory reduction computes as,
\ba{
\label{t22}
\int_{T^2} I_8 = I_6^{\text{SCFT}} + I_6^{\text{flip}} 
} 
The flip fields are crucial properties of the map from the 6d theory on a torus with flux to the well-defined 4d theory: they are needed to have a well-defined puncture-closing operation, and they are needed for the S-duality invariant structure of $\int_{T^2} I_8$. In particular, without the contribution of $I_6^{\text{flip}}$,  \eqref{t22} would not be a function of only geometric data that is compatible with different S-dual frames---as we have emphasized,  different gluings that yield the same surface $T^2$ and same flux must have the same anomalies.

Nonetheless, at large $N$ the flip fields are generally irrelevant in the IR.\footnote{~If one were to insist on keeping the flip couplings relevant and removing the baryons from the interacting ring, the central charges would scale as $N^{3/2}$ for large $N$, a signal that no such interacting fixed point exists for generic fluxes.} A flip coupling $\phi_B B$ is irrelevant when $R_*(B)>\frac43$, which for the $\beta$-type baryons, holds for all $N\geq 3$. For the $\gamma$-type baryons, irrelevance of the flip couplings requires larger $N$ when $N_S\ll N_N$, since $R_*(B_\gamma)/N\to 0$ as $N_S/N_N\to 0$. 
When the flip couplings are irrelevant, the flip fields $\phi_B$ decouple as free singlets, and the baryons $B$ remain in the interacting chiral ring---the equation \eqref{flagship1} captures this perfectly. 
It is clear why this example is a good case study in why the split \eqref{naiveloc} is dynamical in nature: the flips are necessary in the UV, where the relation \eqref{t22} is forced by the geometric consistency of the object $\int_{T^2} I_8$, but irrelevant in the IR, where  the flip fields decouple as free fields, and the baryons $B$ remain.  

On the other hand, there is a range of parameter space $(N,N_N,N_S)$ for which the $\gamma$-flip couplings are relevant, including a range in which the $B_\gamma$-baryon can be argued to decouple as a free field. As we discuss in  detail in Section~\ref{sec:mainresults}, in such cases the identification of the chiral ring of the interacting fixed point is more subtle.

 \paragraph{A dual description from D3-branes} 
 
 These 4d $\CN=1$ SCFTs have a dual description  arising from $N$ D3-branes on $\mathbb{R}^{1,3}\times C(X)$, for $C(X)$ the toric Calabi-Yau cone over the Sasaki-Einstein manifold $X_5=S^2\times S^3$, labeled by two integers $0\leq q\leq p$ and dubbed $Y^{p,q}$ spaces \cite{Gauntlett:2004yd}. The gauge theory on the D3-branes is described by a circular quiver with $2p$ nodes, and a set of bifundamental fields $(U,V,Y,Z)$ connecting the various nodes (as we will review in Section~\ref{sec:ypq}). These are dual at large $N$ to supergravity solutions of the form $AdS_5\times X_5$ with $N$ units of self-dual five-form flux through $X_5$. 

The $Y^{p,q}$ quiver theories are well studied, and in particular their anomalies, central charges, operator dimensions, and chiral rings have all been computed \cite{Martelli:2004wu,Benvenuti:2004dy,Herzog:2004tr,Benvenuti:2005cz}. In the regime of parameter space in which the flip couplings are irrelevant, the anomaly polynomial $I_6^{\text{SCFT}}$ computed from the class $\CS_2$ context can be shown to match exactly with the anomaly polynomial of the fixed point as computed from the $Y^{p,q}$ quiver description \cite{Bah:2022b}, where the flux parameters $N_N$, $N_S$ are related to the $p,q$ integers as,
\ba{
\label{fluxid}
p+q=N_N\,,\qquad p-q=N_S
}
With the identification \eqref{fluxid}, we extend the arguments of \cite{Bah:2022b} to identify $I_6^{\text{flip}}$ in \eqref{flagship1} as precisely minus the contributions of the following dibaryonic operators: the $(p+q)$ $Y$-dibaryons $\det\,Y$, and the $(p-q)$ $Z$-dibaryons $\det\,Z$,
\ba{
-I_6^{\text{flip}}  =(p+q)\,  I_6(\text{det}~Y) + (p-q)\,  I_6(\text{det}~Z) 
}
where $I_6^{\text{flip}}$ was written (schematically) in \eqref{split1}.
In particular, the would-be-flipped baryon operators $B_\beta$,$B_\gamma$ of the necklace quivers are mapped  to the dibaryons of the $Y^{p,q}$ description as $(B_\beta\leftrightarrow \det\,Y, B_\gamma\leftrightarrow \det\,Z)$, with their anomalies matching exactly (in comparing with \eqref{split1}, recall $p+q=n_\beta$ and $p-q=n_\gamma$). To be clear, there are never flip fields present in the $Y^{p,q}$ perspective: from the viewpoint of the $Y^{p,q}$ quivers, there is no reason to partition $I_6^{\text{SCFT}}$ as in \eqref{flagship1} into the inherited piece $\int_{T^2} I_8$, and the localized flip field piece.  Nonetheless, the perspectives are consistent with one another. 

As a byproduct of our analysis, we map the full chiral ring of the $Y^{p,q}$ SCFTs onto that of the IR fixed points of the class $\CS_2$ quivers whenever the flip couplings are irrelevant (in particular, generically at large $N$). We will discuss the case where the flip couplings are relevant in Section~\ref{sec:mainresults}, where we emphasize that there is a regime of $(N,p,q)$ parameter space where the necklace quiver and $Y^{p,q}$ quivers do not flow to the same fixed point.

%

\paragraph{Perspective from wrapped M2-branes} 
The authors of \cite{Bah:2022b}  identify the charges of the operators $B$ which are localized at resolution 2-cycles at the poles in the resolved geometry, corresponding to probe M2-branes wrapping the resolution cycles. (While we continue to refer to these operators as $B$ to reference the fact that in the various quiver descriptions they are baryons,  this is not obvious from this perspective.) These operators contribute  to the anomalies of the fixed point as precisely the term $I_6^{B}=-I_6^{\text{flip}}$ which appears in \eqref{flagship1}.  Partitioning off the  contribution to the anomalies inherited from 6d, it is natural to read their inflow computation as
\ba{
\label{eqinf}
I_6^{\text{SCFT}}=-I_6^{\text{inflow}}\big|_{\text{res}} =\int_{T^2} I_8 + I_6^{B}\,,\qquad I_6^B = - I_6^{\text{flip}}
}
{\it i.e.}~the operators $B$ appear as additional contributions to the naive reduction of the 6d anomaly polynomial localized at the poles.\footnote{~To make contact with the fixed-point computation of $I_6^{\text{inflow}}$ presented in Section~\ref{sec:inflow}, note that the entire computation lives in the first  $\sum_a$ of \eqref{eqsplit}, where the fixed loci $a=1,\dots,4$ are the four poles of the two
resolution cycles (in the Gibbons--Hawking presentation of Section~\ref{sec:inflow}, four
unit-charge Kaluza--Klein monopoles, two per resolution cycle). While the total sum is exactly linear in the fluxes $N_N, N_S$, defining the pole blocks
of \eqref{ns}, this split is a reassembly: the sum of the two northern (southern) monopole
contributions carries a term with $N_S(N_N)$ dependence that cancels against the other pole's terms. } 

Equation~\eqref{eqinf} is precisely the dynamical decomposition of the SCFT anomalies  \eqref{naiveloc}. Said another way, the geometric-engineering computation of $I_6^{\text{SCFT}}$ never sees the flip fields, or equivalently always includes the baryons, and the resolved geometry organizes as in \eqref{naiveloc}. The branes dual to these operators are there and must be counted: from this perspective, the RG irrelevance of the flip fields is the field theory mechanism that implements the compatibility with the geometry.  

As we discussed around \eqref{t2split} the  inherited term of \eqref{flagship1} is itself the result of the inflow computation in the singular background,  in which the $N_{N,S}$ flux is a background for the $U(1)_{\beta,\gamma}\subset SU(2)_{N,S}$ gauge fields supported at the poles, and \eqref{flagship1} reads $[\text{resolved}] - [\text{singular}] = -I_6^{\text{flip}} = I_6^{B}$. 
The physical content is transparent in this form: the W-bosons at each pole---M2-branes wrapping the resolution cycles---are massless in the singular
presentation and massive in the resolved one, and $I_6^{B}$ is precisely the anomaly of the modes whose status changes between the two.
On the other hand, since $\chi({T^2})=0$, we know that the geometric split of \eqref{naiveloc} should only include terms $I_6^{\text{loc-inher},i}$ linear in the fluxes $N_N$ and $N_S$. The contributions $I_6^{B}$ already split into one set localized at the north pole resolution cycles, and one set localized at the south pole cycles: this is evident in \eqref{split1}, where recall $n_\beta = N_N$ and $n_\gamma = N_S$. As anticipated,  the entire anomaly polynomial $I_6^{\text{SCFT}}$ for the torus manifestly splits into terms linear in $N_N$ and $N_S$ as,
\ba{
\label{ns}
I_6^{\text{SCFT}} = I_6^N + I_6^S
}
where $I_6^N$ is proportional to $N_N$, and $I_6^S$ to $N_S$. 

Geometrically, the identification of these would-be-flipped operators as localized contributions $\delta I_6^{\text{loc},i}$ is consistent with their dual description in the $Y^{p,q}$ geometry. The corresponding $\det\,Y$ and $\det\,Z$ operators of the $Y^{p,q}$ quivers are dual to D3-branes wrapping the supersymmetric 3-cycles $\Sigma_1\cong S^3/\mathbb{Z}_{p+q}$ and $\Sigma_2\cong S^3/\mathbb{Z}_{p-q}$ (the loci $y=y_1$ and $y=y_2$ of the metric of \cite{Gauntlett:2004yd} where a circle shrinks), with R-charges  given by the volumes of these cycles \cite{Martelli:2004wu,Benvenuti:2004dy,Herzog:2004tr} equal to those of the $B_\beta$ and $B_\gamma$ baryons. 
As we will derive in Section~\ref{sec:M2}, the operators corresponding to M2-branes wrapping the other calibrated 2-cycles in $M_6$ can be identified with the remaining chiral ring generators of the SCFT, and  mapped to the D3-brane operators discussed in \cite{Benvenuti:2004dy}.

\subsection{Outline}

The rest of this paper is organized as follows.
Section~\ref{sec:background} covers the relevant background. We describe the construction of the 4d $\CN=1$ class $\CS_k$ SCFTs in flux backgrounds, focusing on the torus compactification with cases $k=2$, and then proceed to review the dual $Y^{p,q}$ SCFTs and their quiver gauge theory descriptions, including properties of their chiral rings.  

In Section~\ref{sec:results} we analyze the decomposition of these fixed points' anomalies as in \eqref{naiveloc}, and discuss the origins of the different terms in each of the dual quiver descriptions. As we emphasize, the interpretation of the duality will depend on the values of the fluxes $(N,p,q)$. In much of the parameter space  the class $\CS_k$ and $Y^{p,q}$ quivers flow to the same interacting fixed point, albeit with different decoupled free fields, and in these cases we confirm that their chiral ring operators match as expected. Meanwhile, for some range of the fluxes the fixed points are distinct, in which case we identify the mismatch in chiral operators. These ideas are illustrated in three worked examples: the $p=q=1$ affine quiver (Section~\ref{sec:affine}); the $p=1,q=0$ conifold (Section~\ref{sec:conifold}); and the $p=4,q=3$ case  (Section~\ref{sec:433}).

In Section~\ref{sec:M2} we compute the charges of chiral operators at these fixed points from M2-branes wrapping 2-cycles in the 11d flux background, 
and identify which correspond to each $Y^{p,q}$ chiral ring generator.      

In Section~\ref{sec:inflow} we discuss the application of the BVAB fixed point theorem to class $\CS_k$, and proceed to use equivariant localization to compute the inflow anomalies \eqref{inflow} in both the singular and resolved backgrounds of the internal geometry $M_6$. We conclude that the difference of the two evaluations yields precisely the contributions of the operators that arise from M2-branes wrapping resolution cycles in the geometry, which played a key role in our analysis of decoupling from the quiver perspective in Section~\ref{sec:results}.

Finally, we conclude in Section~\ref{sec:outlook} with brief remarks on interesting future applications of our results. 
In appendices we collect our conventions, and construct representatives of the 2-forms required in Section~\ref{sec:M2}.

\section{Background}
\label{sec:background}
In this section we begin by reviewing the construction of the 4d $\CN=1$ class $\CS_k$ SCFTs from M5-branes, including their corresponding quiver gauge theory description that arises from gluing free trinions as geometric building blocks. Then, we provide an overview of the quiver gauge theories and chiral rings of the $Y^{p,q}$ SCFTs dual to class $\CS_{k=2}$ on the torus. The reader familiar with these constructions may consider skipping ahead to Section~\ref{sec:results}.

\subsection{4d QFTs from M5-branes}
\label{sec:m5intro}

We will consider setups in M-theory which involve a stack of $N$ M5-branes, where the 11d spacetime $M_{11}$ splits into the 6d brane worldvolume  $W_6$ and the five transverse directions to the stack, parameterized by a radial direction $r$ away from the branes and a space $M_4$,
\ba{
\label{m11}
M_{11} \simeq [r] \times M_4 \times W_6
}
At low energies, the dynamics on the branes is described by a six-dimensional QFT living on $W_6$.  When the branes are flat, $M_4\simeq S^4$, and this QFT is the maximally supersymmetric 6d $\CN=(2,0)$ theory of type $A_{N-1}$, whose R-symmetry group $SO(5)_R$ arises from the $SO(5)$ structure group of the normal bundle to the stack of branes, encoding the isometries of the transverse $S^4$.
A class of 6d $\CN=(1,0)$ theories is similarly obtained by replacing $S^4$ with the orbifold $S^4/\Gamma$, for $\Gamma$ an ADE subgroup of $SU(2)$.  

We  further wrap the branes on a  genus-$g$ Riemann surface $\Sigma_g$, so that the worldvolume splits as,
\ba{
W_6 = W_4\times \Sigma_g
}
At energies below the scale set by the size of $\Sigma_g$, the system is described by a 4d QFT living on $W_4$, which in certain circumstances flows to a superconformal field theory. 
It is this SCFT whose symmetries and anomalies we are interested in analyzing, whose anomaly polynomial we denote $I_6^{\text{SCFT}}$.

The symmetries of the 4d QFT are largely determined by the topology of the six-dimensional internal space $M_6$ comprising the transverse directions $M_4$ and the Riemann surface directions,
\ba{
\label{m6}
M_4 \hookrightarrow M_6 \rightarrow \Sigma_g
}
Isometries of  $M_6$ lead to global continuous 0-form symmetries in the 4d QFT, which may admit 't Hooft anomalies. A broader class of symmetries arises from expanding $C_3$ on cohomology classes of $M_6$, which  leads to general $p$-form and possibly discrete global symmetries in 4d \cite{Bah:2020uev}.  Furthermore, the fibration \eqref{m6} encodes the topological twist which is needed to obtain a supersymmetric QFT in four dimensions: wrapping $W_6$ on a curved background in general breaks supersymmetry, and so to preserve supersymmetry in 4d,  we cancel off the contribution of the curvature over the Riemann surface with a background magnetic field over $\Sigma_g$ valued in an abelian subgroup of the R-symmetry group.  The details of this topological twist  depend on the specifics of the setup. 

\paragraph{4d $\CN=1$ SCFTs of class $\mathcal{S}_k$} 
The class  $\mathcal{S}_k$ theories are constructed from M5-branes probing a $\mathbb{C}^2/\mathbb{Z}_k$ singularity \cite{Gaiotto:2015usa,Razamat:2016dpl,Bah:2017gph,Kim:2018lfo}, so that the space $M_4$ defined in \eqref{m11} is
\ba{
M_4=S^4/\mathbb{Z}_k
}
The quotient by $\mathbb{Z}_k$ reduces the structure group of the normal bundle to the branes: the $\mathbb{Z}_k$ group acts on the five transverse directions to the M5-branes, leaving one direction invariant and the remaining four with isometry group $SO(4)\simeq   (SU(2)_R\times SU(2)_L)/\mathbb{Z}_2$. The $SU(2)_R$ is identified as the R-symmetry of the 6d (1,0) SCFT living on  $W_6$, and $SU(2)_L$ is a flavor symmetry rotating the orbifold directions (whose Cartan will be denoted by $U(1)_{{t}}$ below).  The isometry group of $M_4$ corresponds to the subgroup of $ SU(2)_R\times SU(2)_L$ which commutes with the $\mathbb{Z}_k$ action, leading (for $k\geq 3$) to the breaking pattern $SO(5)_R\to SU(2)_R\times U(1)_L$, with $U(1)_L$ the Cartan of $SU(2)_L$. For $k=1,2$, the full $SU(2)_L$ is unbroken (and for $k=1$ the preserved symmetry is all of $SO(5)_R$).\footnote{~For $N=1,2$ the flavor symmetry is further enhanced, and we will therefore restrict to the cases $N>2$ in our expressions. 
}

The north and south poles of the $S^4$ are fixed points of the orbifold action, with each pole locally a $\mathbb{C}^2/\mathbb{Z}_k\equiv A_{k-1}$ singularity, leading to two  additional symmetries $SU(k)_{N}\times SU(k)_{S}$ of the (1,0) theory. These are not isometries of $M_6$, and are localized at the poles; their Cartans arise  from the expansion of $C_3$ onto the 2-forms of the  $\mathbb{C}^2/\mathbb{Z}_k$ loci.

\paragraph{Topological twist}

We preserve $\CN=1$ supersymmetry in four dimensions by twisting with respect to the Cartan of the 6d R-symmetry, $U(1)_R \subset SU(2)_R$. 
Let $V_\Sigma$ denote the unit volume form of the Riemann surface, normalized to satisfy
$\int_{\Sigma_g} V_\Sigma =1$.
Let $c_1(R)$ denote the first Chern class of the $U(1)_R$ symmetry, related to $c_2(SU(2)_R)$ as,
\ba{
c_2(SU(2)_R) = -c_1(R)^2
}
The twist is then implemented by shifting the Chern root of the $SU(2)_R$ bundle as,
\ba{
\label{twist}
c_1(R) = c_1(R') -\frac{\chi_g}{2} V_\Sigma
}
so that $\int_\Sigma c_1(R) = -\chi_g/2$, where  $\chi_g = 2-2g$.
Here $c_1(R')$ is the first Chern class of a $U(1)_{R'}$ global symmetry from the 4d point of view, with the prime emphasizing that it does not generically coincide with the 4d $\CN=1$ superconformal R-symmetry $R_{\CN=1}$, which must be determined via $a$-maximization \cite{Intriligator:2003jj}.
Note that compactification on the flat torus requires no R-symmetry twist ($\chi_1=0$), so that in this case $M_6=M_4\times T^2$ is topologically a product.

\paragraph{Resolved geometry} 

Following \cite{Bah:2019vmq,Bah:2022a,Bah:2022b}, 
we consider M5-branes probing a resolved $\mathbb{C}^2/\mathbb{Z}_k$ singularity. 
The resolved geometry is obtained by blowing up the fixed points of the orbifold action in the singular geometry $S^4/\mathbb{Z}_k$,
\ba{
\label{res}
M_4 = \left[ S^4/\mathbb{Z}_k \right]_{\text{resolved}}
}
The blow-up generates a collection of $(k-1)$ 2-cycles at the north and south poles of $S^4$, breaking the $SU(k)$ symmetry at each pole to its Cartan as,
\ba{
SU(k)_{N} \times SU(k)_{S} \ \to  \ U(1)_{N}^{k-1}\times U(1)_{S}^{k-1}
}
In the 7d gauge theory, blowing up the singularity corresponds to giving a VEV to the adjoint scalars of its vector multiplet, which set the size of the resolution 2-cycles.\footnote{\label{fn:flux}~In the singular background, the geometric data supported on the resolution cycles in the resolved geometry is converted to gauge-theory data.
The singular loci support the 7d $SU(k)_N\times SU(k)_S$ gauge theory, which is broken in the flux background to their Cartan, with field strengths satisfying $\frac{1}{2\pi}\int_{\Sigma_g}F_{\beta_i} = N_{\beta_i}$ and $\frac{1}{2\pi}\int_{\Sigma_g}F_{\gamma_i} = N_{\gamma_i}$ Taking $k=2$, one expands (focusing on the north fixed point) $G_4=2 F_\beta \wedge \omega +\dots$, where $\omega$ is the 2-form dual to the resolution 2-cycle $C_{\rm N}$. The factor of 2 is the charge of an M2-brane wrapping the resolution 2-cycle (a W-boson of $SU(2)_N$); under the $U(1)_\beta$ of \eqref{6dbasis} it has charge $2N$. Then,
\ba{
\int_{C_{\rm N}\times T^2} \frac{G_4}{2\pi} = 2 N_\beta = N_N\,,\qquad \int_{C_{\rm S}\times T^2} \frac{G_4}{2\pi} = 2 N_\gamma = N_S
}
}
We can turn on fluxes for these $U(1)$'s; denoting the 4d symmetries as $U(1)_{\beta_i}$, $U(1)_{\gamma_i}$, and $U(1)_{t}$, the Chern roots $b_i^N$ and $b_i^S$ of the $SU(k)_N$ and $SU(k)_S$ bundles shift as,\footnote{~One can additionally turn on flux for $U(1)_{L}$, see \cite{Bah:2017gph} for details.}
\ba{
\label{fluxes}
c_1(L)= c_1(t) \,,\qquad  b_i^N= N c_1(\beta_i)  + N_{\beta_i}V_\Sigma \,,\qquad  b_i^S = N c_1(\gamma_i) + N_{\gamma_i} V_\Sigma
}
where $i$ runs from 1 to $k$, and $\sum_{i=1}^kb_i^N=\sum_{i=1}^kb_i^S=0$. Here $c_1(L)$ is the first Chern class of the $U(1)_L$ bundle. In particular, the following are constrained quantities:
\ba{
\sum_{i=1}^k c_1(\beta_i) = \sum_{i=1}^k c_1(\gamma_i)=0\,,\qquad \sum_{i=1}^k N_{\beta_i} = \sum_{i=1}^k N_{\gamma_i} = 0
}
The fluxes satisfy $N_{\beta_i},N_{\gamma_i}\in \frac{1}{2}\mathbb{Z}$.

For $k=2$, we can establish a connection to the $AdS_5$ solutions in 11d supergravity by Gauntlett, Martelli, Sparks, and Waldram \cite{Gauntlett:2004zh}, as discussed in \cite{Bah:2019vmq}. The near horizon geometry of the $N$ M5-branes probing the resolved singularity corresponds to a solution of the form,
$AdS_5\times_w M_6^{\text{hol}}$
where the internal space $M_6^{\text{hol}}$ has the topology of \eqref{m6} with $M_4$ regarded as the resolution \eqref{res}, equipped with $G_4$-flux. The $G_4$-flux configuration is specified by three non-negative integers $N,N_N$, and $N_S$ (with $SU(2)_L$ unbroken). While the solutions for $k>2$ are not known explicitly, the topology and $G_4$-flux configuration of $M_6$ admit a natural generalization \cite{Bah:2022a,Bah:2022b}.

\paragraph{Anomalies on the torus for $k=2$}

Restricting to the case of $\Sigma_g = T^2$ and $k=2$, the global symmetries of the 6d SCFT in the resolved geometry are,
\ba{
U(1)_R\times SU(2)_L \times U(1)_N \times U(1)_S
}
These reduce to the following symmetries in 4d,
\ba{
\label{6dbasis}
U(1)_{R'}\times SU(2)_t\times U(1)_\beta \times U(1)_\gamma
}
with Chern classes related as in \eqref{fluxes}. 
Since $c_1(\beta_2)=-c_1(\beta_1)$ and  $N_{\beta_2} = - N_{\beta_1}$, we  denote  
\ba{
c_1(\beta_1) 
= c_1(\beta)\,,\qquad 
c_1(\gamma_1) 
= c_1(\gamma)\,, \qquad
N_{\beta_1}
=N_\beta\,,\qquad
N_{\gamma_1}
=N_\gamma}
with $\beta,\gamma$ labeling the unconstrained $U(1)_\beta\times U(1)_\gamma$.\footnote{~Note the normalization implied by \eqref{fluxes}: $U(1)_\beta$ is the Cartan of $SU(2)_N$ rescaled by $N$, and similarly for $U(1)_\gamma$ and $SU(2)_S$. (Eq.~(3.2) of \cite{Bah:2022b} is printed without this factor of $N$.) In
this normalization the quarks $Q_{1,2}$ of the necklace quivers of
Section~\ref{sec:necklace} carry $q_\beta = \mp 1$, while the W-bosons of $SU(2)_N$ carry $q_\beta = \pm 2N$, precisely the charge of the flipped baryons and their flip fields. (Recall from Footnote~\ref{fn:flux} that $N_N=2N_\beta$ and $N_S=2N_\gamma$.)}

The integral of the anomaly polynomial $I_8$ of the interacting 6d SCFT  on the torus reduces to \cite{Bah:2017gph}, 
\ba{
\label{checkb}
\bs{
\int_{T^2} I_8 &=
\left( N_\beta c_1(\beta) + N_\gamma c_1(\gamma)\right)\left( - \tfrac{1}{6}N p_1(TW_4)   -2  N (N-1) c_1^2(R')  +2    N^2 c_1^2(t)\right)\\
+&2N^2  c_1(\beta) c_1(\gamma) \Big( N_\gamma  c_1(\beta)  +  N_\beta c_1(\gamma)\Big) +2N^2\left(  \tfrac{ 4}{3}N  - 1  \right)\Big(N_\beta c_1(\beta)^3 + N_\gamma c_1(\gamma)^3\Big)
}}
See Appendix \ref{app:conventions} for our conventions for characteristic classes. Subtracting off the following contributions labeled as $I_6^{\text{flip}}$ (for reasons to be made clear in the following sections),
\ba{
\label{i6flip}
\bs{
I_6^{\text{flip}} 
= N_\beta \Bigg(& \frac{1}{3}  c_1(R')^3  + 2 N c_1(R')^2  c_1(\beta) +4 N^2 c_1(R')  c_1(\beta)^2   + \frac{8}{3} N^3 c_1(\beta)^3   \\
\qquad\ &- \left( c_1(R')  + 2 N  c_1(\beta) \right) \frac{ p_1(TW_4)}{12}  \Bigg) \ + \ \left(\beta\to\gamma\right)
}}
 yields the anomalies $I_6^{\text{SCFT}}$ of the interacting 4d fixed point, as in \eqref{flagship1}. The reason for this split and subtleties depending on specific values of $N$ and the fluxes $N_\beta$, $N_\gamma$ will be discussed further in Sections~\ref{sec:mainresults} and \ref{sec:M2}.

\subsection{Class $\CS_{k}$ necklace quivers from free trinions}
\label{sec:necklace}

In this section we review the construction of the quiver gauge theories of class $\CS_{k}$ introduced in \cite{Gaiotto:2015usa}, corresponding to tori with fluxes but no punctures.

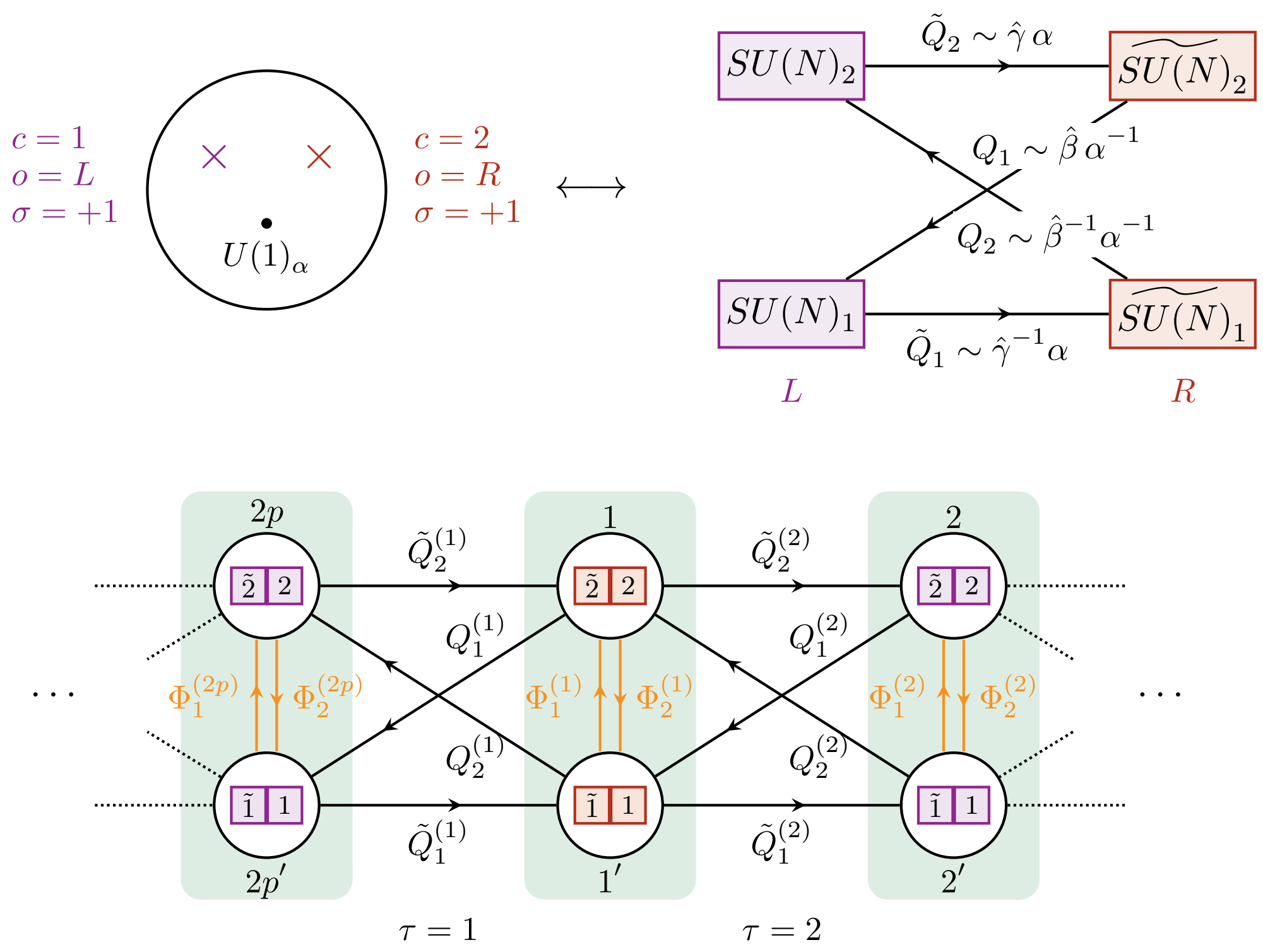
\begin{figure}[t!]
\centering
\begin{tikzpicture}[
  gnode/.style={circle, draw, thick, fill=white, minimum size=11mm, inner sep=0pt},
  mini/.style={rectangle, draw=#1, fill=#1!10, thick, minimum width=3.6mm, minimum height=3.8mm, inner sep=0pt, font=\scriptsize},
  fL/.style={rectangle, draw=Plum, thick, fill=Plum!8, minimum width=15mm, minimum height=7mm, inner sep=2pt, font=\small},
  fR/.style={rectangle, draw=BrickRed, thick, fill=BrickRed!8, minimum width=15mm, minimum height=7mm, inner sep=2pt, font=\small},
  fld/.style={thick, postaction={decorate, decoration={markings, mark=at position #1 with {\arrow{stealth}}}}},
  fld/.default=0.6,
  phi/.style={thick, BurntOrange, postaction={decorate, decoration={markings, mark=at position 0.6 with {\arrow{stealth}}}}},
  lab/.style={font=\footnotesize},
  stub/.style={thick, densely dotted},
]
\begin{scope}[shift={(0,5.3)}]
  \draw[thick] (0,0) circle (1.25);
  \node[Plum]     at (-0.55,0.35) {\large$\times$};
  \node[BrickRed] at ( 0.55,0.35) {\large$\times$};
  \fill (0,-0.35) circle (1.6pt) node[below=1pt, lab] {$U(1)_\alpha$};
  \node[Plum, lab, align=left, anchor=east]     at (-1.4,0.15) {$c=1$\\[-1pt] $o=L$\\[-1pt] $\sigma=+1$};
  \node[BrickRed, lab, align=left, anchor=west] at ( 1.4,0.15) {$c=2$\\[-1pt] $o=R$\\[-1pt] $\sigma=+1$};
  \node at (3.4,0) {$\longleftrightarrow$};
  \node[fL] (L2) at (5.5, 1.3) {$SU(N)_2$};
  \node[fL] (L1) at (5.5,-1.3) {$SU(N)_1$};
  \node[fR] (R2) at (9.6, 1.3) {$\widetilde{SU(N)}_2$};
  \node[fR] (R1) at (9.6,-1.3) {$\widetilde{SU(N)}_1$};
  \draw[fld] (L2) -- node[above, lab] {$\tilde Q_2\sim\hat\gamma\,\alpha$} (R2);
  \draw[fld] (L1) -- node[below, lab] {$\tilde Q_1\sim\hat\gamma^{-1}\alpha$} (R1);
  \draw[fld=0.72] (R2) -- node[pos=0.25, lab, fill=white, inner sep=1pt] {$Q_1\sim\hat\beta\,\alpha^{-1}$} (L1);
  \draw[fld=0.72] (R1) -- node[pos=0.25, lab, fill=white, inner sep=1pt] {$Q_2\sim\hat\beta^{-1}\alpha^{-1}$} (L2);
  \node[Plum, lab]     at (5.5,-2.1) {$L$};
  \node[BrickRed, lab] at (9.6,-2.1) {$R$};
\end{scope}
\foreach \x/\n/\t/\c in {0/0/2p/Plum, 3.6/1/1/BrickRed, 7.2/2/2/Plum} {
  \fill[ForestGreen!12, rounded corners=6pt] (\x-0.9,-2.14) rectangle (\x+0.9,2.14);
  \node[gnode, label={[lab, inner sep=1pt]above:$\t$}]  (o\n) at (\x, 1.15) {};
  \node[gnode, label={[lab, inner sep=1pt]below:$\t'$}] (i\n) at (\x,-1.15) {};
  \node[mini=\c] at (\x-0.19, 1.15) {$\tilde 2$};  \node[mini=\c] at (\x+0.19, 1.15) {$2$};
  \node[mini=\c] at (\x-0.19,-1.15) {$\tilde 1$};  \node[mini=\c] at (\x+0.19,-1.15) {$1$};
  \draw[phi] ([xshift=-3pt]i\n.north) -- ([xshift=-3pt]o\n.south);
  \draw[phi] ([xshift= 3pt]o\n.south) -- ([xshift= 3pt]i\n.north);
  \node[BurntOrange, lab, anchor=east] at (\x-0.12,0) {$\Phi_1^{(\t)}$};
  \node[BurntOrange, lab, anchor=west] at (\x+0.12,0) {$\Phi_2^{(\t)}$};
}
\foreach \a/\b/\t in {0/1/1, 1/2/2} {
  \draw[fld] (o\a) -- node[above, lab] {$\tilde Q_2^{(\t)}$} (o\b);
  \draw[fld] (i\a) -- node[below, lab] {$\tilde Q_1^{(\t)}$} (i\b);
  \draw[fld=0.72] (o\b) -- node[pos=0.35, above, lab] {$Q_1^{(\t)}$} (i\a);
  \draw[fld=0.72] (i\b) -- node[pos=0.35, below, lab] {$Q_2^{(\t)}$} (o\a);
}
\node[lab] at (1.8,-2.45) {$\tau=1$};
\node[lab] at (5.4,-2.45) {$\tau=2$};
\draw[stub] (-1.8, 1.15) -- (o0);  \draw[stub] (-1.8,-1.15) -- (i0);
\draw[stub] (o0) -- (-1.25,0.38);  \draw[stub] (i0) -- (-1.25,-0.38);
\draw[stub] (o2) -- (9.0, 1.15);  \draw[stub] (i2) -- (9.0,-1.15);
\draw[stub] (8.45,0.38) -- (o2);  \draw[stub] (8.45,-0.38) -- (i2);
\node at (-2.2,0) {$\cdots$};  \node at (9.4,0) {$\cdots$};
\end{tikzpicture}
\caption{(top) The free trinion: a sphere with two maximal punctures of colors $c=1,2$, the same sign $\sigma=+1$, and opposite orientations; and one minimal puncture carrying $U(1)_\alpha$. Boxes are $SU(N)$ flavor symmetries, with a tilde for the right-oriented puncture. Each field is labeled by its fugacity, up to a common factor $\hat t^{1/2}$ (Table~\ref{tab:trinion} at $\tau=1$). (bottom) The chain of $2p$ trinions glued into the torus with $2p$ minimal punctures. Each circle is an $SU(N)$ gauge node, corresponding to the diagonal of the two flavor factors it encloses. The colors of the small boxes are those of the glued punctures, which alternate along the chain. Each shaded box is the tube at junction $\tau$. The fields and their charges are listed in Table~\ref{tab:trinion}. \label{fig:trinion}}
\end{figure}

The basic building block is the free trinion: the compactification on a sphere with one minimal and two maximal regular punctures, corresponding to a quiver theory of bifundamental chiral multiplets.
 Each maximal puncture is labeled by a color $c\in \{1,2,\dots, k\}$, a sign $\sigma = \pm 1$, and an orientation $o=\{L,R\}$. The two maximal punctures of a single free trinion share a sign and have colors differing by one, with opposite orientation.  
The flavor symmetry carried by each maximal puncture is $SU(N)^k$, and the minimal puncture carries its own $U(1)_{\alpha}$ symmetry. Additionally, there is an intrinsic global symmetry independent of the punctures,
\ba{
U(1)_{\hat{R}}\times U(1)_{\hat{\beta}} \times U(1)_{\hat{\gamma}}\times U(1)_{\hat{t}} 
}
where $U(1)_{\hat{R}}$ denotes a non-anomalous R-symmetry, 
and where $U(1)_{\hat{t}}$ is enhanced to $SU(2)_{\hat{t}}$ for $k=2$. The relation between the hatted basis $\{\hat{R},\hat{\beta},\hat{\gamma},\hat{t}\}$ natural to the quiver construction, and the unhatted basis $\{R',{\beta},{\gamma},{t}\}$  discussed around \eqref{6dbasis}, is given in Appendix~\ref{sec:master}.  
The field content of the free trinion consists of $2kN^2$ chiral multiplets; for $k=2$ these are organized into the four bifundamental fields $Q_{1,2}$ and $\tilde Q_{1,2}$ depicted in  Figure~\ref{fig:trinion}, with charges given in the top part of  Table~\ref{tab:trinion}.

Trinions can be glued along their maximal punctures to produce a chain of higher genus. Gluing two maximal punctures requires that they share the same color $c$ and have opposite orientation $o$, such that the gluing type is fixed by whether their signs $\sigma$ agree. $\Phi$-gluing applies to same-sign (same-color, opposite-orientation) pairs; it is carried out by 
gauging the diagonal combination of  the two flavor $SU(N)_i$'s with an $\CN=1$ vector multiplet, and adding a bifundamental chiral multiplet $\Phi_i$ (a tube) for each gauged $SU(N)_i$.

The base necklace quiver triangulating the torus then corresponds to a chain of  $\tau=1,\dots,2p$ trinions glued consecutively, incrementing color by one at each step ({\it i.e.},  alternating color for $k=2$), as depicted in Figure~\ref{fig:trinion}. The base necklace has $2p$ ``outside'' nodes $1,\dots,2p$  and $2p$ ``inside'' nodes $1',\dots,2p'$, 
with  bifundamental fields connecting the nodes as (indices mod $2p$): 
\begin{itemize}
\item {Outside loop:} $i \to i+1$ for $i=1,\dots,2p$.
\item {Inside loop:} $i' \to (i+1)'$ for $i=1,\dots,2p$.
\item {Criss-cross A:} $(i+1)' \to i$.
\item {Criss-cross B:} $(i+1) \to i'$.
\end{itemize}
The $\tau$ trinion sits at position $\tau-1$ (L-puncture) to $\tau$
(R-puncture) on the ring, with four fields $Q_{1,2}^{(\tau)}$ and $\tilde Q_{1,2}^{(\tau)}$, listed in Table~\ref{tab:trinion}. 
The $\Phi$-pair at junction $\tau$ is given by $\Phi_{1,2}^{(\tau)}$.
A superpotential couples the $\Phi_i$ fields to the meson matrices of the two glued punctures; for $k=2$ the superpotential for the trinion labeled by $\tau$ is,
 \begin{equation}
 \label{suptau}
W_\tau = \Tr\Big[\Phi_1^{(\tau)}\big(Q_1^{(\tau)}\tilde Q_1^{(\tau)} - \tilde Q_2^{(\tau+1)}Q_1^{(\tau+1)}\big)\Big]
       + \Tr\Big[\Phi_2^{(\tau)}\big(Q_2^{(\tau)}\tilde Q_2^{(\tau)} - \tilde Q_1^{(\tau+1)}Q_2^{(\tau+1)}\big)\Big]
              \end{equation}     
The full superpotential of the base necklace quiver is the sum $\sum_{\tau=1}^{2p} W_\tau$.

\begin{table}[t!]
\begin{center}
\begin{tabular}{l c | c  c c c}
\toprule
Field & Path & $U(1)_{\hat\beta}$ & $U(1)_{\hat\gamma}$ & $U(1)_{\hat t}$ & $U(1)_{\alpha_\tau}$ \\
\midrule
$Q_1^{(\tau)}$ (criss-cross B) &  ${\tau \to (\tau-1)'}$ & $+1$ & $0$ & $+1/2$ & $-1$ \\
$Q_2^{(\tau)}$ (criss-cross A) & ${\tau' \to (\tau-1)}$  &  $-1$ & $0$ & $+1/2$ & $-1$ \\
${\tilde Q_1^{(\tau)}}$ (inside loop)   &  $(\tau-1)' \to \tau'$ & $0$ & $(-1)^\tau$ & $+1/2$ & $+1$ \\
${\tilde Q_2^{(\tau)}}$ (outside loop) & $(\tau-1) \to \tau$   & $0$ & $(-1)^{\tau+1}$ & $+1/2$ & $+1$ \\
 \midrule
${\Phi_1^{(\tau)}}$ &  $ \tau' \to \tau$ & $-1$ & $(-1)^{\tau+1}$ & $-1$ & $0$ \\
${\Phi_2^{(\tau)}}$&   $\tau \to \tau'$ & $+1$ & $(-1)^\tau$ & $-1$ & $0$ \\
\bottomrule
\end{tabular}
\caption{Matter content of the $k=2$ necklace of $\tau=1,\dots,2p$ trinions glued  along their maximal punctures, corresponding to the torus with $2p$ minimal punctures. The $\hat{R}$-charges of the $Q,\tilde{Q}$'s are taken to be $\tfrac{1}{2}$, with $\hat{R}(\Phi)=1$ (this refers to the pre-closure R-symmetry of the trinions, differing from the reference $\hat{R}$ of Appendix~\ref{sec:master} used after the closures). Taking $\tau=1$ matches the trinion depicted in the top of Figure~\ref{fig:trinion};
a puncture of the opposite sign is obtained from this one by shifting the $\hat{\gamma}$-fugacities. 
\label{tab:trinion}}
\end{center}
\end{table}


Closing a minimal puncture is accomplished by giving a vacuum expectation value to a baryonic operator charged under that puncture's $U(1)_{\alpha_\tau}$ symmetry. As the baryon is composed of a bifundamental of two equal-rank $SU(N)$ factors, its VEV identifies the two gauge nodes it connects, Higgsing the gauge group to the diagonal $SU(N)$. As can be seen from the superpotential~\eqref{suptau}, the VEV generates a mass term for one nearby $\Phi$ and one nearby block of bifundamentals, which are integrated out.

There are four candidate baryons for each trinion which can be chosen to close the puncture: two are charged under $U(1)_{\hat \beta}$, in which case we refer to the closure as $\beta$-type, and two are charged under $U(1)_{\hat \gamma}$, which could be chosen to implement a $\gamma$-type closure. 
 At $k=2$ there is exactly one other candidate baryon sharing the same $U(1)_{\alpha_\tau}$ charge as the one involved in the Higgsing; this sibling $B_\beta$ or $B_\gamma$ (depending on the type of closure) must be removed  by a new gauge-singlet flip field $\phi_B$, with $W\supset \phi_B B$, which sets $B=0$ in the chiral ring of the UV Lagrangian. (Whether $B$ stays out of the chiral ring at the fixed point depends on whether this coupling is relevant---see Section~\ref{sec:433}.)
The choice of which baryons receive VEVs maps to a choice of fluxes in six dimensions. 
Each trinion contributes a baseline flux $N_t=\frac{1}{2}$ for $U(1)_t$, and a closure shifts it by $\Delta N_t = - \frac{1}{2}$. A $\beta$-closure shifts by $\Delta N_\beta = +\tfrac{1}{2}$, and a $\gamma$-closure shifts by $\Delta N_\gamma = +\tfrac{1}{2}$.  Higgsing all  $U(1)_{\alpha_\tau}$ minimal puncture symmetries results in a quiver with $2(N_\beta + N_\gamma)= 2p$ gauge nodes and no $N_t$ flux (with $U(1)_t$ enhanced to $SU(2)_t$), corresponding to the torus with no punctures.

\paragraph{Why closing punctures adds flux}

It is instructive to review why closing a minimal puncture by  giving a VEV to a baryon in the necklace quiver adds one unit of flavor flux on the torus. To simplify this discussion, we'll focus on the flux $N_N = 2N_\beta$ associated to the $U(1)_\beta$ symmetry in the 4d field theory; an identical discussion holds upon replacing $N_N\to N_S$ and $U(1)_\beta\to U(1)_\gamma$.
 
 In the class $\CS_k$ construction, 
 there is a $U(1)_\alpha$ symmetry associated to each minimal puncture, under which the quarks in the necklace quiver are charged. The quarks are also each charged under exactly one of the $U(1)_{\hat \beta,\hat \gamma}$ symmetries with charge $\pm 1$. Giving a VEV to the baryon associated to one of these quarks breaks every $U(1)$ under which it is charged, and changes the basis of the surviving generators. The unbroken symmetry is given by combinations (schematically) $\tilde{\beta}= \hat\beta \mp \alpha$, $\tilde{t} = \hat{t} + (\ldots)\alpha$, $\tilde{R} = \hat R + (\ldots)\alpha$ under which the condensing baryon is neutral. 
Functionally speaking, this basis change is identified at the level of anomaly matching as a flux shift: the IR theory has no puncture, and the reorganization of the anomalies in the new basis is evidently equivalent to the anomalies of having shifted the flux $N_N$ by 1 unit \cite{Gaiotto:2015usa, Razamat:2016dpl, Bah:2017gph}.  

Geometrically, the flux is the winding of a  Wilson line around one of the torus cycles. Specifically, call  $w_\beta(x)$ the  B-cycle Wilson line seen by the W-bosons as it traverses the A-cycle,
\ba{
w_\beta(x) = \frac{1}{2\pi} \oint_{S^1_\text{B}} 2A_\beta(x)
}
where $A_\beta(x)$ is the 6d background connection restricted to the B-cycle at position $x$ along $S^1_\text{A}$.\footnote{~Using the dictionary in Appendix \ref{sec:master}, $A_\beta = - A_{\hat{\beta}}- \kappa_\beta A_{\hat{R}}$. As there is no R-symmetry flux on the torus, the shift does not affect the winding argument.} Then, $N_N$ is the winding number of $w_\beta$ as it traverses the A-cycle. 

 Consider first reducing the 6d theory on the B-cycle $S^1_\text{B}$ to obtain the 5d KK theory living on $\mathbb{R}^{1,3}\times S^1_\text{A}$. The punctured torus is a chain of  5d segments (tubes) along $S^1_\text{A}$ (with coordinate $x$), separated by interfaces (free trinions) which carry 4d matter at positions $x_P$ \cite{Kim:2018lfo}. After closing a puncture at $x_P$ by giving some of the matter a VEV and integrating out the massive fluctuations, the junction condition at the interface changes: the combination of background connections is fixed according to the combination under which the condensing baryon is neutral,
\ba{
A_\alpha = \frac{1}{2} A_{\hat t} + \hat{q} A_{\hat \beta} \Big|_{x_P}\,,\qquad \hat{q}=\pm 1
}
So, the theory on the right side of the interface is glued to the one on the left with a relative shift of $w_\beta$ by exactly one winding.

This procedure can be understood as depositing $G_4$-flux as follows.
In the lift to M-theory, one identifies  $2A_\beta = \int_{S^2_{\text{res}}} C_3$, so that 
\ba{
w_\beta(x) = \frac{1}{2\pi}\int_{S^2_{\text{res}}\times S^1_\text{B}} C_3\big|_x
}
The identical Stokes computation gives 
\ba{
\int_{S^2_{\text{res}}\times T^2}  \frac{G_4}{2\pi} = \text{winding of the }C_3\text{-period}
}
so that one unit of $G_4$-flux per closing is deposited.

\subsection{$Y^{p,q}$ quivers and chiral ring}
\label{sec:ypq}

An interesting class of SCFTs arises from  $N$ D3-branes on $\mathbb{R}^{1,3}\times C(X)$ for $C(X)$ the Calabi-Yau cone over the Sasaki-Einstein manifold $X_5$.  The AdS/CFT correspondence relates  the gauge theories on the D-branes to a supergravity solution of the form $AdS_5\times X_5$, with $N$ units of self-dual five-form flux through $X_5$. The CFT central charges are related to the volume of $X_5$ as \cite{Gubser:1998vd}
\ba{
\label{volx}
a=c=\frac{N^2\pi^3}{4\, \text{Vol(X)}}
} 
The authors of \cite{Gauntlett:2004yd} constructed a large class of Sasaki-Einstein metrics on $X_5=S^2\times S^3$ labeled by two  integers $0\leq q \leq p$  called $Y^{p,q}$ spaces, which are dual to 4d $\CN=1$ superconformal field theories. The Calabi-Yau cone is toric, with $U(1)^3$ isometry considered as a subgroup of the full $SU(2)\times U(1)^2$ isometry group of $Y^{p,q}$.

\begin{figure}[t!]
	\centering
\begin{tikzpicture}[
  qn/.style={circle, fill=black, inner sep=0pt, minimum size=5pt},
  nl/.style={font=\footnotesize},
  fl/.style={font=\footnotesize, inner sep=1pt},
  cl/.style={font=\footnotesize, inner sep=0.5pt, fill=white},
  dbl/.style={very thick, #1, postaction={decorate, decoration={markings,
      mark=at position 0.5 with {\arrow{stealth}}, mark=at position 0.5 with {\arrow[xshift=3.5pt]{stealth}}}}},
  sgl/.style={thick, #1, postaction={decorate, decoration={markings, mark=at position 0.55 with {\arrow{stealth}}}}},
  ch/.style={thick, cY, postaction={decorate, decoration={markings, mark=at position #1 with {\arrow{stealth}}}}},
]
\colorlet{cU}{BrickRed}\colorlet{cV}{ForestGreen}\colorlet{cZ}{BurntOrange}\colorlet{cY}{Violet}
\begin{scope}[shift={(-6,3.5)}]
  \node[qn, label={[nl]left:$i{+}3$}]  (a3) at (0,0) {};
  \node[qn, label={[nl]right:$i$}]     (a0) at (4.4,0) {};
  \node[qn, label={[nl]left:$i{+}2$}]  (a2) at (1.1,1.7) {};
  \node[qn, label={[nl]right:$i{+}1$}] (a1) at (3.3,1.7) {};
  \draw[dbl=cU] (a0) -- node[fl, pos=0.3, right=2pt, text=cU] {$U_i$} (a1);
  \draw[dbl=cV] (a1) -- node[fl, above, text=cV] {$V_{i+1}$} (a2);
  \draw[dbl=cU] (a2) -- node[fl, pos=0.7, left=2pt, text=cU] {$U_{i+2}$} (a3);
  \draw[ch=0.72] (a2) -- node[fl, pos=0.84, below left, text=cY] {$Y_i$} (a0);
  \draw[ch=0.72] (a3) -- node[fl, pos=0.16, below right, text=cY] {$Y_{i+1}$} (a1);
\end{scope}
\begin{scope}[shift={(1.6,3.5)}]
  \node[qn, label={[nl]left:$i{+}3$}]  (b3) at (0,0) {};
  \node[qn, label={[nl]right:$i$}]     (b0) at (4.4,0) {};
  \node[qn, label={[nl]left:$i{+}2$}]  (b2) at (1.1,1.7) {};
  \node[qn, label={[nl]right:$i{+}1$}] (b1) at (3.3,1.7) {};
  \draw[dbl=cU] (b0) -- node[fl, pos=0.3, right=2pt, text=cU] {$U_i$} (b1);
  \draw[sgl=cZ] (b1) -- node[fl, above, text=cZ] {$Z_{i+1}$} (b2);
  \draw[dbl=cU] (b2) -- node[fl, pos=0.7, left=2pt, text=cU] {$U_{i+2}$} (b3);
  \draw[sgl=cY] (b3) -- node[fl, below, text=cY] {$\tilde Y_i$} (b0);
\end{scope}
\begin{scope}[shift={(-3.8,0)}]
  \foreach \k in {1,...,8} {
    \node[qn] (A\k) at ({112.5+45*(\k-1)}:2.35) {};
    \node[nl] at ({112.5+45*(\k-1)}:2.72) {$\k$};
  }
  \foreach \a/\b/\f/\s/\c in {1/2/U_1/dbl/cU, 2/3/V_2/dbl/cV, 3/4/U_3/dbl/cU, 4/5/V_4/dbl/cV,
                               5/6/U_5/dbl/cU, 6/7/Z_6/sgl/cZ, 7/8/U_7/dbl/cU, 8/1/Z_8/sgl/cZ} {
    \draw[\s=\c] (A\a) -- (A\b);
    \node[fl, text=\c] at ({135+45*(\a-1)}:2.55) {$\f$};
  }
  \foreach \a/\b/\f/\p/\r in {3/1/Y_1/0.44/0.76, 4/2/Y_2/0.4/0.6, 5/3/Y_3/0.4/0.6, 6/4/Y_4/0.2/0.56,
                              8/5/\tilde Y_5/0.48/0.68, 2/7/\tilde Y_7/0.36/0.55} {
    \draw[ch=\r] (A\a) -- node[cl, pos=\p, text=cY] {$\f$} (A\b);
  }
  \node[font=\small] at (0,-3.2) {$(p,q)=(4,2)$};
\end{scope}
\begin{scope}[shift={(3.8,0)}]
  \foreach \k in {1,...,8} {
    \node[qn] (B\k) at ({112.5+45*(\k-1)}:2.35) {};
    \node[nl] at ({112.5+45*(\k-1)}:2.72) {$\k$};
  }
  \foreach \a/\b/\f/\s/\c in {1/2/U_1/dbl/cU, 2/3/V_2/dbl/cV, 3/4/U_3/dbl/cU, 4/5/V_4/dbl/cV,
                               5/6/U_5/dbl/cU, 6/7/V_6/dbl/cV, 7/8/U_7/dbl/cU, 8/1/Z_8/sgl/cZ} {
    \draw[\s=\c] (B\a) -- (B\b);
    \node[fl, text=\c] at ({135+45*(\a-1)}:2.55) {$\f$};
  }
  \foreach \a/\b/\f/\p/\r in {3/1/Y_1/0.44/0.76, 4/2/Y_2/0.4/0.6, 5/3/Y_3/0.4/0.6, 6/4/Y_4/0.4/0.6,
                              7/5/Y_5/0.4/0.6, 8/6/Y_6/0.15/0.8, 2/7/\tilde Y_7/0.4/0.62} {
    \draw[ch=\r] (B\a) -- node[cl, pos=\p, text=cY] {$\f$} (B\b);
  }
  \node[font=\small] at (0,-3.2) {$(p,q)=(4,3)$};
\end{scope}
\end{tikzpicture}
	\caption{(Top:) The building blocks of the $Y^{p,q}$ quivers. (Bottom:) Example quivers for $(p,q)=(4,2)$ (bottom left) and $(p,q)=(4,3)$ (bottom right), in the frame of Table~\ref{tab:ypqgauge}. Nodes denote $SU(N)$ gauge groups, and double arrowheads denote $SU(2)_\varphi$ doublet fields. \label{fig:ypq}}
	\end{figure}
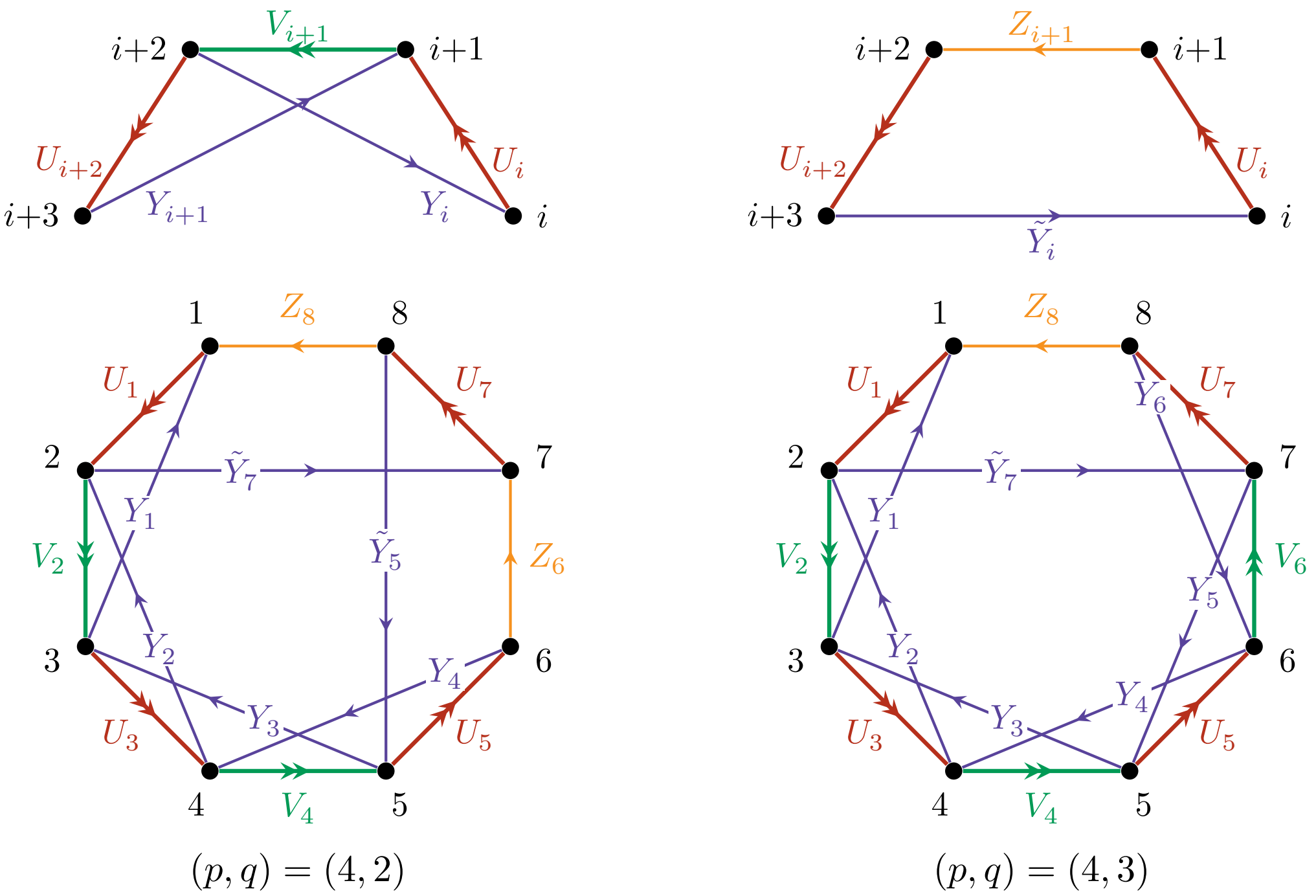

The dual  gauge theories to the $Y^{p,q}$ geometries were constructed in \cite{Benvenuti:2004dy}; they consist of a circular quiver  with $2p$ nodes $SU(N)_{i}$, with indices $i=1,\dots,2p$  defined mod $2p$, and several bifundamental fields connecting various nodes: $p$ $SU(2)$ doublet fields $U^\alpha$;  $q$ $SU(2)$ doublet fields $V^\alpha$;  $(p+q)$ singlets $Y$; $(p-q)$ singlets $Z$.\footnote{~On the D3-branes the gauge group is $U(N)^{2p}$. Its $U(1)$ factors are anomalous or IR free and decouple, and the non-anomalous combination becomes the global $U(1)_B$ \cite{Klebanov:1998hh,Benvenuti:2004dy}.}
The basic building blocks of the quiver are the  $UVY$ triangles and $UZUY$ quadrangles depicted in Figure~\ref{fig:ypq}. 
These fields are coupled together by a superpotential of the form,
\ba{
\label{wpq}
W =    \sum_{\text{tri}}\epsilon_{\alpha\beta}  \tr\left(U_i^\alpha V^\beta_{i+1} Y_i  - V^\beta_{i+1}  U_{i+2}^\alpha {Y}_{i+1} \right) - \sum_{\text{quad}}\epsilon_{\alpha\beta}  \tr\, U^\alpha_{i+1} \tilde{Y}_{i-1} U^\beta_{i-1} Z_i 
} 
We write $Y_{1,\dots,2q}$ for the $Y$ fields in the triangles and $\tilde Y$ for the $p-q$ $Y$ fields in the quadrangles.
Different orderings of the triangles and quadrangles correspond to different Seiberg dual frames; working in the Seiberg-dual frame in which the $2q$ $UVY$ triangles come first, starting at node $i=1$, followed by the $p-q$ $UZU\tilde{Y}$ quadrangles, the bifundamental fields have gauge symmetry charges labeled in Table~\ref{tab:ypqgauge}. In this frame, the sum over triangles in the superpotential~\eqref{wpq} runs over $i=1,3,\dots,2q-1$, and the sum over quadrangles over $i=2q+2,2q+4,\dots,2p$. Two representative examples appear in Figure~\ref{fig:ypq}.

\begin{table}[t!]
\centering
\begin{tabular}{l c c}
\toprule
Field & Path & Range \\
\midrule
$U_i^\alpha$ & $i\to i+1$ & $i=1,3,\dots,2p-1$ \\
$V_i^\alpha$ & $i\to i+1$ & $i=2,4,\dots,2q$ \\
$Z_i$ & $i\to i+1$ & $i=2q+2,2q+4,\dots,2p$ \\
$Y_i$ & $i+2\to i$ & $i=1,2,\dots,2q$ \\
$\tilde Y_i$ & $i+3\to i$ & $i=2q+1,2q+3,\dots,2p-1$ \\
\bottomrule
\end{tabular}
\caption{Bifundamental fields of the $Y^{p,q}$ quiver in the frame of \eqref{wpq}; $i\to j$ denotes the fundamental of $SU(N)_i$ and the antifundamental of $SU(N)_j$.\label{tab:ypqgauge}}
\end{table}

The quiver has four commuting anomaly-free $U(1)$ symmetries, which we take as the  $U(1)_B$  baryonic symmetry, $U(1)_F$ flavor symmetry, $U(1)_{R_0}$ R-symmetry, and $U(1)_{\varphi}\subset SU(2)$ global symmetry.\footnote{~This basis is related to the $U(1)_{a,b,c,d}$ used in {\it e.g.} \cite{Tachikawa:2006thesis} by the following relationship between the charges,
\ba{
\bs{
F &= a-c\,,\quad R_0 =c +  \tfrac{1}{2} (b+d)  \,,\quad \varphi = \tfrac{1}{2}\left(b-d\right)\,,\quad B = (p+q)a - p b + (p-q)c - p d
}}
} The charges of the bifundamental fields under these symmetries are given in Table~\ref{tab:ypqcharge}.
The superconformal $U(1)_{R_*}$ at the fixed point is a linear combination of these symmetries as determined by $a$-maximization, where the $U(1)_\varphi$ Cartan of the non-abelian $SU(2)$ symmetry does not contribute to the mixing. We can thus write the following linear combination for the trial R-charge,\footnote{~The map to the charge basis used in \cite{Benvenuti:2004dy}  is,
\ba{
x = (p+q) \epsilon_B + \epsilon_F\,,\qquad y = 1 + (p-q) \epsilon_B - \epsilon_F\,.
}
}
\ba{
\label{trialR}
R_*(\epsilon_F,\epsilon_B) = R_0 + \epsilon_F F + \epsilon_B B 
}
To perform $a$-maximization, we evaluate the R-charges \eqref{trialR} of all the fields in Table~\ref{tab:ypqcharge}, compute the contributions of these fields and the gauginos to the trial $a$ and $c$ central charges, 
\ba{
\label{atrial}
a_{\text{trial}}&= \frac{3}{32} \left( 3 \tr R_*^3 - \tr R_*\right)\,,\qquad c_{\text{trial}} = \frac{1}{32} \left( 9 \tr R_*^3 - 5\tr R_*\right) 
} 
and solve for the values of $\epsilon_B$ and $\epsilon_F$ that maximize $a_{\text{trial}}$.
For a generic choice of $p$ and $q$, this procedure determines $\epsilon_B$ and $\epsilon_F$ as,
\ba{
\epsilon_B = \frac{-8p^2 + 3q^2 + 4p\sqrt{4p^2-3q^2}}{6pq^2}\,,\qquad \epsilon_F =  \frac{1}{2} - \frac{ -4p^2  +3q^2 +2p\sqrt{4p^2-3q^2}}{6pq}
}
leading to the large-$N$ central charges \cite{Benvenuti:2004dy},
\ba{
\label{accc}
a = c = N^2 \frac{p^2(2p-\sqrt{4p^2-3q^2} )(-2p^2+3q^2+p\sqrt{4p^2-3q^2})}{4q^4}
}
These perfectly match the geometric formula \eqref{volx}. 
As evident by the denominator in \eqref{accc}, the case $q=0$ requires special care; performing $a$-maximization for $q=0$ leads to,
\ba{
\label{qzero}
q=0:\qquad \epsilon_B=0, \qquad \epsilon_F=\frac{1}{2} \quad\Rightarrow\quad a=c= \frac{27}{64}pN^2
}

\begin{table}[t!]
\centering
\begin{tabular}{c|c|cccc|c}
& \text{mult.} &  $U(1)_B$ & $U(1)_F$ & $U(1)_{R_0}$ & $U(1)_\varphi$ & $SU(2)_\varphi$  \\ \hline
$Y$ & $p+q$  & $p-q$ & $-1$ & 1 & 0  & {\bf 1} \\ 
$Z$ & $p-q$   & $p+q$ & 1 & 0 & 0  & {\bf 1}  \\
$U^{\alpha=1,2}$ & $p$  & $-p$  & 0  & $\tfrac{1}{2}$ & $\left(\tfrac{1}{2},-\tfrac{1}{2}\right)$ & ${\bf 2}$  \\
$V^{\alpha=1,2}$ & $q$ & $q$  & 1 &  $\tfrac{1}{2}$ & $\left(\tfrac{1}{2},-\tfrac{1}{2}\right)$  &  ${\bf 2}$
\end{tabular}
\caption{Charges of the fields in the $Y^{p,q}$ quivers under global symmetries. We have listed both the dimension of the representation under $SU(2)_\varphi$, and the weights under the Cartan $U(1)_\varphi \subset SU(2)_\varphi$.  \label{tab:ypqcharge}}
\end{table}

\paragraph{Chiral ring}
The chiral ring contains both mesonic operators, corresponding to traces of closed loops in the quiver, and baryonic operators, the simplest of which are the dibaryons formed from determinants of the microscopic fields \cite{Benvenuti:2004dy,Benvenuti:2005cz}. 
We first list the mesonic ``letters''---the  closed loops of operators that can be built before the trace over gauge indices is taken. The full set of mesonic chiral ring generators is the set of all possible single-trace closures (``words'') built by concatenating the operators enumerated below.\footnote{~For $N>1$, a general single-trace mesonic operator is the trace of a concatenation of the letters, symmetrized to maximal $SU(2)_\varphi$ spin. As the F-terms make the order of the letters irrelevant, and $\CL^+\CL^-\sim\CS^p$ (see \eqref{lls}),  every such operator is of the form
\ba{
T_{s,\ell} = \Tr\big[\CS^s (\CL^\pm)^{|\ell|}\big]_{\rm sym}\,,\qquad s\geq 0\,,\quad s+|\ell|\geq1
}
with $\CL^+$ for $\ell>0$ and $\CL^-$ for $\ell<0$. }
\begin{itemize}
\item {\it Short loops.} Each of the $2q$ triangles and $p-q$ quadrangles in \eqref{wpq} carries two doublet indices, and so decomposes into a spin 0 and spin 1 representation of $SU(2)_\varphi$. The spin-0 parts vanish in the chiral ring by the $Y$ and $\tilde Y$ F-terms.
The F-terms of the $U$, $V$ and $Z$ fields relate the spin-1 parts of short loops based at different nodes. For example, the $V_{i+1}$ F-term, $Y_iU_i^\alpha = U_{i+2}^\alpha Y_{i+1}$, identifies the two triangles that share $V_{i+1}$. All $p+q$ short loops are therefore equal in the chiral ring, 
and define a single triplet of $SU(2)_\varphi$, which we represent by the triangle based at node~1,
\ba{
\CS^{\alpha\beta} = U_1^{(\alpha}V_2^{\beta)}Y_1
}
(for $q=0$, by the quadrangle $U_1^{(\alpha}Z_2U_3^{\beta)}\tilde Y_1$).\footnote{~Ref.~\cite{Benvenuti:2005cz} writes this triplet as the superpotential with $\epsilon_{\alpha\beta}$ replaced by $\sigma^I_{\alpha\beta}$. With the index placement of \eqref{wpq}, that replacement would make the two triangles sharing a $V$ cancel in the chiral ring, so we use a single short loop instead.}

\item {\it Long loops.} There are two types of long loop operators. First there is an operator of length $2p$ that wraps the entire circular quiver, $\CL^+\sim U^p V^q Z^{p-q}$, which carries  $p+q$ $SU(2)_{\varphi}$ doublet indices. The $Y$ and $\tilde Y$ F-terms remove all but the totally symmetric part, which has spin $\frac{p+q}{2}$. 
The second long loop is of the form $\CL^-\sim Y^pU^{p-q}$. The F-terms again symmetrize its $p-q$ doublet indices, leaving the spin-$\frac{p-q}{2}$ representation. The $V$ F-terms allow the $U$ and $Y$ fields to be reordered, so for $p>q$ there is a single such loop.
\item The mesonic operators can be shown to satisfy the relation,
\ba{
\label{lls}
\CL^+\CL^-\sim\CS^p
}
\end{itemize}

The dibaryons are the determinants $\det\,Y$, $\det\,Z$, $\det\,U$ and $\det\,V$. The $SU(2)_\varphi$ indices of $\det\,U$ and $\det\,V$ are symmetrized, giving spin $\frac N2$. The charges of the mesonic letters and dibaryons are collected in Table~\ref{tab:ypqring}.  Four special cases appear below:
\begin{itemize}
\item For $p=q$ there are no $Z$ fields, and there are two distinct loops $\CL^-$, one through the odd nodes and one through the even nodes.
\item For $q=0$ there are no $V$ fields. Note that for $q=0$ one has $\epsilon_B=0$ and $\epsilon_F=\frac12$, so every field has $R_*=\frac12$---then $\CL^\pm$ have $R_*=p$, and all the dibaryons have $R_*=\frac N2$.
\item For $p=1$, $\CS\sim\CL^+\CL^-$ is not an independent letter.
\item For $p=q=1$ the $Y_i$ are adjoints. The $U(N)\times U(N)$ theory on the D3-branes then contains the chiral fields $\Tr Y_i$, whose F-terms set $\mu_B=\epsilon_{\alpha\beta}\Tr[U^\alpha V^\beta]$ to zero. In the $SU(N)\times SU(N)$ SCFT, the $\Tr Y_i$ are absent and $\mu_B$ is a chiral operator (see Section~\ref{sec:affine}).
\end{itemize}

\begin{table}[h!]
\centering
\small
\setlength{\tabcolsep}{4pt}
\renewcommand{\arraystretch}{1.4}
\begin{tabular}{l|ccc|c|c}
\toprule
 & $U(1)_B$ & $U(1)_F$ & $U(1)_{R_0}$ & $SU(2)_\varphi$ & $R_*$ \\ \midrule
$\CS$ & $0$ & $0$ & $2$ & ${\bf 3}$ & $2$ \\
$\CL^+\sim U^pV^qZ^{p-q}$ & $0$ & $p$ & $\frac12(p+q)$ & ${\bf p+q+1}$ & $\frac{p(2p+3q-\sqrt{4p^2-3q^2})}{3q}$ \\
$\CL^-\sim Y^pU^{p-q}$ & $0$ & $-p$ & $\frac12(3p-q)$ & ${\bf p-q+1}$ & $\frac{p(-2p+3q+\sqrt{4p^2-3q^2})}{3q}$ \\ \midrule
$\det\,Y$ (mult.~$p+q$) & $N(p-q)$ & $-N$ & $N$ & ${\bf 1}$ & $\frac{N(-4p^2+2pq+3q^2+(2p-q)\sqrt{4p^2-3q^2})}{3q^2}$ \\
$\det\,Z$ (mult.~$p-q$) & $N(p+q)$ & $N$ & $0$ & ${\bf 1}$ & $\frac{N(-4p^2-2pq+3q^2+(2p+q)\sqrt{4p^2-3q^2})}{3q^2}$ \\
$\det\,U$ (mult.~$p$) & $-Np$ & $0$ & $\frac N2$ & ${\bf N+1}$ & $\frac{2Np(2p-\sqrt{4p^2-3q^2})}{3q^2}$ \\
$\det\,V$ (mult.~$q$) & $Nq$ & $N$ & $\frac N2$ & ${\bf N+1}$ & $\frac{N(-2p+3q+\sqrt{4p^2-3q^2})}{3q}$ \\
\bottomrule
\end{tabular}
\caption{Generators of the chiral ring of operators for the  $Y^{p,q}$ quivers, with charges given for $q>0$ \cite{Benvenuti:2004dy,Benvenuti:2005cz}. The $R_*$ charges follow from \eqref{trialR} with the $a$-maximization values of $\epsilon_{B}$ and $\epsilon_F$ given above.\label{tab:ypqring}}
\end{table}

\section{Decoupling in Field Theory}
\label{sec:results}

In this section, we first discuss the map between the class $\CS_2$ and $Y^{p,q}$ chiral rings, obtained from their quiver descriptions. Then, we analyze how considering different regimes of $(N,p,q)$  result in different conclusions about the IR duality of the two quivers, depending on the relevance/irrelevance of the flip field coupling, and how this is reflected in the splitting of the anomaly polynomials. We proceed to examine three different case studies in detail: $(p=q=1)$, $(p=1,q=0)$, and $(p=4,q=3)$.

\subsection{Map between  class $\CS_2$ and $Y^{p,q}$ chiral rings}
\label{sec:ring}

First, let us discuss how the chiral ring of the $Y^{p,q}$ quivers reviewed in Section~\ref{sec:ypq} maps onto that of the class $\CS_2$ necklace quivers. The generalities discussed below will be carried out explicitly in the cases  $(p,q)=(1,1)$ in Section~\ref{sec:affine},  $(p,q)=(1,0)$ in Section~\ref{sec:conifold}, and $(p,q)=(4,3)$ in Section~\ref{sec:433}.

After closing the minimal punctures, the following light fields of the necklace quiver match onto those of the $Y^{p,q}$ quiver:

\begin{itemize}

 \item The fields $Q^{(\tau)}$ whose baryons are $\beta$-flipped map onto the $Y$ fields, and the fields  $\tilde{Q}^{(\tau)}$ whose baryons are $\gamma$-flipped are the $Z$ fields:
 \ba{\bs{
  Y \leftrightarrow  Q^{(\tau)} \,,\qquad \det\,Y  \leftrightarrow \det\,Q^{(\tau)} =B_\beta\\
  Z \leftrightarrow  \tilde{Q}^{(\tau)} \,,\qquad \det\,Z \leftrightarrow    \det\,\tilde{Q}^{(\tau)}=B_\gamma 
 }}
 The multiplicity $p+q$ of $Y$ bifundamentals is equal to the number of $\beta$-closures $n_\beta$, and the multiplicity  $p-q$ of $Z$ bifundamentals  is equal to the number of $\gamma$-closures $n_\gamma$, with $n_\beta + n_\gamma = 2p$ equaling the total number of trinions involved in the gluing. 
 
 \item Doublets $D_j=(X_j,\Phi_1^{(j)})$ of the $SU(2)_{\hat t}$ whose Cartan is $U(1)_{\hat t}$ map onto the $U$ and $V$ doublets.  
  The  construction via trinions breaks $SU(2)_{\hat{t}}$ to its Cartan $U(1)_{\hat t}$ (and for the conifold $(p,q)=(1,0)$, it also breaks the $SU(2)'$ that rotates $Y$ into $Z$), and operators organize into multiplets of these symmetries only at the fixed point.

 \end{itemize}

 Integrating out the heavy fields, the low-energy superpotential for the necklace quiver reproduces \eqref{wpq}.
 The chiral rings of the two IR fixed points agree, up to the following effect: 
the necklace quiver flip fields $\phi_B$ set the would-be-flipped baryons $B$ to zero in the chiral ring of the UV Lagrangian. At large $N$, these flip couplings are irrelevant, in which case  $\phi_B$ decouples and the baryon returns to the chiral ring as the corresponding $\det\,Y$ or $\det\,Z$. When the flip coupling is relevant, however, the chiral ring might differ from that of the $Y^{p,q}$ SCFT. In the following section we explore this point in detail, including a discussion of how to interpret the split of the anomalies from the points of view of the quiver gauge theories versus 6d reduction.


\subsection{Anomalies and the role of flip fields}
\label{sec:mainresults}

We now analyze how the anomalies split as in \eqref{flagship1}. 
Denote by $I_6(X)$ the anomaly polynomial of one chiral multiplet with the charges of $X$, namely,
\ba{
I_6(X) =\frac{1}{6} \Big((R_X-1)c_1(R)+ q^a_X c_1(F_a) \Big)^3-\Big((R_X-1)c_1(R)+ q^a_X c_1(F_a) \Big)\,\frac{p_1(TW_4)}{24}
}
Let $\phi_\beta$, $\phi_\gamma$ denote the flip fields corresponding to necklace quiver baryons $B_\beta,B_\gamma$, each coupled via a superpotential $W\supset \phi_\beta B_\beta + \phi_\gamma B_\gamma$. Their numbers $n_\beta,n_\gamma$ are determined by the choice of flux as in \eqref{flux1} and \eqref{fluxb},
\ba{
n_\beta = N_N = 2N_\beta = p+q\,,\qquad n _\gamma = N_S=2N_\gamma = p-q\,,\qquad n_\beta + n_\gamma = 2p
}
The following statements hold in general:

\begin{itemize}

\item The inherited term $\int_{T^2} I_8$, given explicitly in \eqref{checkb}, is the anomaly of the UV necklace quiver, including all the flip fields $\phi_\beta$, $\phi_\gamma$.  

\item $I_6^{\text{flip}}$, given in \eqref{i6flip}, is the sum of $I_6(\phi_B)$ over all $n_\beta + n_\gamma$ flip fields $\phi_\beta$, $\phi_\gamma$, or equivalently, minus the sum over the flipped baryons $B_\beta$, $B_\gamma$ (see Footnote~\ref{fn:flipp}):
\ba{
I_6^{\text{flip}} = n_\beta I_6(\phi_\beta) + n_\gamma I_6(\phi_\gamma) = - \left( n_\beta I_6(B_\beta)+n_\gamma I_6(B_\gamma)\right)
}

\end{itemize}

In comparing the anomalies computed from the difference $\int_{T^2} I_8 - I_6^{\text{flip}}$ with those of  the necklace quiver SCFT, we need to differentiate between the following three regimes, depending on the dimension of each baryon $B_\beta$, $B_\gamma$ at the $Y^{p,q}$ fixed point (where recall the dimension is related to the R-charge at the fixed point as $\Delta = \tfrac{3}{2} R_*$, with $R_*$ charges given in  Table~\ref{tab:ypqring}). We will first keep the discussion general to a baryon $B$ that could either be associated with a $\beta$- or $\gamma$-closure, and then specify for which values of the flux parameters each regime applies to  $B_\beta$ versus $B_\gamma$.

\begin{description}

\item[A:]  The coupling of a baryon $B$ to its flip field $\phi_B$ is irrelevant, with $R(B)>4/3$ (equivalently, $R(\phi_B)<2/3$). Then, the flip field decouples as a free singlet, and $B$ remains as a chiral primary in the interacting chiral ring. The necklace quiver flows to the interacting $Y^{p,q}$ SCFT plus free fields $\phi_B$, so that\footnote{~Note that if one were to {\it not} remove the flips from the interacting fixed point, the large-$N$ scaling of the central charges would be contaminated: the $\CO(N^2)$ scaling would be reduced to order $\CO(N^{3/2})$. This is another clear signal that there is no consistent interacting fixed point that includes the flip fields and decouples the baryons for general values of the fluxes.} 
\ba{
I_6^{\text{SCFT}}[\text{necklace}] =I_6^{\text{SCFT}}[Y^{p,q}] = \int_{T^2} I_8 - I_6^{\text{flip}}
}

\item[B:] The flip coupling is relevant, with $2/3 < R(B)< 4/3$.\footnote{~In the special case that $R(B)=4/3$ the coupling is marginally irrelevant \cite{Green:2010da}, and so the behavior is as in  Regime {\bf A}.} Then, the necklace quiver flows to an interacting fixed point in which the baryon $B$ is no longer in the chiral ring (it is removed by $\phi_B$'s F-term, $B \sim \overline{D}^2\overline{\phi}_B$), while $\phi_B$ survives as a chiral operator. In this regime, the necklace quiver flows to an interacting fixed point which is {\it not} the same as the $Y^{p,q}$ SCFT---in particular, the $Y^{p,q}$ quiver does not have the flip field but does include the dibaryon $B\sim \det\,Y$ or $\det\,Z$. 

\item[C:] The flip coupling is relevant, with $R(B)<2/3$. 
In the $Y^{p,q}$ quiver, where there was no flip to begin with, the corresponding dibaryon  apparently violates the unitarity bound, signaling that it becomes free and decouples from the interacting fixed point. In the necklace quiver, the flip acts on (what would be) a free field, and both $\phi_B$ and $B$ are lifted from the spectrum. (One can verify that no other operators have corrected R-charges that fall below the unitarity bound.) Both quivers flow to the same interacting SCFT, which has $B$ removed from the spectrum (and no flip fields $\phi_B$ in the spectrum), although their free sectors differ as the $Y^{p,q}$ quiver includes the decoupled free fields $B$. 

\end{description}

The anomalies of the interacting fixed point obtained from the necklace quiver versus $Y^{p,q}$ quiver are then given by how many flip field couplings fall into regimes {\bf A} or {\bf B},
\ba{
I_6^{\text{SCFT}}[\text{necklace}] = \int_{T^2} I_8-I^{\text{flip}}_{\bf A}\,,\qquad I_6^{\text{SCFT}}[Y^{p,q}] = \int_{T^2} I_8-I^{\text{flip}}_{\bf A}-I^{\text{flip}}_{\bf B}
}
In the context of the discussion from the introduction, which part of \eqref{naiveloc} receives the contribution of the flip fields $\phi_B$ depends on the specific values of $(N,p,q)$.  This is what makes this split \eqref{naiveloc} into the inherited terms $\int_{T^2} I_8$ and localized contributions $\delta I_6^{\text{loc},i}$ dynamical; the localized correction is decided by RG flow.

Next we discuss the conditions on the values of $(N,p,q)$ for the theories to lie in each of the regimes {\bf A}, {\bf B}, {\bf C}.
For all $N\geq 3$, the $\beta$-flip couplings are irrelevant and in regime {\bf A}; as can be deduced from Table~\ref{tab:ypqring},\footnote{~$N=2$ is a special case, with $\det\,Y$ in regime B for $p>q$, and marginally irrelevant for $p=q$. We have generally restricted to $N>2$ in our analysis.}
\ba{
\frac{N}{2}\leq R_*(\det\,Y) \leq \frac{2N}{3} \quad \text{for all }p,q
}
Then, the $\phi_\beta$ flip fields decouple as free singlets from the interacting necklace quiver fixed point, with baryons $B_\beta \sim \det\,Y$ remaining in the chiral ring. 

\begin{figure}[t!]
\centering
\includegraphics[width=0.62\textwidth]{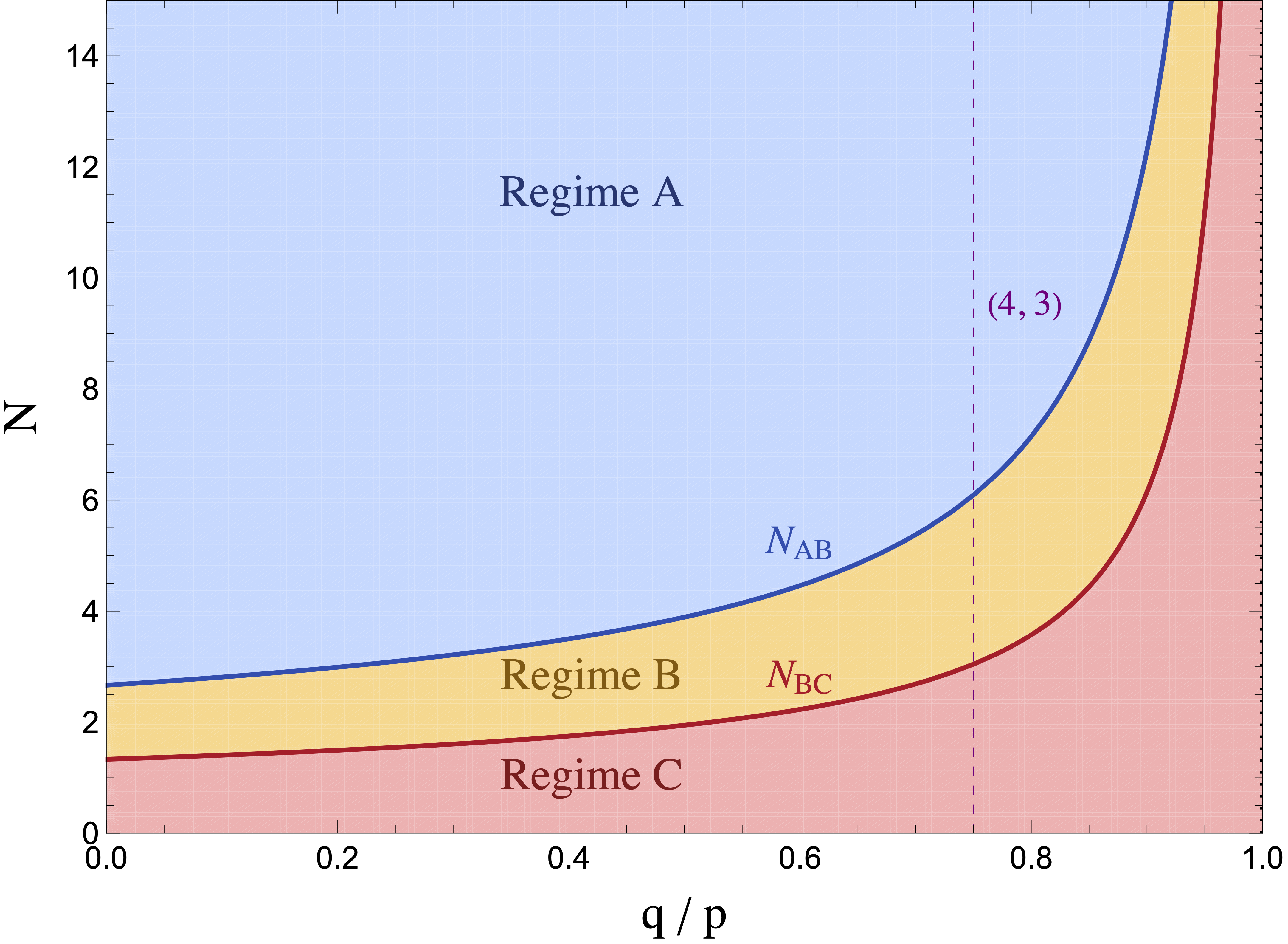}
\caption{Regimes for the $\gamma$-flip coupling in the $(q/p, N)$
plane. Above the blue curve ($N > N_{AB}(x)$), the coupling
$\phi_\gamma \det\,Z$ is irrelevant: $\phi_\gamma$ decouples as a free
singlet and the baryon $\det\,Z$ remains in the interacting chiral
ring (Regime~{\bf A}). Between the two curves ($N_{BC}(x) < N < N_{AB}(x)$), the
flip coupling is relevant and removes $\det\,Z$ from the chiral ring,
while $\phi_\gamma$ survives as an interacting chiral operator
(Regime~{\bf B}). Below the red curve ($N < N_{BC}(x)$), the $Z$-dibaryon
falls below the unitarity bound and is removed from the spectrum
together with its flip (Regime~{\bf C}). }
\label{fig:bounds_gamma_flip}
\end{figure}

The $\gamma$-flips do not necessarily lie in this regime. From
Table~\ref{tab:ypqring}, the R-charge at the $Y^{p,q}$ fixed point of the
$\gamma$-flipped baryon $\det\,Z$ is,
\ba{
\label{RdetZ}
R_*(\det\,Z) = N\, r_Z(q/p)\,, \qquad
r_Z(x) \equiv \frac{-4 - 2x + 3x^2 + (2+x)\sqrt{4 - 3x^2}}{3x^2}
}
where we have used that $r_Z$ is homogeneous of degree zero in $(p,q)$
to express it as a function of the ratio $x = q/p$ alone. As discussed
above, the coupling $\phi_\gamma \det\,Z$ is in Regime~{\bf A} when
$R_*(\det\,Z) > 4/3$, in Regime~{\bf B} when $2/3 < R_*(\det\,Z) < 4/3$, and
in Regime~{\bf C} when $R_*(\det\,Z) < 2/3$. Since $R_*(\det\,Z)$ is linear in
$N$, each threshold translates into a curve in the $(q/p, N)$ plane,
\ba{
\label{NCNA}
N_{AB}(x) = \frac{4/3}{r_Z(x)}\,,\qquad
N_{BC}(x) = \frac{2/3}{r_Z(x)}
}
with $N_{AB}(x)$ separating Regimes~{\bf A} and~{\bf B}, and $N_{BC}(x)$ separating
Regimes~{\bf B} and~{\bf C}. In particular, the $\gamma$-flip lies in Regime~{\bf A} for
$N > N_{AB}(x)$, which is generically the case for large-$N$, and the reason why these considerations do not affect the large-$N$ statement of the duality; in Regime~{\bf B} for $N_{BC}(x) < N < N_{AB}(x)$; and in Regime~{\bf C}
for $N < N_{BC}(x)$. The two curves are plotted as a function of $q/p$ in
Figure~\ref{fig:bounds_gamma_flip}.

Three features of \eqref{NCNA} are worth emphasizing. First, (by construction) the two
boundaries satisfy $N_{AB}(x) = 2\,N_{BC}(x)$ identically: the ratio of the
thresholds $4/3$ and $2/3$ is fixed, and $N$ enters
$R_*(\det\,Z)$ linearly, so the two curves never cross. Second, both
curves depend only on the ratio $x=q/p$, not on $p$ and $q$ separately.
Third, the two limits $q/p \to 0$
and $q/p \to 1$ behave in opposite ways. As $x \to 0$ we have
$r_Z \to 1/2$ matching onto the $q=0$ case~\eqref{qzero}, so that $N_{BC} \to 4/3$ and $N_{AB} \to 8/3$. Then, the
$\gamma$-flip is in Regime~{\bf A} for all $N \geq 3$, and in Regime~{\bf B} for
$N = 2$.
Conversely, $r_Z \to 0$ as $x \to 1$, so both curves
diverge as $q$ approaches $p$. Then, for any fixed $N$, the $\gamma$-flip
is in Regime~{\bf C} for $q/p$ sufficiently close to $1$. This last limit
reflects the degeneration of the formula when the $Z$ fields
themselves disappear at $q = p$; the range of validity of
\eqref{NCNA} is $q < p$.

We explore the interplay of these regimes in the example
$(p=4, q=3)$ of Section~\ref{sec:433}, which contains all three
regimes as $N$ varies. At $q/p = 3/4$ we find
$N_{AB}(3/4) \approx 6.09$ and $N_{BC}(3/4) \approx 3.04$, so that the
$\gamma$-flip is in Regime~{\bf A} for $N \geq 7$, in Regime~{\bf B} for
$N = 4, 5, 6$, and in Regime~{\bf C} for $N = 3$. The vertical dashed line
in Figure~\ref{fig:bounds_gamma_flip} marks $q/p = 3/4$, the value
relevant for this example. This example is useful precisely because
it realizes all three regimes at fixed flux, so that the RG flow of
the $Z$-dibaryon can be read off directly from the value of $N$.


\paragraph{The affine quiver}
The affine quiver $p=q=1$ is a special case, due to the presence of adjoints in the quiver. For all $(p,q)$ the $Y^{p,q}$ SCFT is the $SU(N)^{2p}$ theory: the $U(1)$ factors of the gauge group $U(N)^{2p}$ on the D3-branes are anomalous or IR free, and decouple, and the non-anomalous combination survives as the global $U(1)_B$ under which the dibaryons are charged \cite{Klebanov:1998hh,Benvenuti:2004dy}. The necklace quivers, meanwhile, have $SU(N)$ gauge nodes from the start. For $p=q=1$ the $Y$ fields are adjoints, and the traces $\Tr Y_i$ of the $U(N)\times U(N)$ theory correspond to the singlets $y_\tau$ of Section~\ref{sec:affine}, with both free in the IR. Then, these singlets are counted by $\int I_8-I_6^{\text{flip}}$, but are not present at the  $SU(N)^2$ fixed point. For $N\geq 3$ all flip fields $\phi_B$ are in regime~{\bf A}, so that the necklace quiver flows to the $Y^{1,1}$ SCFT together with the free flip fields and singlets $y_{\tau=1,2}$, and the anomaly polynomials are related as $ I^{\text{SCFT}}[Y^{1,1}] = \int_{T^2} I_8 - I^{\text{flip}} - I_6(y_1) - I_6(y_2)$.\footnote{~Therefore, for the special case $p=q=1$ with $N\geq 3$,  $I_6^{\text{decouple}}$ in \eqref{inflow} is not zero as written in \eqref{t2split}, but satisfies $I^{\text{decouple}}|_{\text{res}}=I_6(y_1)+I_6(y_2)\neq 0$.
\label{fn:pq1}}

\subsection{Case study: the affine quiver  \texorpdfstring{$(p=q=1)$}{(p=q=1)}}
\label{sec:affine}

	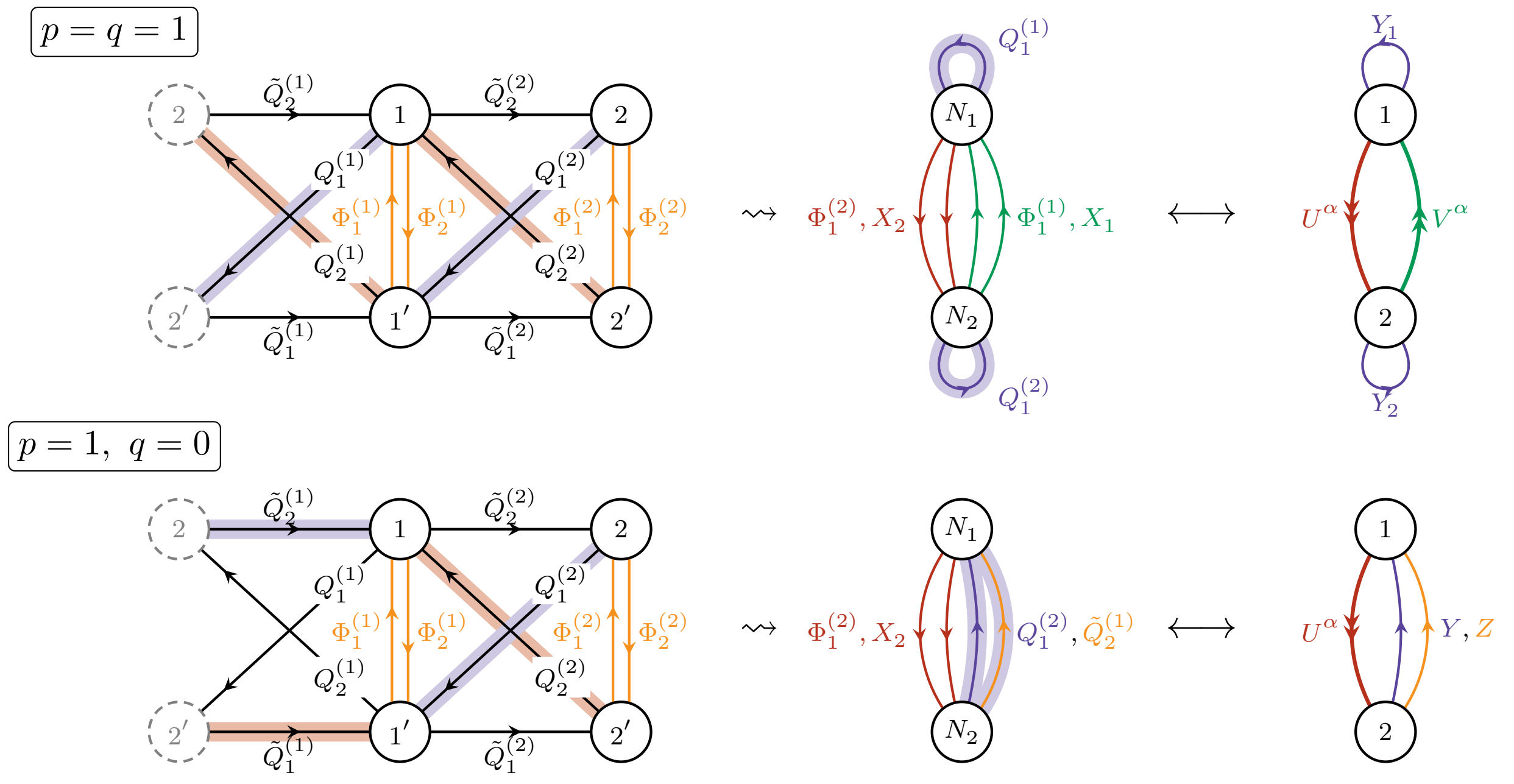
\begin{figure}[t!]
	\centering
\begin{tikzpicture}[
  gn/.style={circle, draw, thick, fill=white, minimum size=6.5mm, inner sep=0pt, font=\scriptsize},
  gc/.style={gn, densely dashed, draw=gray, text=gray},
  fd/.style={thick, postaction={decorate, decoration={markings, mark=at position #1 with {\arrow{stealth}}}}},
  fd/.default=0.55,
  ph/.style={thick, BurntOrange, postaction={decorate, decoration={markings, mark=at position 0.62 with {\arrow{stealth}}}}},
  dbl/.style={very thick, #1, postaction={decorate, decoration={markings,
      mark=at position 0.5 with {\arrow{stealth}}, mark=at position 0.5 with {\arrow[xshift=3.5pt]{stealth}}}}},
  hv/.style={line width=6pt, BrickRed!28, line cap=round},
  hf/.style={line width=6pt, Violet!22, line cap=round},
  lb/.style={font=\scriptsize, inner sep=1pt},
  cl/.style={font=\scriptsize, inner sep=0.6pt, fill=white},
  box/.style={draw, rounded corners=2pt, font=\small, inner sep=3pt},
]
\colorlet{cU}{BrickRed}\colorlet{cV}{ForestGreen}\colorlet{cZ}{BurntOrange}\colorlet{cY}{Violet}
\newcommand{\basenecklace}{%
  \draw[fd] (oL) -- node[lb, above] {$\tilde Q_2^{(1)}$} (o1);
  \draw[fd] (iL) -- node[lb, below] {$\tilde Q_1^{(1)}$} (i1);
  \draw[fd=0.8] (o1) -- node[cl, pos=0.27] {$Q_1^{(1)}$} (iL);
  \draw[fd=0.8] (i1) -- node[cl, pos=0.27] {$Q_2^{(1)}$} (oL);
  \draw[fd] (o1) -- node[lb, above] {$\tilde Q_2^{(2)}$} (o2);
  \draw[fd] (i1) -- node[lb, below] {$\tilde Q_1^{(2)}$} (i2);
  \draw[fd=0.8] (o2) -- node[cl, pos=0.27] {$Q_1^{(2)}$} (i1);
  \draw[fd=0.8] (i2) -- node[cl, pos=0.27] {$Q_2^{(2)}$} (o1);
  \foreach \c in {1,2} {
    \draw[ph] ([xshift=-2.5pt]i\c) -- ([xshift=-2.5pt]o\c);
    \draw[ph] ([xshift=2.5pt]o\c) -- ([xshift=2.5pt]i\c);
  }
  \foreach \c/\x in {1/-5.5, 2/-3.1} {
    \node[lb, text=BurntOrange, anchor=east] at (\x-0.14,0) {$\Phi_1^{(\c)}$};
    \node[lb, text=BurntOrange, anchor=west] at (\x+0.14,0) {$\Phi_2^{(\c)}$};
  }
  \node[gc] at (oL) {$2$}; \node[gc] at (iL) {$2'$};
  \node[gn] at (o1) {$1$}; \node[gn] at (i1) {$1'$};
  \node[gn] at (o2) {$2$}; \node[gn] at (i2) {$2'$};
}
\begin{scope}
  \node[box] at (-8.6,2) {$p=q=1$};
  \coordinate (oL) at (-7.9,1.1); \coordinate (iL) at (-7.9,-1.1);
  \coordinate (o1) at (-5.5,1.1); \coordinate (i1) at (-5.5,-1.1);
  \coordinate (o2) at (-3.1,1.1); \coordinate (i2) at (-3.1,-1.1);
  \draw[hv] (i1) -- (oL); \draw[hv] (i2) -- (o1);
  \draw[hf] (o1) -- (iL); \draw[hf] (o2) -- (i1);
  \basenecklace
  \node at (-1.6,0) {$\rightsquigarrow$};
  \begin{scope}[xshift=6mm]
  \coordinate (N1) at (0,1.1); \coordinate (N2) at (0,-1.1);
  \draw[hf] (N1) .. controls +(50:1.35) and +(130:1.35) .. (N1);
  \draw[hf] (N2) .. controls +(-130:1.35) and +(-50:1.35) .. (N2);
  \draw[fd, cU] (N1) to[bend right=45] (N2);
  \draw[fd, cU] (N1) to[bend right=15] (N2);
  \draw[fd, cV] (N2) to[bend right=15] (N1);
  \draw[fd, cV] (N2) to[bend right=45] (N1);
  \draw[fd, cY] (N1) .. controls +(50:1.35) and +(130:1.35) .. (N1);
  \draw[fd, cY] (N2) .. controls +(-130:1.35) and +(-50:1.35) .. (N2);
  \node[lb, text=cU, left] at (-0.55,0) {$\Phi_1^{(2)},X_2$};
  \node[lb, text=cV, right] at (0.55,0) {$\Phi_1^{(1)},X_1$};
  \node[lb, text=cY, right] at (0.35,1.95) {$Q_1^{(1)}$};
  \node[lb, text=cY, right] at (0.35,-1.95) {$Q_1^{(2)}$};
  \node[gn] at (N1) {$N_1$}; \node[gn] at (N2) {$N_2$};
  \node at (2.6,0) {$\longleftrightarrow$};
  \coordinate (Y1) at (4.6,1.1); \coordinate (Y2) at (4.6,-1.1);
  \draw[dbl=cU] (Y1) to[bend right=35] node[lb, left=2pt, text=cU] {$U^\alpha$} (Y2);
  \draw[dbl=cV] (Y2) to[bend right=35] node[lb, right=2pt, text=cV] {$V^\alpha$} (Y1);
  \draw[fd, cY] (Y1) .. controls +(50:1.35) and +(130:1.35) .. node[lb, above, text=cY] {$Y_1$} (Y1);
  \draw[fd, cY] (Y2) .. controls +(-130:1.35) and +(-50:1.35) .. node[lb, below, text=cY] {$Y_2$} (Y2);
  \node[gn] at (Y1) {$1$}; \node[gn] at (Y2) {$2$};
  \end{scope}
\end{scope}
\begin{scope}[shift={(0,-4.5)}]
  \node[box] at (-8.6,2) {$p=1,\ q=0$};
  \coordinate (oL) at (-7.9,1.1); \coordinate (iL) at (-7.9,-1.1);
  \coordinate (o1) at (-5.5,1.1); \coordinate (i1) at (-5.5,-1.1);
  \coordinate (o2) at (-3.1,1.1); \coordinate (i2) at (-3.1,-1.1);
  \draw[hv] (iL) -- (i1); \draw[hv] (i2) -- (o1);
  \draw[hf] (oL) -- (o1); \draw[hf] (o2) -- (i1);
  \basenecklace
  \node at (-1.6,0) {$\rightsquigarrow$};
  \begin{scope}[xshift=6mm]
  \coordinate (N1) at (0,1.1); \coordinate (N2) at (0,-1.1);
  \draw[hf] (N2) to[bend right=15] (N1);
  \draw[hf] (N2) to[bend right=45] (N1);
  \draw[fd, cU] (N1) to[bend right=45] (N2);
  \draw[fd, cU] (N1) to[bend right=15] (N2);
  \draw[fd, cY] (N2) to[bend right=15] (N1);
  \draw[fd, cZ] (N2) to[bend right=45] (N1);
  \node[lb, text=cU, left] at (-0.55,0) {$\Phi_1^{(2)},X_2$};
  \node[lb, right] at (0.55,0) {$\textcolor{Violet}{Q_1^{(2)}},\textcolor{BurntOrange}{\tilde Q_2^{(1)}}$};
  \node[gn] at (N1) {$N_1$}; \node[gn] at (N2) {$N_2$};
  \node at (2.6,0) {$\longleftrightarrow$};
  \coordinate (Y1) at (4.6,1.1); \coordinate (Y2) at (4.6,-1.1);
  \draw[dbl=cU] (Y1) to[bend right=35] node[lb, left=2pt, text=cU] {$U^\alpha$} (Y2);
  \draw[fd, cY] (Y2) to[bend right=15] (Y1);
  \draw[fd, cZ] (Y2) to[bend right=45] (Y1);
  \node[lb, right] at (5.15,0) {$\textcolor{Violet}{Y},\textcolor{BurntOrange}{Z}$};
  \node[gn] at (Y1) {$1$}; \node[gn] at (Y2) {$2$};
  \end{scope}
\end{scope}
\end{tikzpicture}
	\caption{The map between the necklace and $Y^{p,q}$ quivers for the affine quiver (top) and the conifold (bottom), in the conventions of Sections~\ref{sec:affine} and ~\ref{sec:conifold}. On the left is the base necklace with $2p=2$ trinions, drawn as in Figure~\ref{fig:trinion}; the dashed nodes are the periodic images of $2$ and $2'$, so that trinion~1 joins them to $1$ and $1'$. The fields that get VEVs are shaded red, and the leftover blocks whose baryons are flipped are shaded purple. The quiver of light fields after the closures is depicted in the middle, with nodes $N_1=(1,2')$, $N_2=(2,1')$ for the affine quiver and $N_1=(1,1',2')$, $N_2=(2)$ for the conifold, and the light fields labeled as in \eqref{mapaffine} and \eqref{mapconifold}. Right: the $Y^{1,1}$ and $Y^{1,0}$ quivers, colored to match Figure~\ref{fig:ypq}. \label{fig:quivers}}
	\end{figure}

\paragraph{$Y^{1,1}$ description}
For $p=q=1$ the quiver has two nodes joined by the doublets $U^\alpha$ ($1\to 2$) and $V^\alpha$ ($2\to 1$). The two $Y$ fields are adjoints: $Y_1$ of node~1 and $Y_2$ of node~2. The superpotential \eqref{wpq} becomes
\ba{
\label{Wy11}
W = \epsilon_{\alpha\beta}\Big(\Tr\big[U^\alpha V^\beta Y_1\big] - \Tr\big[V^\beta U^\alpha Y_2\big]\Big)
}
and the charges are those of Table~\ref{tab:ypqcharge} and Table~\ref{tab:ypqring} at $p=q=1$. This is the $\CN=2$ $\hat A_1$ quiver, with $Y^{1,1}=S^5/\mathbb{Z}_2$, where the $Y_i$ are the adjoint chirals of the $\CN=2$ vector multiplets, and $SU(2)_\varphi\times U(1)_B$ is the $U(2)$ flavor symmetry of the two bifundamental hypermultiplets. $R_*=\frac23$ for every field.

Because the $Y_i$ are adjoints, the $U(N)\times U(N)$ theory on the D3-branes must be distinguished from the $SU(N)\times SU(N)$ SCFT. In the $U(N)$ quiver, the $\CL^-$ operators split into two types $(\CL^-)_i= Y_i$, whose traces are  free fields that decouple from the interacting fixed point: $\Tr Y_1+\Tr Y_2$ is the free center-of-mass mode of the brane stack, and the orthogonal combination $\Tr Y_1-\Tr Y_2$ is also IR free. This renders the $Y_i$ traceless at the interacting fixed point, so that the generators $\Tr \CL^-_i=\Tr Y_i$ are absent. The $Y_i$ F-terms then set only the traceless part of $\epsilon_{\alpha\beta}U^\alpha V^\beta$ to zero. Its trace,
\ba{
\mu_B = \epsilon_{\alpha\beta}\Tr\big[U^\alpha V^\beta\big]
}
is the moment map of $U(1)_B$ and remains in the chiral ring, alongside the triplet $\CL^+$ \cite{Gadde:2010zi}.\footnote{~In \cite{Gadde:2010zi}, $\mu_B$ is the twisted-sector $\hat{\mathcal B}_1$ multiplet  of the $\mathbb Z_2$ orbifold (their Table~6), and in  $\CN=2$ superconformal QCD they note that the protection of this operator relies on the gauge group being $SU(N)$ rather than $U(N)$. The mesonic chiral rings of \cite{Benvenuti:2004dy,Benvenuti:2005cz} are those of the $U(N)\times U(N)$ theory, which contain $\Tr Y_i$ and in which $\mu_B=0$.} The generators are listed in Table~\ref{tab:y11ring}.

\begin{table}[t!]
\centering
\renewcommand{\arraystretch}{1.3}
\begin{tabular}{l|ccc|c|c}
\toprule
& $U(1)_B$ & $U(1)_F$ & $U(1)_{R_0}$ & $SU(2)_\varphi$ & $R_*$ \\ \midrule
$\CS^{\alpha\beta} = \Tr\big[U^{(\alpha}V^{\beta)}Y_1\big]$ & $0$ & $0$ & $2$ & ${\bf 3}$ & $2$ \\
$(\CL^+)^{\alpha\beta} = \Tr\big[U^{(\alpha}V^{\beta)}\big]$ & $0$ & $1$ & $1$ & ${\bf 3}$ & $\frac43$ \\
$\mu_B = \epsilon_{\alpha\beta}\Tr\big[U^{\alpha}V^{\beta}\big]$ & $0$ & $1$ & $1$ & ${\bf 1}$ & $\frac43$ \\
$\Tr\, Y_i^k$, $k=2,\dots,N$ & $0$ & $-k$ & $k$ & ${\bf 1}$ & $\frac{2k}{3}$ \\ \midrule
$\det\,Y_{i=1,2}$ & $0$ & $-N$ & $N$ & ${\bf 1}$ & $\frac{2N}{3}$ \\
$\det\,U$ & $-N$ & $0$ & $\frac N2$ & ${\bf N+1}$ & $\frac{2N}{3}$ \\
$\det\,V$ & $N$ & $N$ & $\frac N2$ & ${\bf N+1}$ & $\frac{2N}{3}$ \\
\bottomrule
\end{tabular}
\caption{Basic chiral operators of the $SU(N)\times SU(N)$ $Y^{1,1}$ SCFT. Note that the $\det\,Y_i$ are composites of the $\Tr Y_i^k$.\label{tab:y11ring}}
\end{table}

\paragraph{Necklace quiver description}
Consider now the necklace quiver of Section~\ref{sec:necklace} with $2p=2$ trinions labeled by $\tau=1,2$. Its fields and superpotential are given in Table~\ref{tab:trinion} and Eq.~\eqref{suptau}. The fluxes $N_N=p+q=2$ and $N_S=p-q=0$ require two $\beta$-closures. 
We give VEVs to the $\hat\beta=-1$ criss-cross fields as,
\ba{
\label{alphalockaffine}
\langle Q_2^{(1)}\rangle = v_1\mathbf 1\,,\qquad \langle Q_2^{(2)}\rangle = v_2\mathbf 1\,,\qquad \alpha_{\tau=1,2} = t^{1/2}\hat\beta^{-1}
}
The closures identify nodes $1\sim 2'$ and $2\sim 1'$, leaving the two nodes $N_1=(1,2')$ and $N_2=(2,1')$. The mass terms are
\ba{
W_{\rm mass} = \Tr\Big[\Phi_2^{(1)}\big(v_1\tilde Q_2^{(1)}-v_2\tilde Q_1^{(2)}\big)\Big] + \Tr\Big[\Phi_2^{(2)}\big(v_2\tilde Q_2^{(2)}-v_1\tilde Q_1^{(1)}\big)\Big]
}
The  following combinations remain massless,
\ba{
X_1 = \frac{v_2\tilde Q_2^{(1)}+v_1\tilde Q_1^{(2)}}{n}\,,\qquad X_2 = \frac{v_1\tilde Q_2^{(2)}+v_2\tilde Q_1^{(1)}}{n}\,,\qquad n=\sqrt{v_1^2+v_2^2}
}
Each $X_j$ pairs with $\Phi_1^{(j)}$ into a doublet $D_j=(X_j,\Phi_1^{(j)})$ of the $SU(2)_{\hat{t}}$ whose Cartan is $U(1)_{\hat t}$. The fields $Q_1^{(\tau)}$, whose baryons are flipped, now run from node $N_\tau$ to itself and are thus adjoints. The light fields are listed in Table~\ref{tab:affine}, with charges following from Appendix~\ref{sec:master} with $(\kappa_\beta,\kappa_\gamma)=(\frac13,0)$.

\begin{table}[t!]
\centering
\renewcommand{\arraystretch}{1.2}
\begin{tabular}{l c|cccc|cccc|c}
\toprule
Field & Edge & $\hat\beta$ & $\hat\gamma$ & $\hat R$ & $\hat t$ & $B$ & $F$ & $R_0$ & $\varphi$ & $Y^{1,1}$ \\
\midrule
$Q_1^{(1)}$ & $N_1\to N_1$ & $2$ & $0$ & $\frac23$ & $0$ & $0$ & $-1$ & $1$ & $0$ & $Y_1$ \\
$Q_1^{(2)}$ & $N_2\to N_2$ & $2$ & $0$ & $\frac23$ & $0$ & $0$ & $-1$ & $1$ & $0$ & $Y_2$ \\
$\Phi_1^{(1)}$ & $N_2\to N_1$ & $-1$ & $1$ & $\frac23$ & $-1$ & $1$ & $1$ & $\frac12$ & $\frac12$ & $V^1$ \\
$X_1\sim \tilde{Q}_2^{(1)} + \tilde{Q}_1^{(2)}$ & $N_2\to N_1$ & $-1$ & $1$ & $\frac23$ & $1$ & $1$ & $1$ & $\frac12$ & $-\frac12$ & $V^2$ \\
$\Phi_1^{(2)}$ & $N_1\to N_2$ & $-1$ & $-1$ & $\frac23$ & $-1$ & $-1$ & $0$ & $\frac12$ & $\frac12$ & $U^1$ \\
$X_2\sim \tilde{Q}_1^{(1)} + \tilde{Q}_2^{(2)}$ & $N_1\to N_2$ & $-1$ & $-1$ & $\frac23$ & $1$ & $-1$ & $0$ & $\frac12$ & $-\frac12$ & $U^2$ \\ \midrule
$\phi_\beta^{(\tau)}$, $\tau=1,2$ & -- & $-2N$ & $0$ & $2-\frac{2N}{3}$ & $0$ & $0$ & $N$ & $2-N$ & $0$ & -- \\
\bottomrule
\end{tabular}
\caption{Light fields of the $p=q=1$ necklace quiver after two $\beta$-closures, including the flip fields, with their charges given in the hatted necklace basis as well as the $Y^{p,q}$ quiver basis. Here $\hat R$ is the reference R-symmetry of Appendix~\ref{sec:master}, which at $p=q$ assigns $\hat R=\frac23$ to every light
    field. \label{tab:affine}}
\end{table}

Integrating out the massive pairs sets the massive combinations to zero, since the $\Phi_2^{(\tau)}$ appear only in $W_{\rm mass}$. The remaining terms of \eqref{suptau} give, 
\ba{
\label{eq:Wlowaffine}
\bs{
W_{\rm low} =&\ \frac{v_2}{n}\Tr\Big[Q_1^{(1)}\big(X_2\Phi_1^{(1)}-\Phi_1^{(2)}X_1\big)\Big] + \frac{v_1}{n}\Tr\Big[Q_1^{(2)}\big(X_1\Phi_1^{(2)}-\Phi_1^{(1)}X_2\big)\Big] + W_{\rm flip}\,,\\
W_{\rm flip} =&\ \phi_\beta^{(1)}\det\,Q_1^{(1)} + \phi_\beta^{(2)}\det\,Q_1^{(2)}
}}
Up to an overall normalization and a rescaling of $Q_1^{(2)}$ by $v_1/v_2$, this is the $Y^{1,1}$ superpotential \eqref{Wy11} with
\ba{
\label{mapaffine}
Y_\tau\leftrightarrow Q_1^{(\tau)}\,,\qquad U^\alpha\leftrightarrow\big(\Phi_1^{(2)},\,X_2\big)\,,\qquad V^\alpha\leftrightarrow\big(\Phi_1^{(1)},\,X_1\big)
}
consistent with the charges in Table~\ref{tab:affine}. While the $SU(2)_{\hat{t}}$ is broken to its Cartan in the construction of the necklace quiver, at the fixed point it is restored, with fields organizing into the expected $SU(2)_{\hat{t}}$ multiplets.

Consider in more detail the adjoints $Q_1^{(\tau)}$. Each $Q_1^{(\tau)}$ descends from a bifundamental of the two nodes that the closures identify, so it has $N^2$ components: under the diagonal $SU(N)$ it decomposes into an adjoint $\hat Q_1^{(\tau)}$, which plays the role of $Y_\tau$, and a gauge singlet $y_\tau\equiv\Tr Q_1^{(\tau)}$, as
\ba{
\label{tracee}
Q_1^{(\tau)} = \hat Q_1^{(\tau)} + \frac{y_\tau}{N}\,\mathbf 1\,,\qquad \Tr\hat Q_1^{(\tau)}=0
}
The adjoints $\hat Q_1^{(\tau)}$ enter \eqref{eq:Wlowaffine} exactly as the traceless $Y_\tau$ enter \eqref{Wy11}. Of the singlets, only the combination $v_2y_1-v_1y_2$ appears in the cubic terms, coupled to $\mu_B$,
\ba{
W_{\rm low}\supset -\frac{v_2y_1-v_1y_2}{nN}\,\mu_B\,,\qquad \mu_B \sim \Tr\big[\Phi_1^{(2)}X_1-X_2\Phi_1^{(1)}\big]
}
while the orthogonal combination $v_1y_1+v_2y_2$ drops out. Both traces also appear in $W_{\rm flip}$, since $\det\,Q_1^{(\tau)}$ is the determinant of the full matrix.  The coupling to $\mu_B$ is marginal, and it breaks the $U(1)$ that acts on $v_2y_1-v_1y_2$ alone, so it is marginally irrelevant \cite{Green:2010da}. Meanwhile, for $N\geq 3$, $R_*(\det\,Q_1^{(\tau)})=\frac{2N}{3}>\frac43$, so the flip couplings are irrelevant, and the $\phi_\beta^{(\tau)}$ decouple as free fields while the baryons $\det\,Q_1^{(\tau)}$ remain in the chiral ring, up to terms involving the decoupled $y_\tau$.
Both trace modes are therefore free in the IR, like $\Tr Y_1\pm\Tr Y_2$ in the D3-brane theory, and $\mu_B$ survives in the chiral ring of the interacting SCFT. 

\paragraph{Operator map}
Under \eqref{mapaffine}, the generators of Table~\ref{tab:y11ring} map as follows:
\ba{\bs{
\CS^{\alpha\beta} \quad&\leftrightarrow\quad \Big(\Tr\big[\Phi_1^{(2)}\Phi_1^{(1)}\hat Q_1^{(1)}\big]\,,\ \Tr\big[(\Phi_1^{(2)}X_1+X_2\Phi_1^{(1)})\, \hat Q_1^{(1)}\big]\,,\ \Tr\big[X_2X_1\hat Q_1^{(1)}\big]\Big)\\
(\CL^+)^{\alpha\beta} \quad&\leftrightarrow\quad \Big(\Tr\big[\Phi_1^{(2)}\Phi_1^{(1)}\big]\,,\ \Tr\big[\Phi_1^{(2)}X_1+X_2\Phi_1^{(1)}\big]\,,\ \Tr\big[X_2X_1\big]\Big)\\
\mu_B \quad&\leftrightarrow\quad \Tr\big[\Phi_1^{(2)}X_1-X_2\Phi_1^{(1)}\big]\\
\Tr\,Y_\tau^k \quad&\leftrightarrow\quad \Tr\big[(\hat Q_1^{(\tau)})^k\big]\\
 \det\,Y_\tau \quad&\leftrightarrow\quad \det\,\hat Q_1^{(\tau)}\\
\det\,U \quad&\leftrightarrow\quad \det\,\big(\Phi_1^{(2)},\,X_2\big)\\
 \det\,V \quad&\leftrightarrow\quad \det\,\big(\Phi_1^{(1)},\,X_1\big)
}}
For the triplets we have listed the weight $m=+1,0,-1$ components of $U(1)_\varphi$. $\hat{Q}_1^{(\tau)}$ stands for its traceless part, as in \eqref{tracee}. The two would-be-flipped baryons $\det\,Q_1^{(\tau=1,2)}$ are the two $Y$-dibaryons, up to terms involving the decoupled traces $y_\tau$, in agreement with $n_\beta=N_N=p+q=2$.

\subsection{Case study: the conifold \texorpdfstring{$(p=1,q=0)$}{(p=1,q=0)}}
\label{sec:conifold}

\paragraph{$Y^{1,0}$ description}
For $p=1$, $q=0$ the quiver has two nodes joined by the doublet $U^\alpha$ ($1\to 2$) and the singlets $Y$ and $Z$ ($2\to 1$), where the single $Y$ field sits in the quadrangle ($Y\equiv\tilde Y_1$). The superpotential \eqref{wpq} becomes
\ba{
\label{Wy10}
W = -\epsilon_{\alpha\beta}\Tr\big[U^\alpha Y U^\beta Z\big]
}
and the charges are those of Table~\ref{tab:ypqcharge} at $p=1$, $q=0$, with $R_*=\frac12$ for every field. This is the Klebanov--Witten theory \cite{Klebanov:1998hh}, with $Y^{1,0}=T^{1,1}$. Since $Y$ and $Z$ connect the same pair of nodes, they form a doublet $B^a=(Z,Y)$ of an enhanced $SU(2)'$, and \eqref{Wy10} can be written as $W=\frac12\epsilon_{\alpha\beta}\epsilon_{ab}\Tr\big[U^\alpha B^a U^\beta B^b\big]$.

Since $p=1$, $\CS\sim\CL^+\CL^-$ is not an independent letter, and the letters $\CL^+\sim U^\alpha Z$ and $\CL^-\sim U^\alpha Y$ combine into the bi-doublet $U^\alpha B^a$ of $SU(2)_\varphi\times SU(2)'$.  Besides $\det\,U$, the other dibaryons form a single $SU(2)'$ multiplet of spin $\frac N2$ \cite{Gubser:1998fp},
\ba{
\CO_B^{a_1\cdots a_N} = \epsilon_{i_1\cdots i_N}\,\epsilon^{j_1\cdots j_N}\, (B^{(a_1})_{j_1}{}^{i_1}\cdots (B^{a_N)})_{j_N}{}^{i_N}
}
whose   highest and lowest weights are $\det\,Z$ and $\det\,Y$. The generators are listed in Table~\ref{tab:y10ring}.

\begin{table}[h!]
\centering
\renewcommand{\arraystretch}{1.3}
\begin{tabular}{l|ccc|cc|c}
\toprule
& $U(1)_B$ & $U(1)_F$ & $U(1)_{R_0}$ & $SU(2)_\varphi$ & $SU(2)'$ & $R_*$ \\ \midrule
$\CL^+ = \Tr\big[U^\alpha Z\big]$ & $0$ & $1$ & $\frac12$ & ${\bf 2}$ & \multirow{2}{*}{${\bf 2}$} & $1$ \\
$\CL^- = \Tr\big[U^\alpha Y\big]$ & $0$ & $-1$ & $\frac32$ & ${\bf 2}$ & & $1$ \\ \midrule
$\CO_B\sim Z^kY^{N-k}$, $k=0,\dots,N$ & $N$ & $2k-N$ & $N-k$ & ${\bf 1}$ & ${\bf N+1}$ & $\frac N2$ \\
$\det\,U$ & $-N$ & $0$ & $\frac N2$ & ${\bf N+1}$ & ${\bf 1}$ & $\frac N2$ \\
\bottomrule
\end{tabular}
\caption{Chiral ring generators of the $SU(N)\times SU(N)$ $Y^{1,0}$ SCFT. Note that $\CL^\pm$ are the two $SU(2)'$ components of $\Tr[U^\alpha B^a]$; they are listed separately because $U(1)_F$ and $U(1)_{R_0}$ do not commute with $SU(2)'$.  \label{tab:y10ring}}
\end{table}

\paragraph{Necklace quiver description}
We now consider the necklace of Section~\ref{sec:necklace} with $2p=2$ trinions, $\tau=1,2$. Its fields and superpotential are given in Table~\ref{tab:trinion} and Eq.~\eqref{suptau}. The fluxes $N_N=p+q=1$ and $N_S=p-q=1$ require one $\beta$-closure and one $\gamma$-closure; we give a VEV to the $\hat\gamma=-1$ inside-loop field at $\tau=1$ and to the $\hat\beta=-1$ criss-cross field at $\tau=2$,
\ba{
\label{alphalockconifold}
\langle\tilde Q_1^{(1)}\rangle = w\,\mathbf 1\,,\qquad \langle Q_2^{(2)}\rangle = v_2\mathbf 1\,,\qquad \alpha_1 = t^{-1/2}\hat\gamma\,,\qquad \alpha_2 = t^{1/2}\hat\beta^{-1}
}
The closures identify $1'\sim 2'$ and $1\sim 2'$, leaving the two nodes $N_1=(1,1',2')$ and $N_2=(2)$. The mass terms are,
\ba{
W_{\rm mass} = w\,\Tr\big[\Phi_1^{(1)}Q_1^{(1)}\big] - v_2\Tr\big[\Phi_2^{(1)}\tilde Q_1^{(2)}\big] + \Tr\Big[\Phi_2^{(2)}\big(v_2\tilde Q_2^{(2)}-w\,Q_2^{(1)}\big)\Big]
}
The following combination of fields,
\ba{
X_2 = \frac{w\,\tilde Q_2^{(2)}+v_2Q_2^{(1)}}{\sqrt{v_2^2+w^2}}
}
remains massless, and pairs with $\Phi_1^{(2)}$ into the doublet $D_2=(X_2,\Phi_1^{(2)})$ of  $SU(2)_{\hat{t}}$. The fields $Q_1^{(2)}$ and $\tilde Q_2^{(1)}$, whose baryons are flipped, run from node $N_2$ to $N_1$. The light fields are listed in Table~\ref{tab:conifold}.

\begin{table}[h!]
\centering
\renewcommand{\arraystretch}{1.2}
\begin{tabular}{l c|cccc|cccc|c}
\toprule
Field & Edge & $\hat\beta$ & $\hat\gamma$ & $\hat R$ & $\hat t$ & $B$ & $F$ & $R_0$ & $\varphi$ & $Y^{1,0}$ \\
\midrule
$Q_1^{(2)}$ & $N_2\to N_1$ & $2$ & $0$ & $\frac12$ & $0$ & $1$ & $-1$ & $1$ & $0$ & $Y$ \\
$\tilde Q_2^{(1)}$ & $N_2\to N_1$ & $0$ & $2$ & $\frac12$ & $0$ & $1$ & $1$ & $0$ & $0$ & $Z$ \\
$\Phi_1^{(2)}$ & $N_1\to N_2$ & $-1$ & $-1$ & $\frac12$ & $-1$ & $-1$ & $0$ & $\frac12$ & $\frac12$ & $U^1$ \\
$X_2\sim \tilde{Q}_2^{(2)} + Q_2^{(1)}$ & $N_1\to N_2$ & $-1$ & $-1$ & $\frac12$ & $1$ & $-1$ & $0$ & $\frac12$ & $-\frac12$ & $U^2$ \\ \midrule
$\phi_\beta^{(2)}$ & -- & $-2N$ & $0$ & $2-\frac N2$ & $0$ & $-N$ & $N$ & $2-N$ & $0$ & -- \\
$\phi_\gamma$ & -- & $0$ & $-2N$ & $2-\frac N2$ & $0$ & $-N$ & $-N$ & $2$ & $0$ & -- \\
\bottomrule
\end{tabular}
\caption{Light fields of the $(p,q)=(1,0)$ necklace quiver after one $\gamma$-closure and one $\beta$-closure, and the flip fields, with charges in the hatted necklace basis as well as the $Y^{p,q}$ quiver basis, following from Appendix~\ref{sec:master} with $(\kappa_\beta,\kappa_\gamma)=(\frac14,\frac14)$ \label{tab:conifold}. Here $\hat R$ is the reference R-symmetry of Appendix~\ref{sec:master}, which assigns $\hat R=\frac12$ to every light field.}
\end{table}

Integrating out the massive pairs and keeping terms up to quartic order gives the following low-energy superpotential,
\ba{
\label{eq:Wlowconifold}
\bs{
W_{\rm low} =&\ \frac{1}{\sqrt{v_2^2+w^2}}\,\epsilon_{\alpha\beta}\Tr\big[U^\alpha\,Q_1^{(2)}\,U^\beta\,\tilde Q_2^{(1)}\big] + W_{\rm flip}\,,\qquad U^\alpha = \big(\Phi_1^{(2)},\,X_2\big)\\
W_{\rm flip} =&\ \phi_\beta^{(2)}\det\,Q_1^{(2)} + \phi_\gamma\det\tilde Q_2^{(1)}
}}
Up to an overall normalization, this is evidently the $Y^{1,0}$ superpotential \eqref{Wy10} with
\ba{
\label{mapconifold}
Y\leftrightarrow Q_1^{(2)}\,,\qquad Z\leftrightarrow\tilde Q_2^{(1)}\,,\qquad U^\alpha\leftrightarrow\big(\Phi_1^{(2)},\,X_2\big)
}
consistent with the charges in Table~\ref{tab:conifold}. As in the affine case, $SU(2)_{\hat t}$ is restored at the fixed point. In addition, the two leftover blocks form the $SU(2)'$ doublet $B^a=(Z,Y)\leftrightarrow(\tilde Q_2^{(1)},Q_1^{(2)})$. As the $\beta$- and $\gamma$-closures treat $Q_1^{(2)}$ and $\tilde Q_2^{(1)}$ differently, the enhanced $SU(2)'$ is not manifest in the construction, although the quartic term of \eqref{eq:Wlowconifold} is invariant under it. ($W_{\rm flip}$ breaks $SU(2)'$, but the flip couplings are irrelevant for $N\geq 3$.)

Finally, consider the flip fields $\phi_\beta^{(2)}$ and $\phi_\gamma$. For $N\geq 3$, $R_*(\det\,Q_1^{(2)})=R_*(\det\tilde Q_2^{(1)})=\frac N2>\frac43$, so both flip couplings are irrelevant, and $\phi_\beta^{(2)}$ and $\phi_\gamma$ decouple as free fields while the baryons $\det\,Q_1^{(2)}$ and $\det\,\tilde Q_2^{(1)}$ remain in the chiral ring.

\paragraph{Operator map}
Under \eqref{mapconifold}, the generators of Table~\ref{tab:y10ring} map as follows:
\ba{\bs{
\CL^+ \quad&\leftrightarrow\quad \Big(\Tr\big[\Phi_1^{(2)}\tilde Q_2^{(1)}\big]\,,\ \Tr\big[X_2\tilde Q_2^{(1)}\big]\Big)\\
\CL^- \quad&\leftrightarrow\quad \Big(\Tr\big[\Phi_1^{(2)}Q_1^{(2)}\big]\,,\ \Tr\big[X_2Q_1^{(2)}\big]\Big)\\
\CO_B\sim Z^kY^{N-k} \quad&\leftrightarrow\quad \epsilon\,\epsilon\,\big(\tilde Q_2^{(1)}\big)^k\big(Q_1^{(2)}\big)^{N-k}\,,\qquad k=0,\dots,N\\
\det\,U \quad&\leftrightarrow\quad \det\big(\Phi_1^{(2)},\,X_2\big)
}}
For the doublets we list the weight $m=+\frac12,-\frac12$ components of $U(1)_\varphi$.
The extreme weights $k=N$ and $k=0$ of $\CO_B$ are the two would-be-flipped baryons $\det\tilde Q_2^{(1)}$ and $\det\,Q_1^{(2)}$, one $\gamma$-type and one $\beta$-type, in agreement with $n_\gamma=N_S=1$ and $n_\beta=N_N=1$. The intermediate weights $0<k<N$ have no counterpart among the leftover baryons of the individual closures; they are mixed determinants of the two leftover blocks, consistent with $SU(2)'$ being emergent at the fixed point.

\subsection{Case study: when flips become relevant \texorpdfstring{$(p=4,q=3)$}{(p=4,q=3)}}
\label{sec:433}

In this section we examine the $(p=4,q=3)$ SCFTs, which for various values of $N$ can either lie in regimes~{\bf A}, {\bf B}, or {\bf C}.


The base necklace  obtained by gluing $2p=8$ trinions consists of  $8$ ``outside'' nodes
$1,2,\dots,8$ and $8$ ``inside'' nodes $1',2',\dots,8'$.
In order to obtain the flux values $p=4,q=3$, we implement 7 $\beta$-closures and 1 $\gamma$-closure as follows,
\begin{itemize}
\item $\beta$-closures at $\tau = 2,\dots,8$: vev $\langle Q_2^{(\tau)} \rangle = v_\tau \mathbf 1$
      (the $\hat\beta=-1$ criss-cross field, $\tau' \to \tau-1$);
\item $\gamma$-closure at $\tau = 1$: vev $\langle \tilde Q_1^{(1)} \rangle = w\,\mathbf 1$
      (the $\hat\gamma=-1$ inside-loop field, $8' \to 1'$).
\end{itemize}
The combination of fugacities that leaves the VEV invariant is,
\begin{equation}
\label{alphalock433}
\alpha_\tau = t^{1/2}\hat\beta^{-1}\quad (\tau=2,\dots,8), \qquad \alpha_1 = t^{-1/2}\hat\gamma 
\end{equation}
The seven $\beta$-closures identify nodes $k \sim (k+1)'$ for
$k = 1,\dots,7$, {\it i.e.}\ $1\sim 2'$, $2\sim 3'$, \dots, $7\sim 8'$. The
$\gamma$-closure identifies $8' \sim 1'$. Writing
\begin{equation}
N_k = (k, (k+1)') \quad (k = 1,\dots,6),\qquad
N_7 = (7,8',1'),\qquad N_8 = (8)
\end{equation}
we obtain $8$ nodes.
After substituting the VEVs, the surviving 
fields are given in Table~\ref{tab:433fields}.

\begin{table}[t!]
\begin{center}
\begin{tabular}{l c c c c c}
\toprule
Field & Path & Merged edge & $U(1)_{\hat\beta}$ & $U(1)_{\hat\gamma}$ & $U(1)_{\hat t}$ \\
\midrule
$\tilde Q_2^{(k+1)}$, $k=1,\dots,6$ & $k \to k+1$ & $N_k \to N_{k+1}$ & $-1$ & $(-1)^k$ & $+1$ \\
$\tilde Q_1^{(k+2)}$, $k=1,\dots,6$ & $(k+1)' \to (k+2)'$ & $N_k \to N_{k+1}$ & $-1$ & $(-1)^k$ & $+1$ \\
$\tilde Q_2^{(8)}$ & $7 \to 8$ & $N_7 \to N_8$ & $-1$ & $-1$ & $+1$ \\
$Q_2^{(1)}$ & $1' \to 8$ & $N_7 \to N_8$ & $-1$ & $-1$ & $+1$ \\
$\tilde Q_1^{(2)}$ & $1' \to 2'$ & $N_7 \to N_1$ & $-1$ & $+1$ & $+1$ \\
$Q_1^{(1)}$ & $1 \to 8'$ & $N_1 \to N_7$ & $+1$ & $-1$ & $+1$ \\ \hline
$\Phi_1^{(1)}$ &  $1'\to 1$ & $N_{7}\to N_1$ & $-1$ & $+1$ & $-1$ \\
$\Phi_1^{(j)}$, $j=2,\dots,7$ &  $j'\to j$ & $N_{j-1}\to N_j$ & $-1$ & $(-1)^{j+1}$ & $-1$ \\
$\Phi_1^{(8)}$ &  $8'\to 8$ & $N_{7}\to N_8$ & $-1$ & $-1$ & $-1$ \\
$\Phi_2^{(1)}$ &  $1\to 1'$ & $N_{1}\to N_7$ & $+1$ & $-1$ & $-1$ \\
$\Phi_2^{(j)}$, $j=2,\dots,7$ &  $j\to j'$ & $N_{j}\to N_{j-1}$ & $+1$ & $(-1)^{j}$ & $-1$ \\
$\Phi_2^{(8)}$ &  $8\to 8'$ & $N_{8}\to N_7$ & $+1$ & $+1$ & $-1$ \\
\bottomrule
\end{tabular}
\caption{\label{tab:433fields} The fields of the $(p=4,q=3)$ necklace quiver obtained from gluing $2p=8$ free trinions and implementing the 7 $\beta$-closures and 1 $\gamma$-closure described in this section. In terms of the reference R-symmetry $\hat R$ of Appendix~\ref{sec:master}, the $Y$, $\tilde Y$ and $Z$ fields and both components of each $U$ doublet have $\hat R=\tfrac12$, and both components of each $V$ doublet have $\hat R=1$. }
\end{center}
\end{table}

In the merged-node basis, the eight sibling fields whose baryons  must be flipped are all bifundamentals between distinct nodes.
After shifting the fugacities according to \eqref{alphalock433}, these fields have charges,
\begin{align}
\bs{
Q_1^{(\tau=2,\dots,8)}:\quad
U(1)_{\hat\beta} = +2,\quad U(1)_{\hat\gamma} = 0,\quad
U(1)_{\hat t} = 0 \\ 
\tilde Q_2^{(1)}:\quad U(1)_{\hat\beta} = 0,\quad U(1)_{\hat\gamma} = +2,\quad
U(1)_{\hat t} = 0
}
\end{align}
Each closure introduces one gauge-singlet flip field $\phi_{\beta,\gamma}^{(\tau)}$
coupled to the baryon of the flipped field, as,
\begin{equation}
W_{\rm flip} = \sum_{\tau=2}^{8} \phi_\beta^{(\tau)} \det\,Q_1^{(\tau)} + \phi_\gamma \det\,\tilde Q_2^{(1)} 
\end{equation}
The flip-field charges are opposite to the baryon charges, with
$R[\phi] = 2 - R[\det\,Q]$:
\begin{center}
\begin{tabular}{l c c c}
\toprule
Flip field & $U(1)_{\hat\beta}$ & $U(1)_{\hat\gamma}$ & $U(1)_{\hat t}$ \\
\midrule
$\phi_\beta^{(\tau)}$, $\tau = 2,\dots,8$ & $-2N$ & $0$ & $0$ \\
$\phi_\gamma$ & $0$ & $-2N$ & $0$ \\
\bottomrule
\end{tabular}
\end{center}

\paragraph{Post-closure superpotential and fields}

Substituting the VEVs splits
$W_0$ into mass terms and cubic couplings among the remaining fields, while the
closures add $W_{\rm flip}$. The mass terms are,
\begin{equation}
\label{eq:Wmass}
\begin{aligned}
W_{\rm mass} = {}& w\,\Tr\big[\Phi_1^{(1)} Q_1^{(1)}\big]
- v_2\,\Tr\big[\Phi_2^{(1)}\tilde Q_1^{(2)}\big]
+ \sum_{j=2}^{7}\Tr\Big[\Phi_2^{(j)}\big(v_j\tilde Q_2^{(j)} - v_{j+1}\tilde Q_1^{(j+1)}\big)\Big] \\
&+ \Tr\Big[\Phi_2^{(8)}\big(v_8\tilde Q_2^{(8)} - w\,Q_2^{(1)}\big)\Big]
\end{aligned}
\end{equation}
so the following nine pairs become massive:
\begin{itemize}
\item $(\Phi_1^{(1)},\, Q_1^{(1)})$ and $(\Phi_2^{(1)},\, \tilde Q_1^{(2)})$, on $N_1 \leftrightarrow N_7$;
\item $\Phi_2^{(j)}$ with $v_j\tilde Q_2^{(j)} - v_{j+1}\tilde Q_1^{(j+1)}$, on $N_{j-1}\leftrightarrow N_j$, for $j=2,\dots,7$;
\item $\Phi_2^{(8)}$ with $v_8\tilde Q_2^{(8)} - w\,Q_2^{(1)}$, on $N_7 \leftrightarrow N_8$.
\end{itemize}
 The Higgsed fields do
not appear in the final quiver: $N^2-1$ components of each are eaten by the
broken gauge fields, and their trace modes are gauge-singlet moduli that couple
only to massive fields.

The fields $\Phi_1^{(2)},\dots,\Phi_1^{(8)}$ remain massless. On each edge $N_{j-1}\to N_j$ ($j=2,\dots,8$) the
combination orthogonal to the massive one survives,
\begin{equation}
\label{eq:Xj}
X_j = \frac{v_{j+1}\tilde Q_2^{(j)} + v_j\,\tilde Q_1^{(j+1)}}{n_j}\quad (j=2,\dots,7)\,,
\qquad
X_8 = \frac{w\,\tilde Q_2^{(8)} + v_8\,Q_2^{(1)}}{n_8}
\end{equation}
with $n_j = \sqrt{v_j^2 + v_{j+1}^2}$ and $v_9\equiv w$. The field $X_j$ has
charges $(-1,(-1)^{j+1},+1)$ and $\Phi_1^{(j)}$ has $(-1,(-1)^{j+1},-1)$, so the
two form a doublet $D_j = (X_j,\Phi_1^{(j)})$ of the $SU(2)$ whose Cartan is
$U(1)_{\hat t}$. The $22$ light fields are the $Y^{4,3}$ content, identified as:
\begin{center}
\begin{tabular}{l c c c}
\toprule
Light field & Merged edge & $(\hat\beta,\hat\gamma,\hat t)$ & $Y^{4,3}$ \\
\midrule
$D_j$, $j=2,4,6,8$ & $N_{j-1}\to N_j$ & $(-1,-1,\pm1)$ & $U$ doublets \\
$D_j$, $j=3,5,7$ & $N_{j-1}\to N_j$ & $(-1,+1,\pm1)$ & $V$ doublets \\
$Q_1^{(k)}$, $k=3,\dots,8$ & $N_k\to N_{k-2}$ & $(2,0,0)$ & $Y$ (triangle) \\
$Q_1^{(2)}$ & $N_2\to N_7$ & $(2,0,0)$ & $\tilde Y$ (quadrangle) \\
$\tilde Q_2^{(1)}$ & $N_8\to N_1$ & $(0,2,0)$ & $Z$ \\
\bottomrule
\end{tabular}
\end{center}

Solving the F-terms of the eighteen
massive fields and keeping terms up to quartic order in the light fields gives
\begin{equation}
\label{eq:Wlow}
\begin{aligned}
W_{\rm low} = {}& \sum_{k=3}^{8}\Tr\Big[Q_1^{(k)}\big(a_k\,X_{k-1}\Phi_1^{(k)} - b_k\,\Phi_1^{(k-1)}X_k\big)\Big] \\
&+ \Tr\Big[\tilde Q_2^{(1)}\big(c_2\,\Phi_1^{(2)}Q_1^{(2)}X_8 - c_1\,X_2\,Q_1^{(2)}\Phi_1^{(8)}\big)\Big]
+ W_{\rm flip}
\end{aligned}
\end{equation}
with
\begin{equation}
a_k = \frac{v_{k-1}}{n_{k-1}},\qquad b_k = \frac{v_{k+1}}{n_k},\qquad
c_1 = \frac{v_3}{w\,n_2},\qquad c_2 = \frac{v_8}{v_2\,n_8}
\end{equation}
The six cubic terms are the $UVY$ triangles
$N_{k-2}\to N_{k-1}\to N_k\to N_{k-2}$. The quartic term, the $ZU\tilde YU$ loop
$N_8\to N_1\to N_2\to N_7\to N_8$, is generated only by integrating out the two
massive pairs at the $\gamma$-closure: $(\Phi_1^{(1)},Q_1^{(1)})$ gives the
$c_1$ term and $(\Phi_2^{(1)},\tilde Q_1^{(2)})$ the $c_2$ term. The other seven
pairs simply set the massive combinations to zero.
Since
\begin{equation}
\prod_{k=3}^{8}\frac{b_k}{a_k} = \frac{c_2}{c_1}
\end{equation}
identically, the two components of each $D_j$ can be rescaled so that every
bracket in \eqref{eq:Wlow} becomes an $\epsilon$-contraction. Up to the overall
couplings,
\begin{equation}
W_{\rm low} = \sum_{k=3}^{8}\epsilon_{\alpha\beta}\,\Tr\big[Q_1^{(k)} D_{k-1}^\alpha D_k^\beta\big]
+ \epsilon_{\alpha\beta}\,\Tr\big[\tilde Q_2^{(1)} D_2^\alpha\, Q_1^{(2)} D_8^\beta\big]
+ W_{\rm flip}
\end{equation}
which is precisely the $Y^{4,3}$ superpotential, reproducing the $2q=6$ cubic $UVY$ terms and the $p-q=1$
quartic $ZUYU$ term, plus the added flip couplings.

\paragraph{Decoupling and $a$-maximization}

Whether or not the $\gamma$-flip field $\phi_\gamma$ and associated baryon $B_\gamma = \det\tilde{Q}_{2}^{(1)}\sim \det\, Z$ remain in the necklace quiver chiral ring at the interacting fixed point depends on $N$, as explained in Section~\ref{sec:mainresults}. For $N=3$, the 
$\gamma$-flip lies in Regime~{\bf C}; the  $a$-maximization of
Section~\ref{sec:ypq} gives
\ba{
\label{aunconstr}
R_*(\det\,Z) \approx 0.657 < \frac{2}{3}
}
so $\det\,Z$ violates the unitarity bound. The standard interpretation is that the operator decouples as a free field, so that $a$-maximization must be repeated after subtracting off the contribution of the free operator with R-charge $2/3$ from \eqref{atrial} \cite{Kutasov:2003iy}. Doing so, we find that the $\epsilon_B$, $\epsilon_F$ parameters and central charges are fixed as,
\ba{
\label{a433}
\epsilon_B = -0.0171\,,\qquad
\epsilon_F = 0.3378\,,\qquad
a = 14.951\,,\qquad c =15.430 
}
These can be compared with the values prior to decoupling $\det\, Z$,
$\epsilon_B = -0.0170$, $\epsilon_F = 0.3380$, $a=14.972$, and $c=15.472$.
No other operators violate the unitarity bound after the removal of  $\det\, Z$.
The flip field $\phi_\gamma$ is likewise removed from the spectrum: the coupling $\phi_\gamma \det\,Z$ is still relevant, and both fields are lifted together, so that the would-be contribution of $\det\,Z$ to the anomaly polynomial is precisely canceled by that of $\phi_\gamma$. 
This example illustrates the general statement of Regime~{\bf C}: the interacting fixed point has $\det\,Z$ (and its flip) removed, and the remaining chiral ring generators are identical to those of the $Y^{4,3}$ SCFT with the $Z$-dibaryon subtracted.

\section{Operators from Wrapped M2-branes}
\label{sec:M2}

As first shown in \cite{Bah:2022b}, the terms $\delta I_6^{\text{loc,}i}$ in \eqref{naiveloc} can be computed from M2-branes wrapping resolution 2-cycles in the associated 11d flux background. Adopting the inflow perspective provides some physical intuition here: $\sum_i \delta I_6^{\text{loc,}i}$ is equivalent to the difference of inflow anomaly polynomials for the resolved vs. the singular geometry (see ~\eqref{deltai}, and the derivation in Section~\ref{sec:inflow}). The new degrees of freedom introduced by resolving orbifold singularities with 2-cycles are precisely M2-branes wrapping these cycles. They show up in the external spacetime as point-like operators. Thus we can calculate the charges of wrapped M2-branes and directly predict the 4d QFT anomalies which would be missed in a naive reduction of its parent 6d QFT. 

In the preceding section, we discussed how this same wrapped M2-brane computation allows one to predict the requisite flip fields in the class $\mathcal{S}_2$, necklace quiver description. In this section, we focus on the significance of wrapped M2-brane operators from the $Y^{p,q}$ quiver perspective. First we review how the charges of operators arising from wrapped M2-branes can be computed, illustrating the method not only for branes wrapping resolution cycles, as in \cite{Bah:2022b}, 
but for all the 2-cycles in the internal manifold $M_6$ of the 11d flux background. Each resulting operator is identified with a corresponding chiral ring generator in the $Y^{p,q}$ quiver theories. This identification allows us to see how the generators of the chiral ring of the $Y^{p,q}$ quiver theories are realized geometrically in M-theory. 

The 11d potential $C_3$ couples to the 3d worldvolume of M2-branes. When an M2-brane wraps an internal 2-cycle, the remaining external leg of its $C_3$ contribution can support a 1-form gauge field $A_1$. Integrating the two internal legs over the 2-cycle yields the charge of the wrapped brane under this $U(1)$ gauge field. The associated point-like operator in the external spacetime will be BPS if the 2-cycle in question meets a calibration condition. For the internal space $M_6$ of interest here, the calibration conditions were derived in \cite{Gauntlett:2006ai}. 

Recall that the internal space $M_6$ is the direct product of a resolved $S^4/\mathbb{Z}_2$ space and a torus $T^2$. Adopting the conventions of \cite{Bah:2022b}, there are two $U(1)$s associated to isometries $S_\psi^1$, $S_\varphi^1$, and two additional $U(1)$s introduced by the 2-cycles resolving the orbifold singularities at the north and south poles of $S^4$. We thus have four external U(1) gauge fields $A_1^\psi$, $A_1^\varphi$, $A_1^\text{N}$, and $A_1^\text{S}$.\footnote{~The relation of this symmetry basis to the other bases discussed in this paper is given in Appendix~\ref{sec:master}; this inflow basis is also utilized in Section~\ref{sec:inflow} below.} 
The dependence of $C_3$ on these various external gauge fields can be deduced from the global angular form $E_4$ employed in the anomaly inflow computation \cite{Bah:2022b, Bah:2022a}, as the latter is a closed, equivariant completion of the field strength $G_4=dC_3$ (see Eq.~\eqref{E4res} below).  Consider for example the contribution to $E_4$ linear in the field strength $F_2^\psi=dA_1^\psi$, namely $E_4 \supset \Omega_{2,\psi} \wedge F_2^\psi$. \footnote{~In writing contributions of the external gauge fields to $E_4$ and $C_3$ we suppress $2\pi$ denominators.}  This implies that a wrapped M2-brane would couple to U(1)$_\psi$ via $\delta C_3 \supset \Omega_{2,\psi} \wedge A_1^\psi$. The charge of an M2-brane wrapping, say, the ``north'' resolution cycle under U(1)$_\psi$ can be obtained by integrating the 2-form $\Omega_{2,\psi}$ over this cycle.

Collecting together all such contributions to $\delta C_3$, we have
\begin{equation}\label{C3_expansion}
\delta C_3 \supset \Omega_{2,\psi} \wedge A_1^\psi+\Omega_{2,\varphi} \wedge A_1^\varphi+ \omega_{2,\text{N}} \wedge A_1^\text{N}+ \omega_{2,\text{S}} \wedge A_1^\text{S}
\end{equation}
Here the $\omega_{2,\text{N}}$, $\omega_{2,\text{S}}$ are 2-cohomology representatives in a particular basis, constructed to be invariant under the  $S_\psi^1$, $S_\varphi^1$ isometries and gauged via the replacements $d\psi \rightarrow d\psi+A_1^\psi$, $d\varphi \rightarrow d\varphi+A_1^\varphi$. 
There is an analogous set of 4-cohomology representatives $\Omega_{4,\alpha}$, whose equivariant completion requires the introduction of the 2-forms $\Omega_{2, \psi}$, $\Omega_{2, \varphi}$. Using $I$, $J$ to index the isometries and $\alpha$, $\beta$ to index (co)homology bases, these forms are defined by the conditions
\begin{equation}
\mathcal{L}_I \omega_{2,\alpha}=\mathcal{L}_I \Omega_{2,J}=0, \qquad d \omega_{2,\alpha}=0, \quad d \Omega_{2,I}=-2\pi \, \iota_I  \Omega_{4}
\end{equation}
where $\Omega_4=N^\alpha \Omega_{4,\alpha}$ ($N^\alpha$ are $G_4$ flux quanta on independent 4-cycles, {\it e.g.} $N$, $N_N$, $N_S$). Suitable representatives for the 2-forms $\omega_{2,\alpha}$, $\Omega_{2,I}$ can be found in Appendix \ref{sec:2form}. 
With these forms in hand, the charges of an M2-brane wrapping a given cycle $\mathcal{C}_2$ can be computed from the integrals
\begin{equation}\label{Q_def}
    Q_\alpha=N \int_{\mathcal{C}_2} \omega_{2,\alpha}, \qquad Q_I=\int_{\mathcal{C}_2} \Omega_{2,I}, \qquad I=\psi, \varphi, \quad \alpha=\text{N},\text{S}
\end{equation}

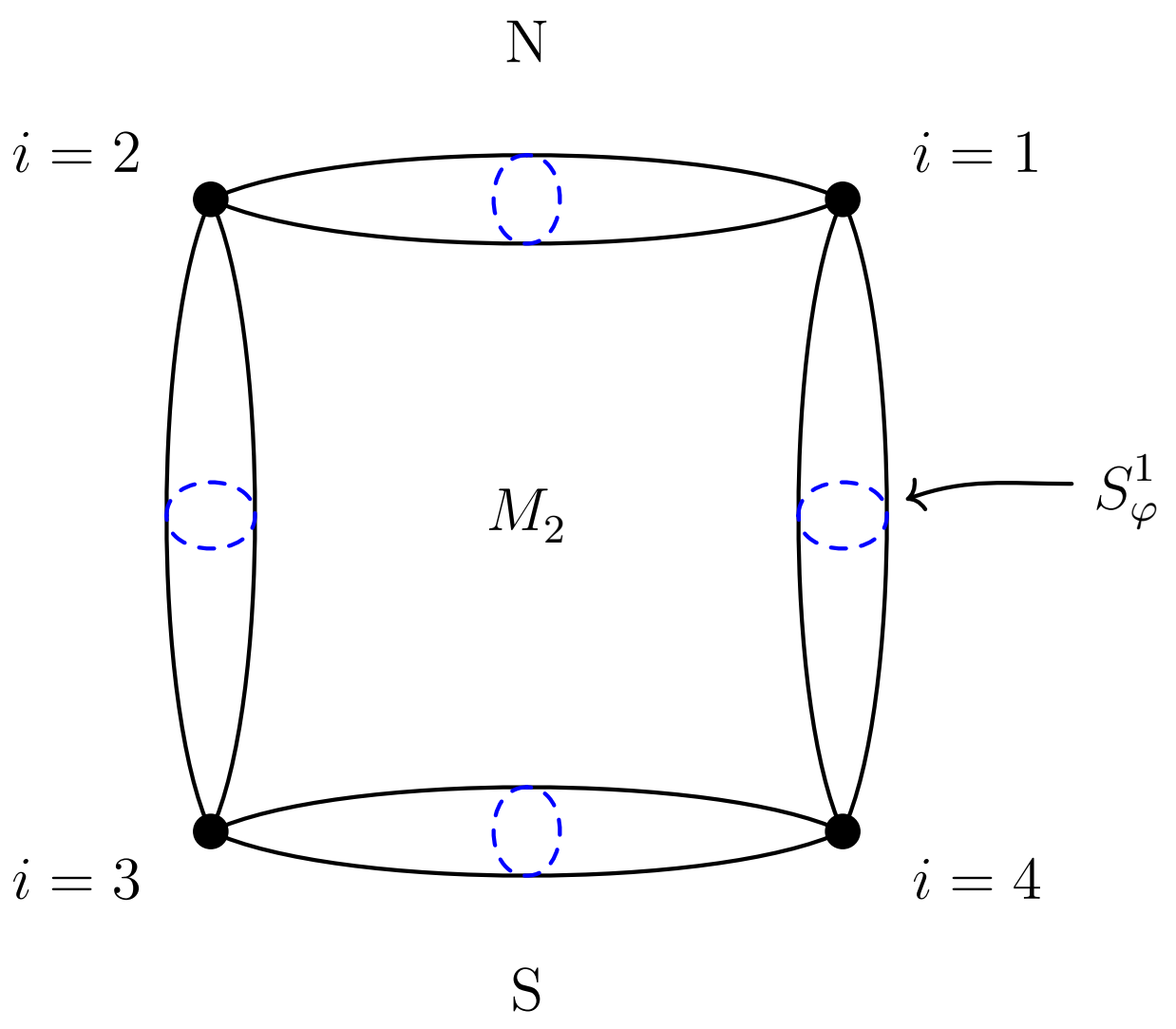
\begin{figure}\label{M4_geometry}
\centering
\begin{tikzpicture}[
    scale=1.1,
    line cap=round,
    line join=round,
    cycle/.style={black, thick},
    circleloop/.style={blue, dashed, thick},
    vertex/.style={circle, fill=black, inner sep=2.5pt}
]

\coordinate (A) at (2,2);
\coordinate (B) at (-2,2);
\coordinate (C) at (-2,-2);
\coordinate (D) at (2,-2);

\node at (0,0) {$M_2$};

\foreach \ang in {0,90,180,270} {
  \begin{scope}[rotate=\ang]
    \draw[cycle]
        (2,-2) .. controls (2.3733,-1.2) and (2.3733,1.2) .. (2,2)
               .. controls (1.6267,1.2) and (1.6267,-1.2) .. (2,-2);
  \end{scope}
}

\draw[circleloop]
    (0,2.28)
    .. controls (-0.28,2.28) and (-0.28,1.72) .. (0,1.72)
    .. controls (0.28,1.72) and (0.28,2.28) .. cycle;

\draw[circleloop]
    (0,-1.72)
    .. controls (-0.28,-1.72) and (-0.28,-2.28) .. (0,-2.28)
    .. controls (0.28,-2.28) and (0.28,-1.72) .. cycle;

\draw[circleloop]
    (-2.28,0)
    .. controls (-2.28,0.28) and (-1.72,0.28) .. (-1.72,0)
    .. controls (-1.72,-0.28) and (-2.28,-0.28) .. cycle;

\draw[circleloop]
    (1.72,0)
    .. controls (1.72,0.28) and (2.28,0.28) .. (2.28,0)
    .. controls (2.28,-0.28) and (1.72,-0.28) .. cycle;

\node[vertex] at (A) {};
\node[vertex] at (B) {};
\node[vertex] at (C) {};
\node[vertex] at (D) {};

\node at (0,3.0) {N};
\node at (0,-3.0) {S};

\node at (2.85,2.3) {$i=1$};
\node at (-2.85,2.3) {$i=2$};
\node at (-2.85,-2.3) {$i=3$};
\node at (2.85,-2.3) {$i=4$};

\node at (3.8,0.15) {$S^1_\varphi$};
\draw[->, thick]
    (3.45,0.2)
    to[out=180,in=20]
    (2.4,0.1);

\end{tikzpicture}
\caption{In the resolved $S^4/\mathbb{Z}_2$ geometry, the $\mathbb{Z}_2$ fixed points at the poles of $S^4$ are replaced  by 2-cycles between pairs of monopoles. The resulting geometry can be conveniently treated as a fibration of two circles $S_\psi^1$, $S_\varphi^1$ over a base space $M_2$. On its boundary the $S_\psi^1$ circle shrinks, but the $S_\varphi^1$ circle supports four 2-cycles, degenerating only at the four monopoles.
\label{fig:M4_geometry}}
\end{figure}

There are $4k$ total 2-cycles in the general $M_6$ space relevant for the 4d class $\mathcal{S}_k$ field theories, so eight for $k=2$. It is helpful to express the resolved $S^4/\mathbb{Z}_2$ geometry as a double fibration of $S_\psi^1$, $S_\varphi^1$ over a 2d base $M_2$, represented schematically in figure \ref{fig:M4_geometry}. The $S_\psi^1$ circle shrinks everywhere on the boundary $\partial M_2$, while $S_\varphi^1$ shrinks only at four monopoles. Locally, this is a Gibbons-Hawking space with unit-charge Kaluza-Klein monopoles. Its boundary consists of a chain of four 2-cycles. If we parameterize $\partial M_2$ with a coordinate $t$, such that the monopoles are located at $t_i$, then these boundary 2-cycles are
\begin{equation}
\mathcal{C}_2^{\partial M_i}=[t_i,t_{i+1}] \times S_\varphi^1, \qquad i=1, \cdots, 4
\end{equation}
In the conventions used here, it is the $\mathcal{C}_2^{\partial M_1}$ and $ \mathcal{C}_2^{\partial M_3}$ cycles which are inserted at the north and south poles of the $S^4$ in the blow-up procedure.
The remaining four 2-cycles consist of the torus $T^2$ at the positions $t_i$ of each monopole,
\begin{equation}
\mathcal{C}_2^{T^2_i}=T^2|_{t_i}
\end{equation}
From the standpoint of 2-homology, many of these cycles are degenerate--- in general only $2k-1$ are independent, or three in this case. But we can wrap M2-branes on all eight total 2-cycles, yielding eight distinct 4d operators. Using \eqref{Q_def} to compute the charges of each yields Table \ref{tab:M2s_11d_basis}.

\begin{table}[h!]
\centering
\renewcommand{\arraystretch}{1.2}
\begin{tabular}{c|c|c|c|c}
\toprule
Wrapped cycle & $U(1)_\psi$ & $U(1)_\varphi$ & $U(1)_\text{N}$ & $U(1)_\text{S}$ \\
\midrule
$\mathcal{C}_2^{\partial M_1}$ & $N_N$ & $0$ & $2N$ & $0$ \\
$\mathcal{C}_2^{\partial M_2}$ & $-\frac{1}{2}(N_N+N_S+N)$ & $-\frac{N}{2}$ & $-N$ & $N$ \\
$\mathcal{C}_2^{\partial M_3}$ & $N_S$ & $0$ & $0$ & $-2N$ \\
$\mathcal{C}_2^{\partial M_4}$ & $-\frac{1}{2}(N_N+N_S+N)$ & $\frac{N}{2}$ & $-N$ & $N$ \\
\midrule
$\mathcal{C}_2^{T^2_1}$ & $\frac{1}{2N}(N_S^2-N_N^2-N_NN)$ & $-\frac{1}{2}N_N$ & $-N_N$ & $-N_S$ \\
$\mathcal{C}_2^{T^2_2}$ & $\frac{1}{2N}(N_S^2-N_N^2-N_NN)$ & $\frac{1}{2}N_N$ & $-N_N$ & $-N_S$ \\
$\mathcal{C}_2^{T^2_3}$ & $\frac{1}{2N}(N_S^2-N_N^2+N_SN)$ & $-\frac{1}{2}N_S$ & $-N_N$ & $-N_S$ \\
$\mathcal{C}_2^{T^2_4}$ & $\frac{1}{2N}(N_S^2-N_N^2+N_SN)$ & $\frac{1}{2}N_S$ & $-N_N$ & $-N_S$ \\
\bottomrule
\end{tabular}
\caption{Charges of wrapped M2-branes under the four $U(1)$ gauge symmetries in 4d, in the inflow basis.}
\label{tab:M2s_11d_basis}
\end{table}
The rows for $\mathcal{C}_2^{\partial M_1}$ and $ \mathcal{C}_2^{\partial M_3}$ are equivalent under a change of basis to eq. (4.4) in \cite{Bah:2022b}, for $k=2$. Since these are the M2-branes wrapping resolution cycles, they are `blow-up modes' which are massless in the singular geometry but massive in this resolved geometry. If we use the dictionary in Appendix \ref{sec:master} to translate their charges into the $Y^{p,q}$ quiver basis, one finds they match the charges of the chiral ring generators $\text{det } Y$ and $\text{det } Z$ in Table \ref{tab:ypqring} up to an overall sign, set by a choice of orientation. As an aside, recall that the operator $\text{det } Z$ approaches the unitarity bound in the limit $q \rightarrow p$, signaling decoupling. Since $N_S=p-q$, from the 11d flux background perspective this is the degenerate limit where the $G_4$-flux wrapping one of the resolution cycles goes to zero.

Implementing the same change of basis for the operators associated to the other six wrapped M2-branes yields Table \ref{tab:M2s_Ypq_basis}. 
\begin{table}[h!]
\centering
\renewcommand{\arraystretch}{1.2}
\begin{tabular}{c|c|c|c|c|c}
\toprule
Wrapped cycle & $U(1)_B$ & $U(1)_F$ & $U(1)_{R_0}$ & $U(1)_\varphi$ & $Y^{p,q}$ quiver operator \\
\midrule
$\mathcal{C}_2^{\partial M_1}$ & $N(q-p)$ & $N$ & $-N$ & $0$ & $-\det\,Y$ \\
$\mathcal{C}_2^{\partial M_2}$ & $Np$ & $0$ & $-\frac{1}{2}N$ & $-\frac{1}{2}N$ & $-\det\,U_{\text{top}}$ \\
$\mathcal{C}_2^{\partial M_3}$ & $-N(p+q)$ & $-N$ & $0$ & $0$ & $-\det\,Z$ \\
$\mathcal{C}_2^{\partial M_4}$ & $Np$ & $0$ & $-\frac{1}{2}N$ & $\frac{1}{2}N$ & $-\det\,U_{\text{bot}}$ \\
\midrule
$\mathcal{C}_2^{T^2_1}$ & $0$ & $-p$ & $-\frac{1}{2}(p+q)$ & $-\frac{1}{2}(p+q)$ & $-\mathcal{L}^+_{\text{top}}$ \\
$\mathcal{C}_2^{T^2_2}$ & $0$ & $-p$ & $-\frac{1}{2}(p+q)$ & $\frac{1}{2}(p+q)$ & $-\mathcal{L}^+_{\text{bot}}$ \\
$\mathcal{C}_2^{T^2_3}$ & $0$ & $-p$ & $\frac{1}{2}(3p-q)$ & $-\frac{1}{2}(p-q)$ & $\mathcal{L}^-_{\text{bot}}$ \\
$\mathcal{C}_2^{T^2_4}$ & $0$ & $-p$ & $\frac{1}{2}(3p-q)$ & $\frac{1}{2}(p-q)$ & $\mathcal{L}^-_{\text{top}}$ \\
\bottomrule
\end{tabular}
\caption{Charges of wrapped M2-branes under $U(1)$ symmetries in the $Y^{p,q}$ quiver basis. The chiral ring generators with these charges are identified in the last column.}
\label{tab:M2s_Ypq_basis}
\end{table}
Note that every wrapped M2-brane yields some generator of the chiral ring of the $Y^{p,q}$ quiver theories. The top and bottom of each $SU(2)_\varphi$ multiplet arise from branes wrapping distinct 2-cycles. In particular, the top and bottom of the $\mathcal{L}^+$ multiplet are reproduced by M2-branes wrapping the torus at the $i=1,2$ monopoles, located at the ``north'' end of the geometry. Similarly the top and bottom of the $\mathcal{L}^-$ multiplet are reproduced by branes wrapping the torus at monopoles on the ``south'' end of the geometry. On the other hand, the $\text{det } U$ multiplet is reproduced instead by branes wrapping the boundary cycles connecting north and south. 

Every chiral ring generator in Table \ref{tab:ypqring} with non-trivial charges under the baryonic and flavor symmetries has a geometric realization as an M2-brane wrapping a 2-cycle in the 11d flux background.
The $\text{det } V$ field does not appear directly in Table \ref{tab:M2s_Ypq_basis}, but at the level of charges, $-\text{det } V_\text{top}$ is reproduced by an M2 wrapping the linear combination $\mathcal{C}_2^{\partial M_2}+\mathcal{C}_2^{\partial M_3}$, while $-\text{det } V_\text{bot}$ is reproduced by wrapping $\mathcal{C}_2^{\partial M_3}+\mathcal{C}_2^{\partial M_4}$. 

Thanks to T-duality, we can compare this wrapped M2-brane accounting of the generators in 11d to a wrapped D3-brane accounting in 10d, such as in \cite{Benvenuti_2005}. There the generators $\text{det } Y$, $\text{det } Z$, and $\text{det } U$ arise from D3-branes wrapping 3-cycles with topology $S^3/\mathbb{Z}_{p-q}$, $S^3/\mathbb{Z}_{p+q}$, and $S^3/\mathbb{Z}_{p}$, respectively, while $\text{det } V$ is associated to a linear combination of two of these 3-cycles, just as seen here for M2-branes wrapping 2-cycles. Note that an enumeration of the 2-cycles of $M_6$ or 3-cycles of $Y^{p,q}$ does not appear sufficient to predict the number of independent chiral ring generators (even though each admits an straightforward interpretation in terms of wrapped branes). 

\section{Anomaly Inflow and Equivariant Localization}
\label{sec:inflow}

In this section we present the equivariant localization evaluation of the anomaly inflow integral $I_6^{\text{inflow}}$, both in the singular and resolved backgrounds. The computation in the resolved background reproduces the results of \cite{Bah:2022a,Bah:2022b}, where the fiber integrals were evaluated directly rather than with a fixed point theorem. Comparing the two allows one to isolate the localized contributions $\delta I_6^{\text{loc},i}$ in \eqref{deltai}.

\subsection{The BVAB theorem and its application to class $\CS_k$}
   
 \paragraph{BVAB theorem} 
 
 The mathematical tool we will utilize is Berline-Vergne-Atiyah-Bott (BVAB) equivariant localization \cite{BerlineVergne:1982,Atiyah:1984px}. Let us first review the main formula and set our notation (we generally follow the presentation of \cite{BenettiGenolini:2023ndb}).
 
Let $M$ be a compact, oriented manifold or orbifold of dimension $2n$ without boundary, acted on by a compact torus  $\mathbb{T} = U(1)^r$.\footnote{~The theorem can be generalized to noncompact $M$ in certain cases, depending on the convergence properties of $\Phi$. See {\it e.g.}~\cite{Martelli:2023oqk}.} Denote the degree-two group parameters by $\epsilon^{a=1,\dots,r}$, and generating vector fields by $\xi_a=2\pi \partial_{\phi_a}$, where $\phi_a\sim \phi_a+2\pi$ are angles on $\mathbb{T}$. One constructs an equivariant exterior derivative,
 \ba{
 d_{\mathbb{T}} = d + \epsilon^a \iota_{\xi_a}
 }
 which squares to the Lie derivative as $d_{\mathbb T}^2 = \epsilon^a\mathcal{L}_{\xi_a}$.
 An equivariant form $\Phi$ is  a $\mathbb{T}$-invariant sum of ordinary forms of mixed degree, $\Phi = \Phi_{(0)} + \Phi_{(2)}  + \dots$, each a polynomial in the group parameters $\epsilon$. The condition for its equivariant closure is,
 \ba{
 d_{\mathbb{T}}\Phi=0\quad\Leftrightarrow\quad d\Phi_{(2j)} = -\epsilon^a \iota_{\xi_a} \Phi_{(2j+2)}
 }

The BVAB theorem reduces the integral of equivariantly closed $\Phi$ over $M$ to contributions from the fixed-point set $F$ of $\mathbb{T}$,
  \ba{
  \label{BVAB}
 \int_M \Phi = \sum_{F\subset M^\mathbb{T}}\frac{1}{|\Gamma_F|}\int_F \frac{\iota^*_F \Phi}{ e_{\mathbb{T}} (N_F)}
 }
where $F$ runs over the connected components of the fixed point locus,
$M^{\mathbb T} = \{x \in M : g\cdot x = x \ \text{for every } g\in\mathbb{T}\}$, and $\iota_F: F \hookrightarrow M$ is its embedding in $M$.  
The denominator $e_{\mathbb{T}}(N_F)$ is the
equivariant Euler class of the normal bundle,  and $|\Gamma_F|$ is the order of the local orbifold group  in cases where $F$ sits at an orbifold singularity, in which case the weights in $e_{\mathbb{T}}(N_F)$ are computed on the local covering space \cite{Meinrenken:1998} (and otherwise $|\Gamma_F|=1$).
Since the identity \eqref{BVAB} holds for all $\epsilon$,  all poles in $\epsilon$ on the right-hand-side must cancel to yield a polynomial.  The $\epsilon$ parameters may then be  promoted to 2-form background fields, $\epsilon^a\to F^a/(2\pi)$, with the isometries gauged as $D\phi^a = d\phi^a + A^a$, so that the right-hand-side outputs anomaly polynomials directly as a function of the background curvatures.

Each component of $F$ is a closed submanifold of $M$ of even codimension $2m$, and the right-hand-side of \eqref{BVAB} consists of a sum over these different components. Its normal bundle $N_F$ splits into a sum of complex line bundles $L_1\oplus \dots \oplus L_m$, so that by the Whitney sum rule the equivariant Euler class is,
\ba{
\label{eta}
e_{\mathbb{T}} (N_F) = \prod_{i=1}^m \left[ c_1(L_i) -  w_i(\epsilon) \right]
}
with torus weights $w_i$ defined by $\epsilon^a\CL_{\xi_a} z_i = 2\pi i\, w_i(\epsilon)z_i$, for  $z_i$ a fiber coordinate on $L_i$.\footnote{~The orientations of the $L_i$ fix the signs of the $w_i$, and are chosen such that the orientations of $F$ together with $L_1,\dots,L_m$ reproduce the orientation of $M$. We orient $M_4$ as in \cite{Bah:2022b}. The sign of $E_4$ is then fixed by $\int_{M_4}E_4=+N$.}
In particular,  if all fixed points $p$ are isolated ({\it i.e.}~codimension equal to $\text{dim}(M)=2n$), only the 0-form parts of $\Phi$ contribute, and the integral reduces to,
 \ba{
 \label{BVAB2}
\int_M\Phi= (-1)^n  \sum_p \frac{\Phi_{(0)}(p)}{|\Gamma_p|\prod_{i=1}^n w_i(p;\epsilon)}
}
where $w_i(p;\epsilon)$ are the oriented weights at $p$. 

 \paragraph{Application to class $\CS_k$ on a torus}
 
 In our context, the internal manifold $M_6$ is the fibration $M_4\hookrightarrow M_6 \to T^2$. Write $\mathbb{R}^5=\mathbb{C}^2\oplus \mathbb{R}$ with coordinates $(z_1,z_2,y)$, so that $M_4=S^4/\mathbb{Z}_k$ is the unit sphere $|z_1|^2 + |z_2|^2+y^2=1$ with identification $(z_1,z_2)\to (e^{2\pi i/k}z_1,e^{-2\pi i/k} z_2)$. We take $\mathbb{T}$ to be the Cartan of $SO(4)\cong (SU(2)_L\times SU(2)_R)/\mathbb{Z}_2$, which rotates the $\mathbb{C}^2$ coordinates as $z_j\to e^{i\phi_j}z_j$ while leaving $y$ fixed, and commutes with $\mathbb{Z}_k$. Denoting the equivariant parameters by $(\epsilon_1,\epsilon_2)$, we identify the Cartan of $SU(2)_R$ with the direction $\epsilon_1=\epsilon_2$, and the $U(1)_L\subset SU(2)_L$ with $\epsilon_1=-\epsilon_2$. Following \cite{Bah:2022a}, we parameterize the torus $\mathbb{T}/\mathbb{Z}_k$ that acts effectively on $M_4$ by the angles $\psi = \arg(z_1{z}_2)$ and $\varphi = \frac{k}{2} \arg(z_1\bar{z}_2)$. The corresponding equivariant parameters $\epsilon_\psi$, $\epsilon_\varphi$ and the 4d background classes are related to $(\epsilon_1,\epsilon_2)$ as (see Appendix~\ref{sec:master} for the full dictionary between bases),
 \ba{
\label{dict}
\epsilon_\psi=\epsilon_1+\epsilon_2\,,\qquad \epsilon_\varphi=\frac{k}{2}(\epsilon_1-\epsilon_2)\,,\qquad
c_1(R')=\frac12(\epsilon_1+\epsilon_2)\,,\qquad c_1(t)=-\frac12(\epsilon_1-\epsilon_2)
}

The fixed points of $M_4$ are the two poles $P=\text{N},\text{S}$, at $z_1=z_2=0$ and $y=\pm 1$, each of which is an orbifold point $\mathbb{C}^2/\mathbb{Z}_k$ with covering weights $(\epsilon_1,\epsilon_2)$, and $|\mathbb{Z}_k|=k$. 
The fixed loci of $M_6$ consist of the torus at each of these points, $F_P=P\times T^2$. These are codimension-4 components with trivial normal bundles (no $c_1(L_i)$ corrections),
so $e_{\mathbb{T}}(N_{F_P})$ is the product of the two oriented weights at $P$: $e_{\mathbb{T}}(N_{F_P})=\eta_P\,\epsilon_1\epsilon_2$, with $\eta_{\rm N}=-1$ and $\eta_{\rm S}=+1$.\footnote{\label{fn:orient}~The relative minus sign appears because the orientation of $(z_1,z_2)$ agrees with that of $M_4$ (fixed below \eqref{eta}) at S, and is opposite to it at N. With this choice the numerators of \eqref{genres} are the monopole charges $n_i=-\eta_P$ of \cite{Bah:2022a}.} 

Several simplifications occur for $\Sigma = T^2$.  
As there is no twist due to $\chi(T^2)=0$, the fibration reduces to a direct product $M_6=M_4\times T^2$, and $\mathbb{T}$ acts trivially on $T^2$. Then, the integral over $M_6$  reduces to an integral over $M_4$,
\ba{
\int_{M_6}\Phi=\int_{M_4}\Phi^\Sigma\,,\qquad \Phi^\Sigma\equiv\int_{T^2}\Phi
}
The fiber integration over $T^2$ commutes with the action of the equivariant derivative $d_{\mathbb{T}}$, hence $\Phi^\Sigma$ is equivariantly closed on $M_4$.
The codimension-4 fixed loci on $M_6$,  $F_P=P\times T^2$, reduce to isolated fixed points on $M_4$, and so we may apply \eqref{BVAB2} directly, which yields,
 \ba{
 \label{ns2}
 \int_{M_6} \Phi = \frac{1}{k \epsilon_1 \epsilon_2} \left[ \Phi_{(0)}^{\Sigma} (\rm S) - \Phi_{(0)}^{\Sigma} (\rm N)\right]\,,\qquad \Phi_{(0)}^{\Sigma}(P) = \int_{T^2} \Phi\big|_{P\times T^2}
 }
The simplified form \eqref{ns2} involves only data at isolated points. (This simplification does not occur for general genus, where $M_6$ is no longer topologically a product; one should use \eqref{BVAB} on $M_6$ to generalize to $\chi\neq0$.)

 \paragraph{Resolved background}
 
 From here on, we set $k=2$ (see Footnote \ref{foot:Phi_gen_k} for generalization to $k>2$). 
 Blowing up each pole replaces each $\mathbb{C}^2/\mathbb{Z}_2$ by the total space of $\CO(-2)\to \mathbb{P}^1$ \cite{Eguchi:1978xp}. 
The action of $\mathbb{T}$ on each exceptional $\mathbb{P}^1$ has two fixed points: $p_1\equiv $ the $z_1$-axis, and $p_2\equiv $ the $z_2$-axis. The four fixed loci of $M_6^{\rm res}$ are then the tori $p_a^P\times T^2$, for $a=1,2$ and $P=\rm N,\rm S$, and similarly to \eqref{ns2}, these loci reduce to isolated fixed points on $M_4$. Near the fixed point $p_1$ the $\mathbb{Z}_2$-invariant coordinates are $(z_2/z_1,\,z_1^2)$, and near $p_2$ they are $(z_1/z_2,\,z_2^2)$.
The weights of these fixed points $(w_1,w_2)$, defined below \eqref{eta}, are then,
\ba{
\label{EHweights}
(w_1,w_2)\big|_{p_1}=\big(\epsilon_2-\epsilon_1,\ 2\epsilon_1\big)\,,\qquad (w_1,w_2)\big|_{p_2}=\big(\epsilon_1-\epsilon_2,\ 2\epsilon_2\big)
}
The equivariant Euler class at each point $p_a^P$ is therefore $e(p_a^P)=\eta_P\,w_1w_2|_{p_a}$, with $\eta_{\rm N}=-1$, $\eta_{\rm S}=+1$ (the relative minus sign is the same as in Footnote~\ref{fn:orient})
and the integral  reduces  to a sum over isolated fixed points,
\ba{
\label{resint}
\int_{M_6^{\rm res}}\Phi=\sum_{P=\text{N,S}}\eta_P\left[\frac{\Phi^\Sigma_{(0)}(p_1^P)}{2\epsilon_1(\epsilon_2-\epsilon_1)}+\frac{\Phi^\Sigma_{(0)}(p_2^P)}{2\epsilon_2(\epsilon_1-\epsilon_2)}\right],\qquad \eta_{\rm N}=-1\,,\quad\eta_{\rm S}=+1
}
where  $\Phi^\Sigma_{(0)}(p_a^P)=\int_{T^2}\Phi\big|_{p_a^P\times T^2}$.
Note that \eqref{resint} reduces to \eqref{ns2} upon pulling back the integrand from the singular space, due to the identity
\ba{
\label{recomb}
\frac{1}{2\epsilon_1(\epsilon_2-\epsilon_1)}+\frac{1}{2\epsilon_2(\epsilon_1-\epsilon_2)}=\frac{1}{2\epsilon_1\epsilon_2}
}
Any difference between the singular and resolved evaluations, therefore, arises from differences in the integrands themselves.

It is useful to translate to the notation of \cite{Bah:2022a,Bah:2022b}, in order to compare with their results. The resolved $M_4$ is locally a multi-center Gibbons-Hawking space, expressed as a fibration $S^1_\varphi \hookrightarrow M_4 \to S^1_\psi \times M_2$, with unit-charge monopoles at positions $t_1,\dots,t_4$ on the boundary of $M_2$. The $t_i$ correspond to the four fixed points $p_a^P$ as
\ba{
t_1=p_1^{\rm N}\,,\qquad t_2=p_2^{\rm N}\,,\qquad t_3=p_2^{\rm S}\,,\qquad t_4=p_1^{\rm S}
}
The boundary intervals $(t_1,t_2)$ and $(t_3,t_4)$ on which $\partial_\psi$ degenerates correspond with the exceptional $\mathbb{P}^1$'s, while the intervals $(t_2,t_3)$ and $(t_4,t_1)$ are the 2-spheres at $\{z_1=0\}$ and $\{z_2=0\}$ that connect the north and south poles, on which $\partial_\psi \pm \partial_\varphi$ degenerates.
Using 
 \eqref{dict} at $k=2$, the weights become
\ba{
(w_1,w_2)\big|_{p_1}=\big(-\epsilon_\varphi,\ \epsilon_\psi+\epsilon_\varphi\big)\,,\qquad (w_1,w_2)\big|_{p_2}= \big(\epsilon_\varphi,\ \epsilon_\psi-\epsilon_\varphi\big)
}
Then,  \eqref{resint} is rewritten as,\footnote{\label{foot:Phi_gen_k}~The generalization to $k>2$ is,
\ba{
\label{genres}
\int_{M_6^{\rm res}}\Phi=\sum_{i=1}^{2k}\frac{n_i\,\Phi^\Sigma_{(0)}(t_i)}{(\epsilon_\varphi-\ell_{i-1}\epsilon_\psi)(\epsilon_\varphi-\ell_i\epsilon_\psi)}\,,\qquad n_i=\ell_i-\ell_{i-1}=\pm1
}
where the $2k$ fixed points are identified with monopole locations $t_i$ as $t_{j}=p_j^{\rm{N}}$ and $t_{2k+1-j}=p_j^{\rm{S}}$ for $j=1,\dots,k$, and the $i$'th interval $(t_i,t_{i+1})$ has slope $\ell_i=i-\frac k2$ for $1\le i\le k$ and $\ell_i=\frac{3k}{2}-i$ for $k<i\le 2k$, with $\ell_0\equiv\ell_{2k}$.
}
\ba{
\label{resmono}
\int_{M_6^{\rm res}}\Phi=\frac{\Phi^\Sigma_{(0)}(t_1)-\Phi^\Sigma_{(0)}(t_4)}{\epsilon_\varphi(\epsilon_\varphi+\epsilon_\psi)}+\frac{\Phi^\Sigma_{(0)}(t_2)-\Phi^\Sigma_{(0)}(t_3)}{\epsilon_\varphi(\epsilon_\varphi-\epsilon_\psi)}
}


\paragraph{Application to inflow}

  The anomalies of the 4d theories obtained by compactification of M5-branes descend from the topological couplings of M-theory. By descent, these are encoded in the 12-form characteristic class $\CI_{12}$ \cite{Freed:1998tg,Harvey:1998bx,Bah:2019rgq}, 
  \ba{
  \label{I12}
  I_6^{\text{inflow}} = \int_{M_6}\CI_{12}\,,\qquad \CI_{12} = - \frac{1}{6} E_4^3 - E_4 X_8\,,\qquad X_8 = \frac{1}{192} \left[ p_1^2(TM_{11}) -4 p_2(TM_{11}) \right] 
  }
where $E_4$ is the closed, globally defined 4-form that encodes $G_4/(2\pi)$ near the stack of M5-branes, normalized so that $\int_{M_4}E_4=N$.  Anomaly inflow requires that $I_6^{\rm inflow}$ cancel the anomaly of all degrees of freedom on the branes, including decoupled fields, which are not present for $\chi_g=0$ in the resolved background.  Both $E_4$ and $X_8$ are equivariantly closed under gauging the isometries of $M_6$, and so the integral can be evaluated using \eqref{ns2} in the singular background, or \eqref{resmono} in the resolved background. (In the singular background, \eqref{I12} is supplemented by terms localized at the orbifold loci, given in \eqref{I12sing} below.)  We next turn to the evaluation of these integrals, first in the singular background.

\subsection{Computation of {$I_6^{\text{inflow}}$} in the singular background}

Each pole of $M_4=S^4/\mathbb{Z}_2$ is an orbifold singularity, on which M-theory supports a 7d $SU(2)_{P=\rm N,S}$ vector multiplet \cite{Witten:1995ex}. 
In the flux background, the $SU(2)_P$ bundle has Chern roots $\pm b_1^P$ given in \eqref{fluxes}.\footnote{~After resolving the singularity, this becomes the contribution to the $G_4$-flux $-\eta_P\,b_1^P\wedge\omega_{2,P}$; see \eqref{E4res}.} 
The four-form flux near the brane stack, at a small radius $r=r_0$, is,
\ba{
\label{G4sing}
\frac{G_4}{2\pi}\big|_{r=r_0}=E_4=\bar E_4+f_\Sigma\wedge V_\Sigma
}
Here $\bar{E}_4$ is $N$ times the global angular form of the normal bundle, satisfying $\int_{M_4} \bar{E}_4=N$. It is independent of the fluxes $N_{\beta,\gamma}$, and  has no legs along $T^2$. The second term arises from reducing $C_3$ on $V_\Sigma$, the only harmonic 2-form of the singular $M_6$; this yields a  4d $U(1)$  gauge field with field strength denoted $f_\Sigma$, whose value is fixed below by the St\"uckelberg mechanism.

We  will need the evaluation of $\bar{E}_4$ at the poles in order to apply the fixed point formula \eqref{ns2}. Applying \eqref{BVAB2} to $\int_{M_4} \bar{E}_4=N$, together with the Bott--Cattaneo relation $\int_{M_4}\bar E_4^2=0$, fixes the evaluation of the bulk term at the poles as \cite{Hosseini:2020vgl,Hristov:2026uxs},\footnote{~Applying \eqref{BVAB2} to $\int_{M_4}\bar{E}_4=N$ fixes only the difference $\bar E_4|_{\rm S}-\bar E_4|_{\rm N}=2N\epsilon_1\epsilon_2$, since the equivariant completion of $\bar E_4$ is defined up to an additive constant. This constant is fixed by the fact that $\bar E_4$ satisfies the relation $\int_{M_4}\bar E_4^2=0$ \cite{Bott:1997va}. By \eqref{BVAB2} this requires $(\bar E_4|_{\rm S})^2=(\bar E_4|_{\rm N})^2$, thus requiring that $\bar E_4|_{\rm S}=-\bar E_4|_{\rm N}$.}
\ba{
\label{poleflux}
\bar{E}_4\big|_{\rm N} =- \bar{E}_4\big|_{\rm S}  = - N \epsilon_1 \epsilon_2 
}
To evaluate $X_8$, we note that at the fixed loci the tangent bundle $TM_{11}$ splits as,
\ba{
TM_{11}\big|_{P\times T^2}  = TW_4 \oplus TT^2 \oplus \mathbb{R}_r \oplus \mathbb{C}^2
}
where  $\epsilon_1,\epsilon_2$ are the covering weights on $\mathbb{C}^2$, and $TT^2$ and $\mathbb{R}_r$ are trivial. The total Pontryagin class is then $p(TM_{11})|_P=p(TW_4)\,(1+\epsilon_1^2)(1+\epsilon_2^2)$, so that $X_8$ is given by,
\ba{
\label{X8P}
X_8\big|_P=\frac{1}{192}\Big[p_1^2(TW_4)-4p_2(TW_4)-2p_1(TW_4)\big(\epsilon_1^2+\epsilon_2^2\big)+\big(\epsilon_1^2-\epsilon_2^2\big)^2\Big]
}
In particular, it has no legs along $T^2$.

As explained in \cite{Ohmori:2014kda}, there are additional terms in the integrand localized at the orbifold loci which are not captured by the geometry of $M_4$.
The 7d gauge fields couple to $C_3$ through the 7d coupling $C_3\wedge J_4^P$, with
\ba{
\label{J4}
J_4^P=\frac{\chi_{\mathbb{Z}_2}}{48}\Big(4c_2(SU(2)_R)+p_1(TW_4)\Big)+\frac14{\rm Tr}F_P^2\,,\qquad \chi_{\mathbb{Z}_2}=\frac32\,,\qquad \frac14{\rm Tr}F_P^2=-\big(b_1^P\big)^2
}
The boundary conditions of the 7d vector multiplets at the M5-branes also each contribute half of a 6d $\CN=(1,0)$ vector multiplet, $\tfrac{1}{2}I^{\rm vec}$. Including these contributions, the total integrand is,
\ba{
\label{I12sing}
\CI_{12}^{\rm sing}=-\frac16 E_4^3- E_4\wedge X_8-\sum_{P={\rm N,S}}\delta_P\wedge\Big(\eta_P\,E_4\wedge J_4^P-\frac12I^{\rm vec}(F_P)\Big)
}
The first two terms are \eqref{I12}, and the coefficient of $J_4^P$ is $\eta_P E_4|_P=N\epsilon_1\epsilon_2+\eta_P f_\Sigma V_\Sigma$, using \eqref{poleflux}.\footnote{~The sign $\eta_P$ arises because the 7d coupling is defined with the complex orientation of the transverse $\mathbb{C}^2/\mathbb{Z}_2$, which agrees with the orientation of $M_4$ at $P=$S and is opposite to it at $P=$N (see footnote~\ref{fn:orient}).}
Here  $\delta_P$ is the Poincar\'e dual of $P\times T^2$ in $M_6$ ({\it i.e.} the equivariant Thom class of the fixed locus), normalized so that $\int_{M_4}\delta_P=1$. Its restriction to $P\times T^2$ is $|\Gamma_P|\,e_{\mathbb{T}}(N_{P\times T^2})$, which cancels the denominator in \eqref{BVAB}, so that $\int_{M_6}\delta_P\wedge\Phi=\int_{T^2}\Phi|_{P\times T^2}$.

Our task is to evaluate the integral \eqref{ns2} over $T^2$, with integrand $\CI_{12}^{\rm sing}|_{P\times T^2}$. It is illuminating to first evaluate the integral setting $f_\Sigma=0$. The only nonzero contributions come through the $b_1^{\rm N,S}\supset N_{\beta,\gamma} V_{\Sigma}$ terms, and the result is,
\ba{
\label{singres}
\bs{
-I_6^{\text{inflow}}\Big|_{\rm sing}^{f_\Sigma=0}=&\left(N_\beta c_1(\beta)+N_\gamma c_1(\gamma)\right)\left(-\tfrac16Np_1(TW_4)-2N(N-1)c_1^2(R')+2N^2c_1^2(t)\right)\\
&+\tfrac83N^3\Big(N_\beta c_1(\beta)^3+N_\gamma c_1(\gamma)^3\Big)
}}
where we used \eqref{dict} to translate the weights $\epsilon_1,\epsilon_2$ to the background curvatures of the 4d symmetries.

To compare with \eqref{checkb}, consider the eight-form anomaly polynomial for the brane system before reducing on the torus. Recall that the inflow anomaly  must cancel the anomaly of the full brane system, including the decoupled center-of-mass tensor multiplet with its Green--Schwarz coupling \cite{Ohmori:2014kda},
\ba{
\label{I8tot}
I_8^{\rm tot}=I_8+I_8^{\rm com}\,,\qquad I_8^{\rm com}=I_8^{\rm tensor}+\frac{1}{2N}\Big(\frac14{\rm Tr}F_{N}^2-\frac14{\rm Tr}F_{S}^2\Big)^2
}
where $I_8$ is the anomaly polynomial of the interacting 6d SCFT, whose torus reduction is \eqref{checkb}.
The free tensor integrates to zero on the torus, while the Green--Schwarz term yields a nonzero contribution,
\ba{
\label{singcom}
\int_{T^2}I_8^{\rm com}=2N^2\big(c_1(\beta)^2-c_1(\gamma)^2\big)\big(N_\beta c_1(\beta)-N_\gamma c_1(\gamma)\big)
}
One verifies that \eqref{singres} equals the reduced anomaly that includes this center-of-mass sector,
\ba{
\label{singcheck}
-I_6^{\text{inflow}}\Big|_{\rm sing }^{f_\Sigma=0}=\int_{T^2}I_8+\int_{T^2}I_8^{\rm com}
}
The result \eqref{singcheck} is an anomaly polynomial that factors into a sum of two pole contributions,  with
each pole contributing a term linear in its own flux $N_\beta$, $N_\gamma$ that involves only on its own  class $c_1(\beta)$ or $c_1(\gamma)$. This factorization only arose after the Green--Schwarz contribution removed the cross-terms  $2N^2c_1(\beta)c_1(\gamma)(N_\gamma c_1(\beta)+N_\beta c_1(\gamma))$. 

This center-of-mass sector is removed by  $f_\Sigma$, which satisfies the following constraints \cite{Bah:2019vmq,Bah:2020uev,Bah:2022a}. In general, we could include a shift  $E_4\to E_4+h_4$ to  \eqref{G4sing}, with $h_4$ the field strength of the component $c_3$ of $C_3$ with all legs along the external spacetime. $I_6^{\rm inflow}$ picks up a term linear in $h_4$, wedged with a 2-form linear in the field strengths of $U(1)_\beta$, $U(1)_\gamma$, and the $U(1)$ of $f_\Sigma$. This term reflects the 5d topological mass term which gives a St\"uckelberg mass to one combination of these three $U(1)$'s.  Integrating out the massive fields imposes the $c_3$ equation of motion,
\ba{
\label{fSigmasing}
\frac{\partial(-I_6^{\rm inflow})}{\partial h_4}=0\quad\Rightarrow\quad f_\Sigma=-2\big(N_\beta c_1(\beta)-N_\gamma c_1(\gamma)\big)
}
The bulk terms linear in $f_\Sigma$ vanish; including $f_\Sigma$ with this constraint adds $\sum_P\eta_P f_\Sigma\int_{T^2}V_\Sigma\,J_4^P$ to the computation of $-I_6^{\rm inflow}$, which is precisely equal to $-\int_{T^2}I_8^{\rm com}$. The final result is,
\ba{ \label{singfinal}
-I_6^{\rm inflow}\big|_{\rm sing}=\int_{T^2}I_8
}

\subsection{Computation of $I_6^{\text{inflow}}$ in the resolved background}

In the resolved background there are no orbifold loci, and the integrand is \eqref{I12} with
\ba{
\label{E4res}
E_4=\bar E_4-\sum_{P={\rm N,S}}\eta_P\, b_1^P\wedge\omega_{2,P}+f_\Sigma\wedge V_\Sigma
}
Here $\bar E_4$ is the same bulk four-form as the singular background, which takes the values  given in \eqref{poleflux} at the fixed points.
In the second set of terms, $\omega_{2,P}$ is the  2-form of \eqref{C3_expansion} dual to $C_P$ (see the footnote in Section~\ref{sec:m5intro}), anti-self-dual with respect to the complex orientation of the patch of $\mathbb{C}^2/\mathbb{Z}_2$ near $P$, and normalized as $\int_{C_P}\omega_{2,P}=-2\eta_P$ on the exceptional curve $C_P\cong\mathbb{P}^1$ at $P$. Its equivariant completion restricts to the  fixed points as,
\footnote{\label{fn:CPorient}~These values follow from applying \eqref{BVAB} to $\int_{C_P}\omega_{2,P}=-2\eta_P$ with the tangent weights $\pm(\epsilon_2-\epsilon_1)$ of \eqref{EHweights}, where $C_P$ followed by its complex-oriented normal bundle is oriented like $M_4$ near $P$. Reversing the orientation of both $C_{P=\rm{N},\rm{S}}$ is equivalent to $c_1(R')\to-c_1(R')$.}
\ba{
\label{omres}
\omega_{2,P}\big|_{p_a^P}=-2\epsilon_a\,,\qquad \omega_{2,P}\big|_{p_a^{P'}}=0\qquad(P'\neq P)
}


The last term in \eqref{E4res} arises from reducing $C_3$ on $V_\Sigma$, which as in the singular background gives a third $U(1)$  with field strength denoted $f_\Sigma$, in addition to the two from $\omega_{2,P}$. As in the discussion around \eqref{fSigmasing}, one combination of the three $U(1)$'s gets a mass through the St\"uckelberg mechanism, and integrating out the massive fields imposes the $c_3$ equation of motion, which here reduces to $\int_{M_6}E_4^2=0$ since $\int_{M_6}X_8=0$.  As a result, $f_\Sigma$ is constrained as in  \eqref{fSigmasing} \cite{Bah:2022a,Bah:2022b}.\footnote{~This is eq.~(A.3) of \cite{Bah:2022b}, $\sum_\alpha N_\alpha F_2^\alpha=0$.}
Note that $f_\Sigma\wedge V_\Sigma$ agrees with $\frac1N\big(\frac14{\rm Tr}F_{\rm N}^2-\frac14{\rm Tr}F_{\rm S}^2\big)$, the combination in the Green--Schwarz term of $I_8^{\rm com}$, up to a constant with no legs on $M_6$ and that does not contribute to \eqref{I12}.
At the fixed points, \eqref{E4res} restricts to\footnote{~Up to a different choice of constants in the moment maps, this agrees with the four-form of \cite{Bah:2022b} (their (C.7)) at the monopoles $t_i$, after their change of basis (C.22), with a factor of $1/N$ restored in the terms of (C.22) that mix $c_1(\beta),c_1(\gamma)$ with $c_1(R')$.}
\ba{
\label{E4fixed}
E_4\big|_{p_a^P\times T^2}=\bar E_4\big|_P+2\eta_P\,\epsilon_a\,b_1^P+f_\Sigma V_\Sigma\,,\qquad \eta_P\bar E_4\big|_P=N\epsilon_1\epsilon_2
}

Inserting \eqref{E4fixed} into \eqref{I12} and applying \eqref{resint}, the integral reduces to integrals over $T^2$ at the four fixed points,
\ba{
\label{resint2}
-I_6^{\rm inflow}\big|_{\rm res}=\sum_{P={\rm N,S}}\ \sum_{a=1,2}\frac{\eta_P}{(w_1w_2)|_{p_a}}\int_{T^2}\Big(\frac16E_4^3+E_4\wedge X_8\Big)\Big|_{p_a^P\times T^2}
}
where $X_8|_{p_a}$ is given by \eqref{X8P} with $(\epsilon_1,\epsilon_2)$ replaced by the weights $(w_1,w_2)|_{p_a}$ of \eqref{EHweights}. Only terms linear in $V_\Sigma$ survive the integral over $T^2$.  The sums over the two fixed points $p_{a=1,2}$ at each pole that appear are,
\ba{
\label{fpsums}
\bs{
&\sum_a\frac{1}{w_1w_2}=\frac{1}{2\epsilon_1\epsilon_2}\,,\quad \sum_a\frac{\epsilon_a}{w_1w_2}=0\,,\quad \sum_a\frac{\epsilon_a^2}{w_1w_2}=-\frac12\,,\quad \sum_a\frac{\epsilon_a^3}{w_1w_2}=-c_1(R')\\
&\sum_a\frac{\epsilon_a\,X_8|_{p_a}}{w_1w_2}=-\frac{c_1(R')}{24}\Big[4c_1(R')^2-p_1(TW_4)\Big]
}}
with $w_1w_2$ evaluated at $p_a$. The result is,
\ba{
\label{resblocks}
\bs{
-I_6^{\rm inflow}\big|_{\rm res}=\;\sum_{P={\rm N,S}}\int_{T^2}\Bigg(&-N\epsilon_1\epsilon_2\big(b_1^P\big)^2-\frac43c_1(R')\big(b_1^P\big)^3\\
&-\frac{c_1(R')}{12}\Big[4c_1(R')^2-p_1(TW_4)\Big]b_1^P\Bigg)-\int_{T^2}I_8^{\rm com}
}}
where the $I_8^{\rm com}$ piece came from the terms involving $f_\Sigma V_\Sigma$, which give $N^2f_\Sigma\big(c_1(\beta)^2-c_1(\gamma)^2\big)=-\int_{T^2}I_8^{\rm com}$ by \eqref{fSigmasing} and \eqref{singcom}.
Using \eqref{fluxes}, this evaluates to
\ba{
\label{resres}
\bs{
-I_6^{\rm inflow}\big|_{\rm res}=\;&-2N^2N_\beta\Big(\big[c_1(R')^2-c_1(t)^2\big]c_1(\beta)+2c_1(R')\,c_1(\beta)^2\Big)+\big(\beta\to\gamma\big)\\
&-2N^2\big(c_1(\beta)^2-c_1(\gamma)^2\big)\big(N_\beta c_1(\beta)-N_\gamma c_1(\gamma)\big)\\
&-\tfrac{1}{12}\big(N_\beta+N_\gamma\big)\,c_1(R')\big[4c_1(R')^2-p_1(TW_4)\big]
}}
The first line comes from the pole terms of $E_4^3$; the second line mixing  $\beta$ and $\gamma$  comes from the center-of-mass term involving $f_\Sigma$; and the third line from $E_4\wedge X_8$. This result is precisely $\int_{T^2}I_8-I_6^{\rm flip}$, reproducing eq.~(5.5) of \cite{Bah:2022b}.  

The computation of \cite{Bah:2022b} was organized rather differently: 
they compute the integral over $M_4$  directly on its Gibbons--Hawking presentation, as a sum over the intervals $(t_i,t_{i+1})$ between monopoles on the boundary of the base $M_2$ (see Figure~\ref{fig:M4_geometry}). 
Summation by parts rewrites \eqref{resmono} as a sum over the same intervals, which agrees with theirs in total but not term-by-term.

\paragraph{Comparison}

 Comparing \eqref{resres} with \eqref{singfinal},
\ba{
\label{diffthm}
-I_6^{\rm inflow}\big|_{\rm res}=-I_6^{\rm inflow}\big|_{\rm sing}-I_6^{\rm flip}=\int_{T^2}I_8-I_6^{\rm flip}
}
Neither evaluation splits into separate contributions of the two poles, but because the $f_\Sigma$ terms are identical in the two backgrounds (both give $-\int_{T^2}I_8^{\rm com}$), the difference does. In particular, subtracting the $f_\Sigma$-independent contribution of pole $P$ in \eqref{resblocks} from that in the singular evaluation,
\ba{
\label{flipW}\bs{
\int_{T^2}\Big(& N\epsilon_1\epsilon_2J_4^P-\frac12I^{\rm vec}(F_P)\Big)\\
&-\int_{T^2}\Big(-N\epsilon_1\epsilon_2\big(b_1^P\big)^2-\frac43c_1(R')\big(b_1^P\big)^3-\frac{c_1(R')}{12}\Big[4c_1(R')^2-p_1(TW_4)\Big]b_1^P\Big)\\
&=\int_{T^2}I_8^{\rm W}\big(c_1(R')+2b_1^P\big)
}}
where $I_8^{\rm W}(x)=\frac1{24}x^4-\frac1{48}p_1\,x^2+\frac{7p_1^2-4p_2}{5760}$ is the anomaly polynomial of a 6d Weyl fermion coupled to a line bundle with first Chern class $x$.  Since $2b_1^{\rm N}$ ($2b_1^{\rm S}$) has flux $2N_\beta$ ($2N_\gamma$) through $T^2$, the index theorem implies that \eqref{flipW} describes $2N_\beta$ ($2N_\gamma$) 4d Weyl fermions with $q_{R'}=1$ and $q_\beta=2N$ ($q_\gamma=2N$). This is precisely the contribution of pole $P$ to $I_6^{\rm flip}$ in \eqref{i6flip}: the flip fields which couple to operators arising from M2-branes wrapping $C_P$ \cite{Bah:2022b}.

\section{Outlook}
\label{sec:outlook}



The technology of equivariant localization used in Section \ref{sec:inflow} could be used to compute finite-$N$ anomalies for many other wrapped brane systems, localizing $\CI_{12}$ (or the equivalent) on the fixed loci of the internal space. The most natural generalizations would be to: class $\CS_{k>2}$ on general genus-$g$ Riemann surfaces; branes probing $\mathbb{C}^2/\Gamma$ singularities, whose field theories consist of ADE conformal matter (see {\it e.g.} \cite{DelZotto:2014hpa,Kim:2018lfo}); and the reinterpretation of anomaly inflow for class $\CS$ geometries involving  regular punctures as presented in \cite{Bah:2018jrv,Bah:2019jts}, in order to generalize the recent large-$N$ equivariant localization analysis of \cite{Couzens:2025nxw,Couzens:2026qne}. 

Our results suggest that it will be fruitful to compare the reductions $\int_{\Sigma} I_8$ (or its equivalent) to inflow computations in different backgrounds, in order to gain insight into the decoupled sectors of SCFTs with no known quiver gauge theory description. The flagship example of this work enjoyed two separate quiver descriptions, which we utilized to verify our interpretation of the relation between anomaly polynomials computed from different methods. This interpretation could provide a starting point in situations where no quiver gauge theory is known, such as the spindle geometries of {\it e.g.}~\cite{Ferrero:2021wvk} (for which the large-$N$ anomalies have already been computed via equivariant localization in \cite{BenettiGenolini:2023kxp,BenettiGenolini:2023ndb}), offering a prediction of decoupled sectors and shedding light on the physics of strongly-interacting fixed points.

We also expect the 4d $\CN=1$ theories of class $\CS$ involving regular punctures
\cite{Bah:2011vv,Bah:2012dg,Gadde:2013fma,Agarwal:2014rua,Agarwal:2015vla,Nardoni:2016ffl,Fazzi:2016eec} to provide a rich set of case studies to help build out the perspective developed in this paper, as in those cases the quiver gauge theories feature iterated flips and unitarity bound crossings.

\subsection*{Acknowledgements}


We would like to thank Ibrahima Bah for interesting related conversations, and gratefully acknowledge support from the Vassar Undergraduate Research Summer Institute, which made this work possible.

{\it Statement of AI use:} Claude was used to perform checks of certain derivations, simplify steps in some computations, verify consistency across conventions, and to help create figures. All text was written by the authors, and all of this paper's contents reflect the authors’ own ideas and derivations.

\appendix



\section{Conventions}
\label{app:conventions}
In computing the anomalies of QFT  from the field content, we will use the following conventions:
\ba{
\label{Igen}
I_{d+2} =  \left[ \hat{A}(R) \text{ch}(F) \right]_{d+2}
}
where the $(d+2)$ subscript instructs us to extract the $(d+2)$-form part of the expansion of the curvatures. 
The $A$-roof genus $\hat{A}(R)$ is defined in terms of the Pontryagin classes of the spacetime tangent bundle $TW_d$ with  curvature denoted $R$, 
\ba{
\hat{A}(R)&=1 - \frac{p_1(TW_d)}{24} + \frac{7p_1(TW_d)^2-4p_2(TW_d)}{5760} + \dots \\
p_1(TW_d) &= - \frac{1}{2(2\pi)^2}\tr (R^2)\,,\qquad p_2(TW_d) = \frac{1}{8 (2\pi)^4} \left[ (\tr (R^2))^2 - 2 \tr (R^4) \right]
}
The Chern character $\text{ch}(F)$  is defined in terms of the Chern classes $c_k(G)$ of a complex $G$-vector bundle with Hermitian field strength $F$ as,
\ba{
\text{ch}(G) &= \tr_{r_i} e^{\frac{1}{2\pi} F^a T^a_{r_i} }= \text{dim}(r_i) + \frac{1}{2\pi} \tr F +\frac{1}{2(2\pi)^2}\tr(F^2)  + \frac{1}{6(2\pi)^3}\tr (F^3)  + \dots \\
c_1(G) &= \frac{\tr F}{2\pi}\,,\qquad c_2(G) =-  \frac{1}{2(2\pi)^2} \left[ \tr (F^2) - (\tr F)^2\right] \\ c_3(G) &= \frac{1}{6(2\pi)^3} \left[ 2 \tr (F^3) - 3 \tr (F^2) \tr F + (\tr F)^3 \right]
}
For example, for an $SU(N)$ bundle $\tr F_{SU(N)}=0$, so $c_1(SU(N))=0$ and $c_2(SU(N)) = -\tfrac{1}{8\pi^2} \tr F^2_{SU(N)}$. For a $U(1)$ bundle, $c_1(U(1)) = \tfrac{1}{2\pi} F_{U(1)}$. In these conventions, the variation of the effective action $\delta S$ is related to $I_{d+2}$ as,
\ba{
\delta S = 2\pi \int_{M_d} I_d\,,\qquad I_{d+2} = dI_{d+1}\,,\qquad \delta I_{d+1} = dI_d
}

\section{Relations Between 6d, Inflow, and Quiver Bases}
\label{sec:master}

The flagship examples of this paper, the 4d $\mathcal{N}=1$ SCFTs obtained from class $\mathcal{S}_{k=2}$ on $T^2$, have four continuous anomaly-free symmetries: an $SU(2)$ symmetry (which is enhanced from $U(1)$ from the perspective of the class $\CS_2$ necklace quiver); and three $U(1)$ symmetries. 
There are four natural  bases for presenting these symmetries which are mentioned throughout this work: firstly there is the basis natural from the reduction of the 6d symmetries via $\int I_8$, discussed around \eqref{6dbasis}, using labels $\{R', \beta,\gamma,t\}$; secondly there is the basis natural in the necklace quiver construction discussed in Section~\ref{sec:necklace}, which we denote $\{\hat{R}, \hat{\beta},\hat{\gamma},\hat{t}\}$; there is the basis used in the $Y^{p,q}$ quivers of Section~\ref{sec:ypq}, $\{R_0,B,F,\varphi\}$; and there is the basis used in the inflow computation of $\int \CI_{12}$ in Sections~\ref{sec:M2} and~\ref{sec:inflow}, $\{\psi, \mathrm N,\mathrm S, \varphi \}$. The last entries of these lists refer to the Cartan of the non-abelian $SU(2)$.
The relations between the first Chern classes in the different bases are given in the table below. We have defined the symbol,
\ba{
(\kappa_\beta,\kappa_\gamma) = (\tfrac{1}{4},\tfrac{1}{4} )\ \text{generically}, \quad  (\tfrac{1}{3},0)\ \text{at }p=q
}
The distinction between $p\neq q$ and  $p=q$  is an artifact of the reference R-symmetry $\hat{R}$ in the necklace quiver frame, whose assignment changes when the quartic superpotential terms disappear at $p=q$. The fluxes are related as,
\ba{
\label{fluxb}
N_N = 2 N_\beta = p + q\,,\qquad N_S = 2 N_\gamma = p-q
}

It is useful to note what these changes of basis do at the level of the symmetry
generators. Translating the relations between Chern classes into
relations between charges, the $\int I_8$, necklace, and inflow bases share the same
flavor generators up to sign and normalization,
\ba{
q_{\rm N} = q_\beta = -\hat q_\beta\,,\qquad
q_{\rm S} = -q_\gamma = \hat q_\gamma\,,\qquad
q_\varphi = -\tfrac12 q_t = -\tfrac12 \hat q_t
}
and differ only in their choice of reference R-symmetry,
\ba{
\hat q_R = q_{R'} - \kappa_\beta\, q_\beta - \kappa_\gamma\, q_\gamma\,,\qquad
q_\psi = \tfrac12 q_{R'} + \tfrac{1}{N}\big(N_\beta\, q_\beta + N_\gamma\, q_\gamma\big)
}
In particular $U(1)_{\rm N}$ and $U(1)_\beta$ are the same symmetry: the relation
$c_1(\mathrm N) = c_1(\beta) - \tfrac{2N_\beta}{N}\,c_1(R')$ in the table arises entirely because the inflow basis uses the geometric R-symmetry $U(1)_\psi$ which mixes with the flavor symmetries. The $Y^{p,q}$ basis in addition rotates the flavor generators into $(B,F)$.

\newgeometry{top=0.05in, bottom=0.02in, left=1in, right=1in}
\begin{landscape}

\begin{small}

\setlength{\tabcolsep}{0.15em} 
\renewcommand{\arraystretch}{1.2}%
\begin{tabular}{|c||l| l|l|l|}
\hline
  & \qquad $\int I_8$ & \qquad $\CS_2$ necklace & \qquad $Y^{p,q}$ quiver & \qquad $ \int \CI_{12}$ inflow 
\\ \hline \hline
 $\begin{aligned} &c_1(R') \\ &c_1(\beta)  \\  &c_1(\gamma) \\ &c_1(t)  \end{aligned}$ & 
    & 
    $\begin{aligned}  &= c_1(\hat{R}) \\  &= - c_1(\hat{\beta}) -\kappa_\beta c_1(\hat{R})   \\ &=  - c_1(\hat{\gamma})  - \kappa_\gamma c_1(\hat{R})  \\  &=  c_1(\hat{t})\end{aligned}$  & 
    $\begin{aligned}  &=  c_1(R_0) \\   &=+\tfrac{1}{2} c_1(F)  - \tfrac{(p-q) }{2} c_1(B)- \tfrac{1}{2} c_1(R_0) \\ &=  - \tfrac{1}{2} c_1(F) - \tfrac{(p+q)}{2} c_1(B) \\  &=  - \tfrac{1}{2} c_1(\varphi)\end{aligned}$   
   & $\begin{aligned}  &= \tfrac{1}{2} c_1(\psi) \\  &=  \tfrac{N_\beta}{N} c_1(\psi) +c_1(\mathrm N) \\  &=   \tfrac{N_\gamma}{N} c_1(\psi)- c_1(\mathrm S)  \\  &= - \tfrac{1}{2} c_1(\varphi) \end{aligned}$  
 \\ \hline
 $\begin{aligned} &c_1(\hat{R})  \\ &c_1(\hat{\beta})  \\ &c_1(\hat{\gamma}) \\ &c_1(\hat{t})  \end{aligned}$ &  
  $\begin{aligned}  &= c_1(R')\\  &=- c_1(\beta) -\kappa_\beta  c_1(R')   \\ &= - c_1(\gamma)-\kappa_\gamma c_1(R') \\ &= c_1(t) \end{aligned}$  & 
  & 
  $\begin{aligned}
&= c_1(R_0) \\
&= -\tfrac{1}{2} c_1(F) + \tfrac{(p-q)}{2} c_1(B)\\
&\quad
   + \big(\tfrac{1}{2} - \kappa_\beta\big)\, c_1(R_0) \\
&= +\tfrac{1}{2} c_1(F) + \tfrac{(p+q)}{2} c_1(B)
   - \kappa_\gamma\, c_1(R_0) \\
&= -\tfrac{1}{2} c_1(\varphi)
\end{aligned}$
  & 
$\begin{aligned}
&= \tfrac{1}{2} c_1(\psi) \\
&= - c_1(\mathrm N) - \big(\tfrac{\kappa_\beta}{2} + \tfrac{N_\beta}{N}\big)\, c_1(\psi) \\
&= \;\;\, c_1(\mathrm S) - \big(\tfrac{\kappa_\gamma}{2} + \tfrac{N_\gamma}{N}\big)\, c_1(\psi) \\
&= -\tfrac{1}{2} c_1(\varphi)
\end{aligned}$
  \\ \hline
 $\begin{aligned}& c_1(R_0)  \\ &c_1(B)  \\ &c_1(F) \\ &c_1(\varphi) \end{aligned}$  &  
 $\begin{aligned} &=c_1(R') \\ &=-\tfrac{1}{p} \left( \tfrac{1}{2} c_1(R') +  c_1(\beta) +c_1(\gamma)\right)   \\ &= \tfrac{(p+q)}{p} \left(\tfrac{1}{2}  c_1(R') + c_1(\beta)\right) - \tfrac{(p-q)}{p} c_1(\gamma)\\ &= -2c_1(t) \end{aligned}$  &   
$\begin{aligned}
&= c_1(\hat R) \\
&= \tfrac{1}{p}c_1(\hat\beta) + \tfrac{1}{p} c_1(\hat\gamma)\\
&\quad 
   + \tfrac{1}{p}\big(\kappa_\beta + \kappa_\gamma - \tfrac{1}{2}\big)\, c_1(\hat R) \\
&= 
       - \tfrac{p+q}{p}\, c_1(\hat\beta) + \tfrac{p-q}{p}\, c_1(\hat\gamma)\\
       &\quad +\left( \tfrac{(p+q)}{2p}\big(1-2\kappa_\beta\big)
               + \tfrac{(p-q)}{p}\,\kappa_\gamma \right)\, c_1(\hat R)\\
&= -2\, c_1(\hat t)
\end{aligned}$&
   &  
$\begin{aligned}
&= \tfrac{1}{2} c_1(\psi) \\
&= -\big(\tfrac{1}{4p} + \tfrac{1}{N}\big) c_1(\psi) - \tfrac{1}{p} c_1(\mathrm N) \\
&\quad + \tfrac{1}{p} c_1(\mathrm S) \\
&= \big(\tfrac{p+q}{4p} + \tfrac{2q}{N}\big) c_1(\psi) + \tfrac{p+q}{p} c_1(\mathrm N) \\ &\quad + \tfrac{p-q}{p} c_1(\mathrm S) \\
&= c_1(\varphi)
\end{aligned}$
  \\ \hline
 $\begin{aligned} &c_1(\psi) \\ &c_1(\mathrm N)  \\ &c_1(\mathrm S) \\ &c_1(\varphi)  \end{aligned}$  &  
 $\begin{aligned} &= 2 c_1(R') \\  &= c_1(\beta) - \tfrac{2 N_\beta}{N} c_1(R') \\ &= - c_1(\gamma) + \tfrac{2N_\gamma}{N} c_1(R')\\ &=  -2c_1(t)\end{aligned}$ &  
$\begin{aligned}
&= 2\, c_1(\hat R) \\
&= -\big(\kappa_\beta + \tfrac{2N_\beta}{N}\big)\, c_1(\hat R) - c_1(\hat\beta) \\
&= \;\;\,\big(\kappa_\gamma + \tfrac{2N_\gamma}{N}\big)\, c_1(\hat R) + c_1(\hat\gamma) \\
&= -2\, c_1(\hat t)
\end{aligned}$&   
$\begin{aligned}
&= 2 c_1(R_0) \\
&= \tfrac{1}{2} c_1(F) - \tfrac{(p-q)}{2} c_1(B) \\
&\quad - \big(\tfrac{1}{2} + \tfrac{p+q}{N}\big) c_1(R_0) \\
&= \tfrac{1}{2} c_1(F) + \tfrac{(p+q)}{2} c_1(B) + \tfrac{(p-q)}{N} c_1(R_0) \\
&= c_1(\varphi)
\end{aligned}$
 & 
 \\ \hline  
\end{tabular}

\end{small}

\end{landscape}
\restoregeometry

\section{2-form Representatives}
\label{sec:2form}
This appendix addresses the construction of suitable representatives of the 2-forms $\Omega_{2,I}$ and $\omega_{2,\alpha}$ discussed in Section \ref{sec:M2}, as well as how they can be used to compute the charges of M2-branes wrapping 2-cycles in the internal space $M_6$. The results are summarized in Table \ref{tab:M2s_11d_basis}. We generally adopt the conventions of appendix C in \cite{Bah:2022b}.

\paragraph{Parameterization of $\Omega_{2,I}$} Representatives of the 2-forms $\Omega_{2,I}$ can be written
\begin{align} \label{eq:2-forms}
\Omega_{2,\psi}&=(dX-L T_1) \wedge d\psi -T_1 \wedge D\varphi +U_\psi V_\Sigma \\
\Omega_{2,\varphi} &=(dU_\varphi +L dY+T_1) \wedge d\psi+dY \wedge D\varphi +U_\varphi V_\Sigma \nonumber
\end{align}
where $X$, $Y$, $U_\psi$, $U_\varphi$, and $L$ are scalar functions of the coordinates on $M_2$, and $T_1$ is a 1-form on this space. We have suppressed $2 \pi$ denominators everywhere the 1-forms $d\psi$ and $D\varphi$ appear. Here $D\varphi=d\varphi-L \, d\psi$, with $L$ piecewise-constant on the boundary of $M_2$, changing by one across each monopole. In particular, for $k=2$ and the indexing in Figure \ref{fig:M4_geometry},
\begin{equation}
L |_{[ t_i, t_{i+1} ]}\equiv \ell_i= 0,1,0,-1 \quad \text{for } i=1,2,3,4
\end{equation}
The 2-forms \eqref{eq:2-forms} are constructed in the equivariant completion of 4-cohomology representatives $\Omega_4^\alpha$, which in this parameterization satisfy
\begin{equation}
\Omega_4 \equiv N_\alpha \Omega_4^\alpha =dT_1 \wedge d\psi \wedge D\varphi +(dU_\psi+L dU_\varphi)\wedge d\psi \wedge V_\Sigma+dU_\varphi \wedge D\varphi  \wedge V_\Sigma
\end{equation}
While the precise dependence of these various scalar functions on the coordinates of $M_2$ is under-determined, we can constrain their values at the monopoles.
The integral of $\Omega_4$ over the various 4-cycles gives the $G_4$-flux quanta 
\begin{equation}\label{eq:flux_quanta}
N = \int_{M_4} \Omega_4= \int_{\partial M_2}T_1,  \quad N_i=\int_{\mathcal{C}_4^{i}} \Omega_4= \int_{t_i}^{t_{i+1}} dU_\varphi
\end{equation}
where $\mathcal{C}_4^i=\mathcal{C}_2^{\partial M_i} \times T^2$. Abbreviating $U_\varphi(t_i)$, as $U_\varphi^i$, we see that $U_\varphi^{i+1}-U_\varphi^i=N_i$. The $U_\varphi^i$ are thus determined up to a single reference value,
\begin{equation}
U_\varphi^i=U_\varphi^1+\sum_{j=1}^{i-1} N_j 
\end{equation}
Since $S_\psi^1$ shrinks on $\partial M_2$, any contribution to $\Omega_4$ with a $d\psi$ leg must vanish on each boundary segment. This regularity condition imposes further constraints. In particular, we must have
\begin{equation}
(dU_\psi+L dU_\varphi)|_{\partial M_2}=0 \quad \Rightarrow \quad U_{\psi}^{i+1}-U_\psi^i=\ell_i(U_\varphi^{i+1}-U_\varphi^i)=\ell_i N_i
\end{equation}
Thus the $U_\psi^i$ are similarly determined up to a reference value,
\begin{equation}
		U_\psi^i=U_\psi^1-\sum_{j=1}^{i-1} \ell_j N_j
\end{equation}
We fix these reference values as in \cite{Bah:2022a, Bah:2022b} by imposing the normalizations
\begin{equation}
	\int_{M_6}  \Omega_4 \wedge \Omega_{2,\psi}=\int_{M_6}  \Omega_4 \wedge \Omega_{2,\varphi}=0
\end{equation}
As described in \cite{Bah:2022a}, the geometry of $M_6$ leads to various sum rules in homology, which in turn constrain the flux quanta. In the $k=2$, $\Sigma = T^2$ case,
\begin{equation}
	\sum_{i=1}^4 N_i=0, \qquad  \sum_{i=1}^4 \ell_i N_i=0
\end{equation}
Eliminating $N_2$, $N_4$ to express our answers in terms of the north and south ``resolution fluxes" $N_N \equiv N_1$, $N_S \equiv N_3$, one finds
\begin{align}
	U_\psi^1=U_\psi^2=\frac{N_S^2-N_N^2-N_N N}{2N}, & \qquad U_\psi^3=U_\psi^4=\frac{N_S^2-N_N^2+N_S N}{2N},
	\\
	U_\varphi^1= - U_\varphi^2 = -\frac{N_N}{2}, & \qquad U_\varphi^3=-U_\varphi^4=-\frac{N_S}{2}
\end{align}
From the requirement that the 2-forms $\Omega_{2,I}$ are regular on $\partial M_2$, we have also that 
\begin{equation}\label{eq:2-form_reg}
	(dX- L T_1)|_{\partial M_2}=0, \qquad (dU_\varphi+L dY+T_1)|_{\partial M_2}=0 
\end{equation}
Integrating over $\partial M_2$ and using \eqref{eq:flux_quanta}, the latter implies 
\begin{equation}
	N=\sum_{i=1}^4 \ell_i (Y^{i+1}-Y^i)
\end{equation}
A convenient and consistent choice of the values $Y^i$ is
\begin{equation} \label{eq:Y_vals}
	Y^1=Y^2=\frac{N}{4}, \quad Y^3=Y^4=-\frac{N}{4}
\end{equation}
The integrals of $T_1$ can then be evaluated over the intervals $[t_i,t_{i+1}]$ with the aid of \eqref{eq:2-form_reg},
\begin{equation}
-\int_{i}^{i+1} T_1=(U_\varphi^{i+1}-U_\varphi)+\ell_i (Y^{i+1}-Y^i)
\end{equation}

\paragraph{M2 charges under isometries} recall we can compute charges of wrapped M2-branes under the $S_\psi^1$, $S_\varphi^1$ isometries from
\begin{equation}
	Q_I=\int_{\mathcal{C}_2}\Omega_{2,I}, \quad I=\psi, \varphi
\end{equation}
Comparing to \eqref{eq:2-forms}, we can write these wrapped M2-brane charges as follows
\begin{align}
	\text{M2 wrapping } [t_i,t_{i+1}] \times S_\varphi^1 \, & : \qquad Q_\psi^{\partial M_i}=-\int_{i}^{i+1} T_1
	, \qquad Q_\varphi^{\partial M_i}=\int_i^{i+1} dY_{}
	\\
	\text{M2 wrapping } T^2|_{t_i} \, & : \qquad Q_\psi^{T^2_i}=U_\psi^i \qquad Q_\varphi^{T^2_i}=U_\varphi^i
\end{align}
For example, an M2-branes wrapping the cycle $\mathcal{C}_2^{\partial M_i}=[t_i,t_{i+1}] \times S_\varphi^1$ will carry charge under the $S_\varphi^1$ isometry
\begin{equation}
Q_\varphi^{\partial M_i}=\int_i^{i+1} dY_{}=Y^{i+1}-Y^i
\end{equation}
Using \eqref{eq:Y_vals}, we find $Q_\varphi^{\partial M_1}=Q_\varphi^{\partial M_3}=0$ and $Q_\varphi^{\partial M_2}=Q_\varphi^{\partial M_4}=-N/2$. The charges $Q_\psi$, $Q_\varphi$ of M2-branes wrapping all eight 2-cycles in $M_6$ can be found in Table \ref{tab:M2s_11d_basis}.

\paragraph{Parameterization of $\omega_{2,\alpha}$} 2-cohomology representatives on $M_6$ can be written
\begin{equation}\label{eq:2-cohom_rep}
	\omega_{2,\alpha}=\left(dW_{\alpha,\psi}+L dW_{\alpha,\varphi} \right) \wedge d\psi+dW_{\alpha, \varphi} \wedge D\varphi+W_{\alpha,\Sigma} V_\Sigma
\end{equation}
for some scalar functions $W_{\alpha,\psi}$, $W_{\alpha,\varphi}$, $W_{\alpha,\Sigma}$ on $M_2$, such that 
\begin{equation}
	(dW_{\alpha,\psi}+L dW_{\alpha,\varphi})|_{\partial M_2}=0 
\end{equation}
For $k=2$ there are three independent elements each in the 4-homology and 2-homology of $M_6$. As in \cite{Bah:2022a,Bah:2022b}, we adopt a 4-homology basis labeled by  $\alpha = \{ \text{N}, \text{S}, \text{C} \}$, such that 
\begin{equation}
\mathcal{C}_4^\text{N}=\mathcal{C}_2^{\partial M_1} \times T^2, \qquad \mathcal{C}_4^\text{S}=\mathcal{C}_2^{\partial M_3}\times T^2, \qquad \mathcal{C}_4^\text{C}=M_4
\end{equation}
Normalization of the 2-(co)homology basis and Poincar\'e duality,
\begin{equation}
	\int_{\mathcal{C}_2^\alpha} \omega_{2,\beta}= \delta_{\alpha,\beta}, \qquad \int_{M_6} \Omega_{4,\alpha} \wedge \omega_{2,\beta}= \delta_{\alpha,\beta}
\end{equation}
can be used to fix the values of the functions $W_{\alpha,\psi}$, $W_{\alpha,\varphi}$ at the monopoles as
 \begin{align}\label{eq:W_vals}
	W_{\alpha,\psi}^1=W_{\alpha,\psi}^2&=-\delta_{\alpha}^\text{N}, \qquad W_{\alpha,\psi}^3=W_{\alpha,\psi}^4=-\delta_{\alpha}^\text{S},
	\\
	W_{\alpha,\varphi}^1=-W_{\alpha,\varphi}^2&=-\delta_{\alpha}^\text{N}, \qquad W_{\alpha,\varphi}^3=-W_{\alpha,\varphi}^4=\delta_{\alpha}^\text{S},\nonumber
	\\
	&W_{\alpha, \Sigma}^i =\delta_\alpha^\text{C} \quad \text{for all } i\nonumber
\end{align}

\paragraph{M2 charges under flavor symmetries} as discussed in Section \ref{sec:M2} the charge of an M2-brane wrapping a 2-cycle $\mathcal{C}_2$ under the 4d field strengths $F_2^\alpha$ is
\begin{equation}
    Q_\alpha=N\int_{\mathcal{C}_2} \omega_{2, \alpha}, \qquad \alpha=\text{N, S, C}
\end{equation}
Using \eqref{eq:2-cohom_rep}, the possible charges are related to the parameterization of the $\omega_{2,\alpha}$ as below.
\begin{align}
	\text{M2 wrapping } [t_i,t_{i+1}] \times S_\varphi^1 \, & : \qquad Q_\alpha^{\partial M_2^i}=N \int_i^{i+1} dW_{\alpha, \varphi}
	\\
	\text{M2 wrapping } T^2|_{t_i} \, & : \qquad Q_\alpha^{T^2_i}=N W_{\alpha,\Sigma}^i
\end{align}
So for M2-branes wrapping the 2-cycles on the boundary of $M_2$, $Q_\alpha^{\partial M_2^i}=N (W_{\alpha,\varphi}^{i+1}-W_{\alpha,\varphi}^i)$, which we can evaluate using \eqref{eq:W_vals}
$$Q_{\alpha}^{\partial M_2^1}=2N \delta_\alpha^\text{N}, \qquad Q_{\alpha}^{\partial M_2^2}=Q_{\alpha}^{\partial M_2^4}=N(\delta_\alpha^\text{S}- \delta_\alpha^\text{N}), \qquad Q_{\alpha}^{\partial M_2^3}=-2N \delta_\alpha^\text{S}$$
For branes wrapping the torus at each monopole $t_i$ we find $Q_\alpha^{T^2_i}=N \delta_\alpha^\text{C}$. However, as discussed in \cite{Bah:2022a}, a St\"uckelberg mechanism leads to a constraint relating the field strengths $F_2^\alpha$, namely $N_\alpha F_2^\alpha=0$, or $
    F^\text{C}_2=-\frac{1}{N}\left( N_N F_2^\text{N}+N_S F_2^\text{S}\right) $
Thus we can alternately express these charges
$$Q_\alpha^{T^2_i}=-(N_N \delta_\alpha^\text{N}+N_S \delta_\alpha^\text{S} ) \quad \text{for all } i$$


\bibliographystyle{utphys}
\bibliography{ursi2026}


\end{document}